# Raisins in a Hydrogen Pie: Ultrastable Cesium and Rubidium Polyhydrides

Di Zhou[1, †, *], Dmitrii Semenok[1, †, *], Michele Galasso[2], Frederico Gil Alabarse[3], Denis Sannikov[6], Ivan A. Troyan[4], Yuki Nakamoto[5], Katsuya Shimizu[5], and Artem R. Oganov[6,*]

[1] Center for High Pressure Science & Technology Advanced Research, Bldg. #8E, ZPark, 10 Xibeiwang East Rd, Haidian District, Beijing, 100193, China

[2] Institute of Solid State Physics, University of Latvia, 8 Kengaraga str., LV-1063 Riga, Latvia

[3] Elettra Sincrotrone Trieste, Trieste 34149, Italy

[4] Shubnikov Institute of Crystallography, Federal Scientific Research Center Crystallography and Photonics, Russian Academy of Sciences, 59 Leninsky Prospekt, Moscow 119333, Russia

[5] KYOKUGEN, Graduate School of Engineering Science, Osaka University, Machikaneyamacho 1-3, Toyonaka, Osaka, 560-8531, Japan

[6] Skolkovo Institute of Science and Technology, 121205, Bolshoy Boulevard 30, bld. 1, Moscow, Russia

[†] These authors contributed equally to this work

Corresponding authors: Di Zhou (di.zhou@hpstar.ac.cn), Dmitrii Semenok (dmitrii.semenok@hpstar.ac.cn), Artem R. Oganov (a.oganov@skoltech.ru)

## Abstract

We proposed a new method for synthesis of metal polyhydrides via high-pressure thermal decomposition of corresponding amidoboranes in diamond anvil cells. Within this approach, we synthesized molecular semiconducting cesium ($P4/nmm$-$CsH_7$, $P1$-$CsH_{15+x}$) and rubidium ($RbH_{9-x}$) polyhydrides with a very high hydrogen content reaching 93 at%. Preservation of $CsH_7$ at near ambient conditions, confirmed both experimentally and theoretically, represents a significant advance in the stabilization of hydrogen-rich compounds. In addition, we synthesized two crystal modifications of $RbH_{9-x}$ with pseudo hexagonal and pseudo tetragonal structures identified by synchrotron X-ray diffraction and Raman measurements. Both phases are stable at 8-10 GPa. This is an unprecedented low stabilization pressure for polyhydrides. These discoveries open up possibilities for modifying existing hydrogen storage materials to increase their efficiency.

## Introduction

Stabilizing polyhydrides is crucial to developing hydrogen batteries and environmentally friendly vehicles based on them. [1]. Preservation of a H-rich shell can be achieved by the local electric field of a metal atom in neutral and charged clusters such as $ThH_5^-$ [2], $LaH_8^-$ [3] (Figure 1a, c), in organic complexes due to field of ligands (Figure 1b), and via applying external pressure to metal polyhydrides, for instance, $YH_6$ [4] and $CeH_{9-10}$ [5] (Figure 1d). Frankly speaking, metal polyhydrides could be a perfect material for hydrogen storage batteries, but the synthesis of most of them requires using of ultrahigh pressure of millions of atmospheres. When pressure is reduced, polyhydrides usually decompose irreversibly. However, there is an exception to this rule: molecular polyhydrides of alkali and alkaline earth elements are much more stable and require only tens of GPa for their synthesis [6,7]. The aim of this paper is to examine this important exception.

In general, the synthesis and stabilization pressure of polyhydrides decreases with increasing ionization potential and atomic radius of alkali metal atoms. At least four insulating lithium hydrides were synthesized by compression of LiH with hydrogen up to 130–182 GPa including theoretically predicted $LiH_2$ and $LiH_6$ [8,9]. The formation of sodium polyhydrides already requires significantly lower pressures. Compression of NaH and hydrogen to 40 GPa with consequent laser heating leads to formation of $NaH_3$ and molecular $NaH_7$ [10]. The later phase contains $H^-$, linear and bent $H_3^-$, and $H_2$ units at the same time. Formation of $KH_5$ and $RbH_9$ under relatively low pressure was also discussed [11,12]. Continuing with this series, as far as we know, no experimental attempts have been made to synthesize Cs polyhydrides. Insufficient experimental research on Cs and Rb hydrides prompted us to take on this work. Moreover, Cs and Rb are especially convenient for studies under pressure since they are rather heavy elements and crystal structure of their hydrides can be easily determined using X-ray diffraction.

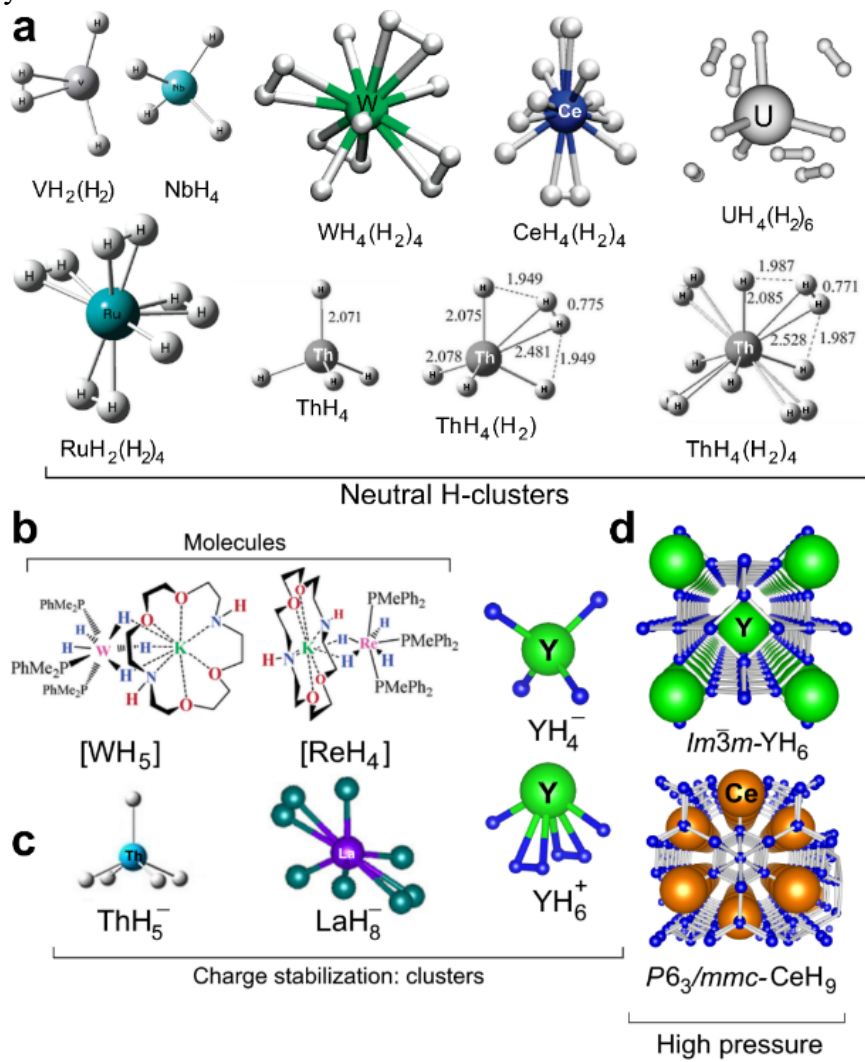

**Figure 1.** Examples of polyhydrides stabilization: (a) in neutral clusters (gas phase) [13-18], (b) in complex organic compounds (solutions) [19], (c) in charged clusters (plasma) [2,3], (d) in three-dimensional crystal structures of polyhydrides stabilized under high pressure [20].

Cesium and rubidium are very active alkaline metals, which, according to ab initio calculations, react with $H_2$ to form polyhydrides already at low pressures of 10-30 GPa [21]. Theoretical crystal structural searches show that $RbH_5$, $RbH_9$, $CsH_3$, $CsH_7$, and $CsH_9$ are the most thermodynamically stable phases at high pressures [21-23]. Great chemical reactivity of these metals and their monohydrides (CsH, RbH), low melting points of Cs and Rb, and high surface tension in the liquid state, seriously complicate study of the reactions of Cs and Rb with hydrogen. There is a popular method for the synthesis of polyhydrides in diamond anvil cells (DACs): the heating of metal precursors with ammonia borane ($NH_3BH_3$, AB), which at high temperatures serves as a source of hydrogen. In fact, this method cannot be directly applied to Cs and Rb, since both of them react violently with $NH_3BH_3$ to form corresponding amidoboranes: CsAB and RbAB [24]. Remarkably, these amidoboranes do not undergo further reactions with metallic Cs and Rb and can be used as both hydrogen and metal sources for high-pressure high-temperature synthesis of polyhydrides.

We proposed a new approach to metal polyhydrides via the thermal decomposition of their amidoboranes in high-pressure DACs. The applicability of this method is illustrated by the examples of Cs and Rb polyhydrides. New $CsH_7$, $CsH_{15+x}$ and two modifications of $RbH_{9-x}$ were synthesized at pressures of 10-20 GPa. The later phases remain stable below 10 GPa, while $CsH_7$ can be decompressed to near-ambient pressure. These properties of cesium and rubidium polyhydrides open up possibilities for modifying existing hydrogen storage materials to increase their efficiency.

## Results

*Synthesis of cesium polyhydrides: CsH$_7$ and CsH$_{15+x}$*

Ammonia borane is a very convenient hydrogen source for synthesis of polyhydrides in DACs [20]. Cs and Rb can easily react with ammonia borane to form transparent salts called amidoboranes. We used two methods to synthesize amidoboranes at ambient temperature and pressure in an argon glovebox: (1) solid state mixing Cs or Rb with AB; (2) reaction of both metals with saturated solution of ammonia borane in tetrahydrofuran (THF). The solid-state reaction is accompanied by excessive heating and yields the high-temperature (HT) crystal modifications of CsAB and RbAB, known from literature [24]. Whereas the synthesis in THF solution leads to the low-temperature (LT) modifications of corresponding amidoboranes. Raman spectra of obtained amidoboranes can be found in Supporting Figures S9-10, S13. A more controlled reaction in THF solution is preferred.

As we observed, ultraviolet light (365 nm UV laser, see Supporting Figures S6-S8) cannot decompose amidoboranes of alkali metals with release of hydrogen, therefore we used infrared (1.04 μm) laser heating to promote decomposition of amidoboranes using gold or Cs/Rb as an absorber. Amidoboranes CsAB and RbAB are transparent in the visible and near-IR range, which makes it impossible to continue laser heating as soon as the metal particle has reacted. To improve the laser heating process, we added a gold foil as an absorber of IR laser radiation. Follow-up Raman spectroscopy indicates that laser heating of CsAB and RbAB leads to their decomposition with emission of hydrogen (Supporting Figures S11, S15).

In the first experiment, the Cs/LT-CsAB mixture was loaded into a Mao-type symmetric diamond anvil cell DAC X1 with an anvil culet diameter of 250 μm and a Re gasket opening of 150 μm (Figure 2b). After increasing the pressure to 41 GPa, intense laser heating of the sample was carried out (see [25]). As a result, almost all metallic Cs has reacted. After laser heating, we examined the resulting products using synchrotron X-ray diffraction (XRD) (Figure 2a). Comparison of XRD patterns (Figure 2a, Supporting Figure S33) allowed to detect characteristic diffraction pattern of cesium monohydride *Cmcm*-CsH and diffraction peaks of the new compound, *P4/nmm*-CsH$_7$, predicted as a part of evolutionary crystal structure search for stable Cs-H phases at 30 GPa (Figure 2e) [26-28]. In the following decompression experiment, CsH undergoes a phase transformation to $Pm\bar{3}m$-CsH below 15 GPa, in accordance with the literature data [29]. Experimental unit cell parameters of CsH and CsH$_7$ are given in Supporting Tables S2, S3. The ratio of the CsH and CsH$_7$ is close to 1:1 (Figure 2a), which speaks in favor of a probable disproportionation reaction:

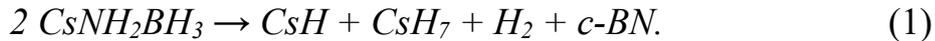

$$2\ CsNH_2BH_3 \rightarrow CsH + CsH_7 + H_2 + c\text{-}BN. \qquad (1)$$

Hydrogen produced in this reaction can in some cases be detected by Raman spectroscopy (Supporting Figures S11, S15), but in many other cases, it reacts with cesium or its hydrides forming new compounds (e.g., CsH$_{15+x}$).

Tetragonal *P4/nmm*-CsH$_7$ (Figure 1c) is a typical molecular-ionic polyhydride with a bandgap of about 1.77 eV at 20 GPa (Supporting Figure S38). Its unit cell contains one H$^-$ ion and three hydrogen molecules ($d_{HH}$ = 0.81 Å at 30 GPa). Presence of isolated hydride anions indicates a high potential ionic conductivity of the compound as observed for strontium [7], barium [30] and chlorine hydrides [31]. Cs-sublattice in this compound is deformed $Pm\bar{3}m$ structure, similar to the previously studied sodium polyhydride *Cc*-NaH$_7$ [10]. Surprisingly, during decompression of DAC X1, we found that the main series of XRD peaks of possibly distorted CsH$_7$ remains in the XRD pattern until the DAC is almost opened, indicating the extraordinary stability of this compound down to about 1 GPa (Supporting Figure S34-S36). Comparison with the theoretically predicted equation of state *V(P)* (Figure 1d), shows that despite of possible structural distortion, CsH$_7$ does not lose a lot of hydrogen at low pressures. Ab initio calculations performed in the harmonic approximation confirm that at zero temperature (0 K) *P4/nmm*-CsH$_7$ remains thermodynamically stable down to 5 GPa, and dynamically stable even at ambient pressure (Supporting Figures S40, S42-S45).

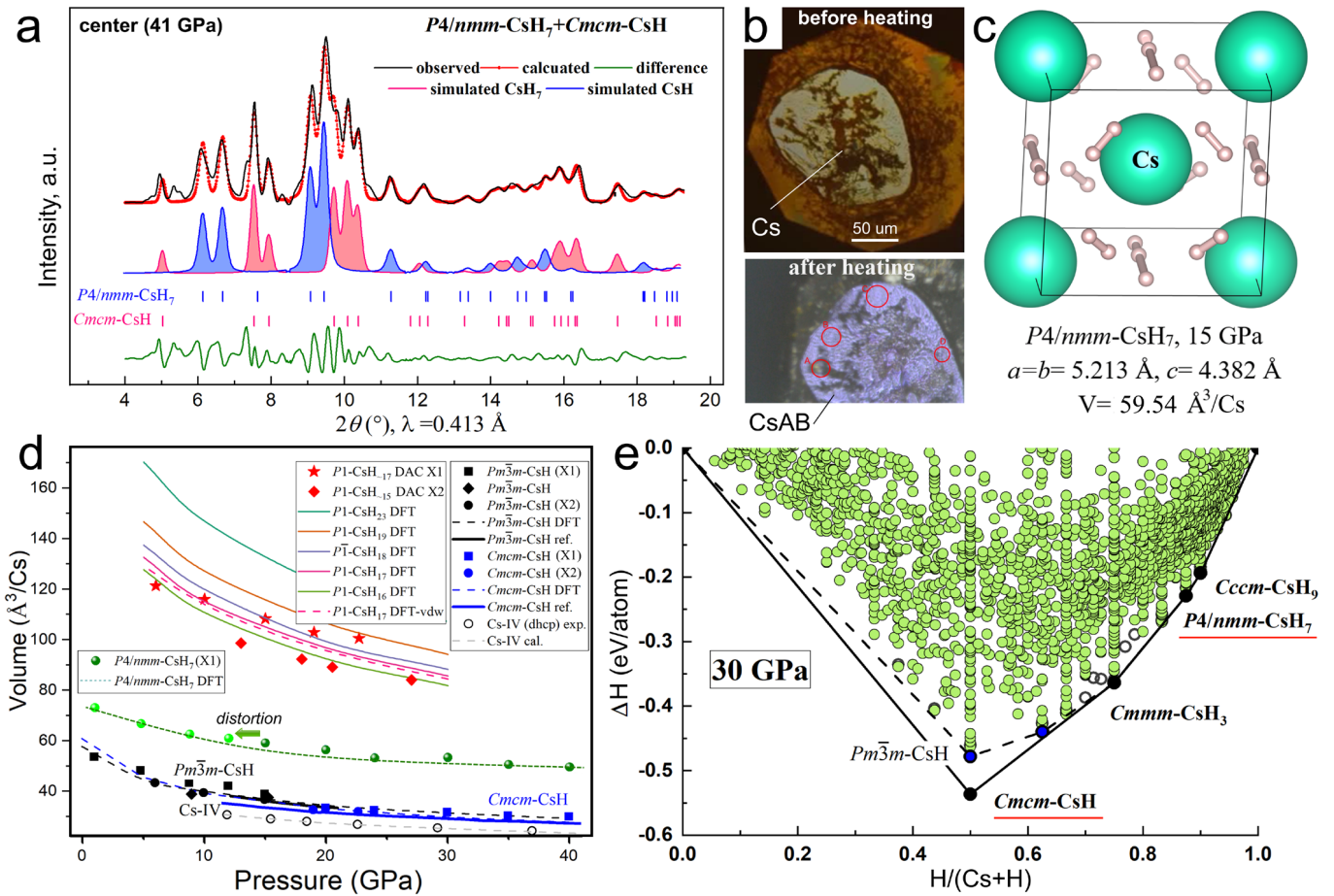

**Figure 2.** Synthesis of cesium hydride $P4/nmm$-CsH$_7$ at 41 GPa. When heated, cesium almost instantly reacts with formation of transparent products. (a) Experimental XRD pattern ($\lambda$ = 0.413 Å, SPring-8) of reaction products after the laser heating of Cs/CsAB at 41 GPa. Black line is the experimental XRD pattern, the red one is the Le Bail refinement [32], the green one – is the difference. Blue shaded spectrum is the predicted XRD pattern of CsH$_7$ (by Mercury 2021.2.0 code [33]), the pink one – the same for $Cmcm$-CsH. (b) Photographs of the Cs/CsAB sample before and after laser heating. (c) Optimized unit cell of $P4/nmm$-CsH$_7$ at 15 GPa (PAW PBE, VASP code [34-36]). (d) Experimental pressure vs. unit cell volume diagram for various studied cesium hydrides. Theoretical calculations are indicated by continuous lines and the "DFT" (density functional theory) marks. (e) Thermodynamic convex hull of Cs-H system at 30 GPa and 0 K. The diagram is dominated by H-rich compounds. Stable phases are CsH, CsH$_3$, CsH$_7$ and CsH$_9$ (black circles).

This first experiment demonstrated that cesium hydrides, CsH and CsH$_7$, can be obtained as the main products from Cs/CsAB mixture via the laser heating. However, it may happen that not all CsAB decomposes. This likely situation could lead to mixing and misinterpretation of the diffraction peaks from the cesium polyhydrides CsH$_x$ and the residual amidoborane CsAB. Surprisingly, a separate experiment shows that compressed CsAB undergoes rapid and irreversible amorphization above 10 GPa (see Supporting Figure S31). Therefore, residual CsAB has no effect on the X-ray diffraction patterns of Cs polyhydrides. This fact greatly simplifies the synthesis and identification of polyhydrides in the proposed approach.

In the second experiment (DAC X2), we investigated a mixture of cesium amidoborane and pieces of gold foil (Au/LT-CsAB) heated at pressure of 22 GPa. X-ray diffraction patterns of the synthesis products were studied at the Elettra synchrotron radiation facility, Xpress beamline (Trieste, Italy), and are shown in Figure 3. This time, in addition to monohydride CsH ("dotted" diffraction lines, Figure 3d), we found a series of broad diffuse diffraction lines corresponding to the new compound with rather large unit cell volume. Using the results of structural search calculations in USPEX code [26-28] (Figure 2e, Supporting Tables S8-S9), we interpreted this novel compound as a non-stoichiometric molecular polyhydride $P1$-CsH$_{15+x}$ (where x = 0…2) formed by the decomposition of three or four neighboring CsAB molecules. The unit cell volume of semiconducting CsH$_{17}$ is 85.3 Å$^3$/Cs at 22 GPa. According to the evolutionary structural search, a prototype of this phase is $P1$-CsH$_{17}$, staying just slightly above (+3 meV/atom) the convex hull at 30 GPa. By the way, a similar series of diffuse diffraction lines can also be found in a detailed analysis of the first experiment with DAC X1. Using the information obtained from DAC X1, we established the hydrogen content in the CsH$_{15+x}$ and the experimental unit cell volume as ≈ 100 Å$^3$/Cs at 23 GPa, although with less accuracy (Figure 2d).

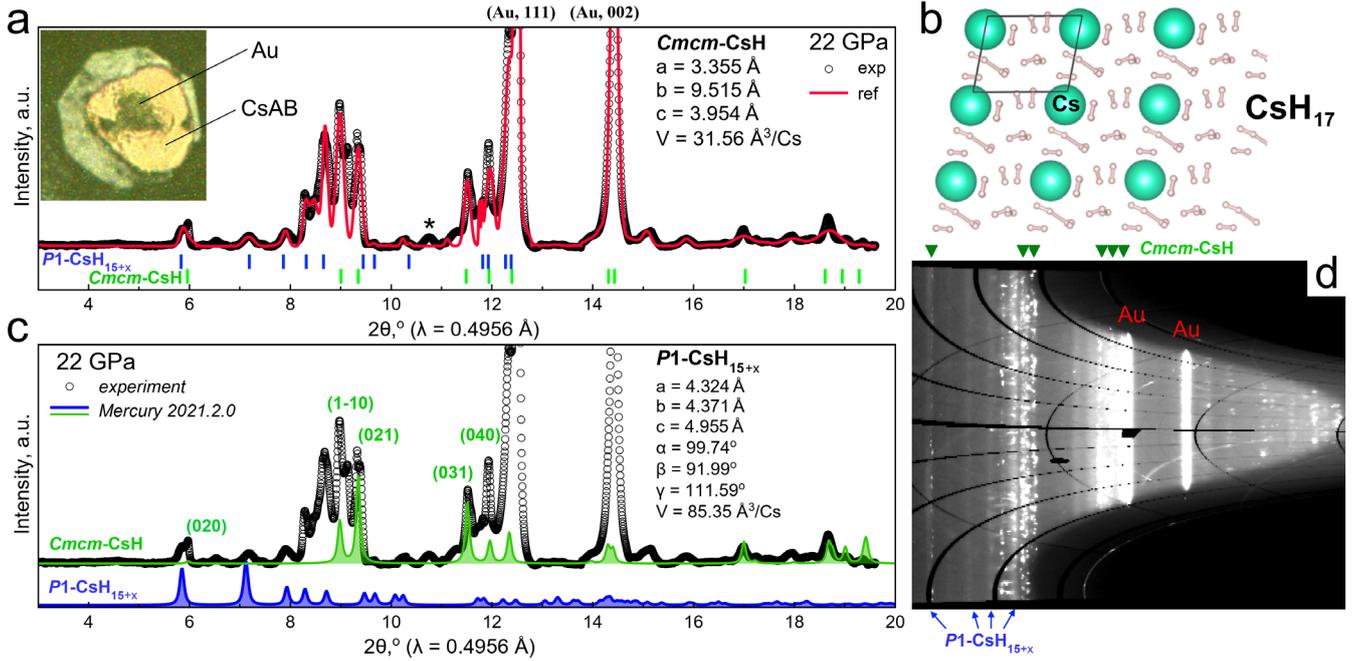

**Figure 3.** Formation of cesium polyhydride $P$1-CsH$_{15+x}$ from CsAB/Au below 27 GPa in DAC X2. (a) Powder X-ray diffraction pattern of the reaction products in the vicinity of a gold target (black curve, λ = 0.4956 Å, Elettra). The black line is the experimental XRD pattern, the red line is the Le Bail refinement of the unit cell parameters of CsH$_{15+x}$, $Cmcm$-CsH and gold. Inset: photo of DAC's chamber at 22 GPa. (b) Optimized structure of CsH$_{17}$ at 20 GPa (PAW PBE, VASP code [34-36]). (c) Comparison of experimental powder diffraction and predicted XRD pattern of $P$1-CsH$_{15}$ and $Cmcm$-CsH at 22 GPa. (d) Diffraction image ("cake") of CsAB decomposition products. Coarse-crystalline phase corresponds to CsH, while the disordered CsH$_{15+x}$ phase corresponds to a series of broad diffuse diffraction lines with low intensity.

Estimation of the hydrogen content in CsH$_{15+x}$ can also be approached from the other side, using the known volume of the H atom in pure hydrogen: 4.24 Å$^3$/H at 22 GPa [37]. Taking the volume of Cs$^+$ from the equation of state of CsH$_7$, we can calculate the ratio of hydrogen and cesium H : Cs in our compound, which is between 13 and 17. Such a high hydrogen content is not surprising given the huge density of various stable and metastable solid solutions of Cs$^+$ in molecular hydrogen (Figure 2e). Similar compounds, a kind of "hydrogen sponges", can be obtained in the Sr-H [7] and Ba-H [6] systems at high pressure.

*Synthesis of rubidium polyhydrides: RbH$_{9-x}$*

Synthesis of rubidium polyhydrides was carried out starting from LT-RbAB/Rb mixture (Supporting Figure S13) via pulsed laser heating above 1000°C at 12-12.5 GPa using a gold target in the DAC Y. The experiment was followed by decompression of the reaction products to 1.7 GPa. At pressures of 8-12 GPa, X-ray diffraction patterns of the reaction mixture (Figure 4) indicate the formation of a monohydride $Pm\bar{3}m$-RbH [38] and two crystal modifications of rubidium polyhydride RbH$_{9-x}$: pseudo hexagonal ($Cccm$) and pseudo tetragonal ($Cm$), according to the chemical reaction between two RbAB molecules:

$$2\ RbNH_2BH_3 \rightarrow RbH + RbH_9 + c\text{-}BN. \qquad (2)$$

Pseudo hexagonal RbH$_{9-x}$ is somewhat less stable from the point of view of thermodynamics and lies above the convex hull by 2.3 meV/atom (Figure 5c, Supporting Tables S6-S7). This is an unprecedentedly low synthesis pressure for compounds of this stoichiometry. Just for the sake of comparison, to obtain nonahydrides MH$_9$ of the neighboring elements, Sr and Y, pressures above 200 GPa [39] and 80 GPa [7,40] are required, respectively (Figure 5a,b). Comparison of experimental unit cell volumes of obtained RbH$_{9-x}$ phases with the calculated ones (Supporting Tables S4-S5) shows that we are dealing with non-stoichiometric compounds having about 8-9 hydrogen atoms per each rubidium atom (x = 0…1). A comparison with the data of Kuzovnikov et al. [11], who used direct reaction of RbH with molecular hydrogen, leads to a similar conclusion.

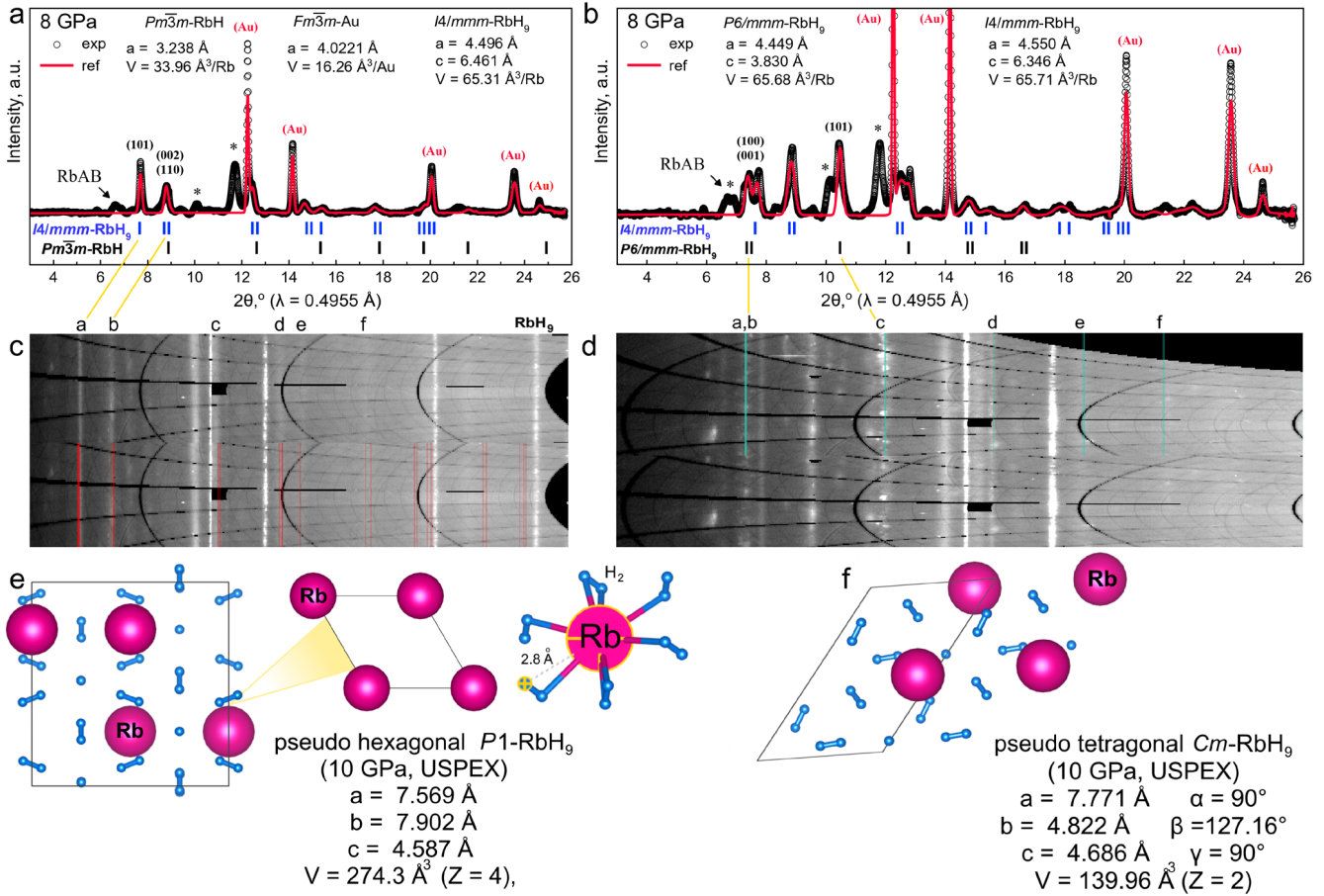

**Figure 4.** Formation of rubidium polyhydrides, pseudo hexagonal and pseudo tetragonal $RbH_{9-x}$, from RbAB in DAC Y. (a) Experimental X-ray diffraction pattern (black curve, λ = 0.4955 Å, Elettra) and the Le Bail refinement (red curve) of unit cell parameters of pseudo $I4/mmm$-$RbH_{9-x}$ and $Pm\bar{3}m$-RbH at 8 GPa. There is a broad peak of residual RbAB marked by arrow. (b) The same in another place of the sample, where there is also a pseudo $P6/mmm$-$RbH_{9-x}$. Gold (Au) was used as a pressure sensor. (c, d) Diffraction images ("cake") of decomposition products of RbAB. Red and azure lines show diffraction lines (diffuse and spotted) corresponding to two structural modifications $RbH_{9-x}$. (e, f) Crystal structure of rubidium polyhydrides obtained using USPEX [26-28] and VASP [34-36] codes. The calculated parameters of the unit cells of stoichiometric $RbH_9$, the sublattice of Rb atoms, and the coordination sphere of Rb atoms in pseudo $P6/mmm$-$RbH_9$ are also given.

Both obtained rubidium polyhydrides are semiconductors with relatively wide bandgaps of 2.15 - 2.7 eV. Molecular hydrogen sublattice in both $RbH_{9-x}$ phases has low symmetry ($P1$), which, however, can be increased by symmetrization (tolerance is 0.2) to $Cccm$ and $Cm$. After that, the sublattice of rubidium atoms has a much higher space group number, $P6/mmm$ (sh) in the case of pseudo hexagonal $RbH_{9-x}$, and $I4/mmm$ in the case of pseudo tetragonal phase. The same is observed for $BaH_{12}$ [6] and $SrH_9$ [7]. Metal sublattice of pseudo tetragonal $RbH_{9-x}$ transforms to cubic ($Fm\bar{3}m$) after refinement, and this is one of the remarkable findings of this work, since this phase was not found among the products of direct reaction of RbH with hydrogen [11].

Simple hexagonal ($P6/mmm$) sh-$RbH_9$ was recently synthesized by Kuzovnikov et al. from RbH and pure hydrogen at 24-56 GPa [11]. This confirms the feasibility of our synthetic approach to rubidium polyhydrides starting from the corresponding amidoborane. During decompression, the authors found that sh-$RbH_9$ decomposes below 10 GPa. At the edge of the stability region, the volume of $P6/mmm$-$RbH_9$ unit cell is about 72 Å$^3$/Rb [11], in reasonable agreement with theoretical calculations (Figure 4e, 69.3 Å$^3$/Rb at 10

GPa, PAW PBE). We synthesized, however, a slightly different compound with smaller unit cell volume of $V_{hex}(10\ GPa) = 64.4\ Å^3/Rb$. This speaks in favor of lower hydrogen content: $RbH_{9-x}$ instead of $RbH_9$, since the volume per hydrogen atom at 10 GPa is approximately 5.83 $Å^3$/H [37].

It is important to note that, according to theoretical calculations at 0 K, pseudo tetragonal $RbH_9$ remains stable at ambient pressure (Supporting Figures S51-S54), whereas pseudo hexagonal $RbH_9$ becomes dynamically unstable below 5 GPa (Supporting Figures S48-S50). Experimental decomposition is observed below 8 GPa. The pseudo tetragonal $RbH_{9-x}$ probably remains stable even at 3.3 GPa undergoing a structure distortion.

*Raman spectroscopy*

Raman scattering of cesium hydrides was studied at room temperature at excitation laser wavelengths of 532 and 633 nm. Before the laser heating, samples of amidoboranes in DACs exhibit a set of Raman signals consistent with the literature data for RbAB and CsAB (Supporting Figures S9-S10, S13). After the laser heating of CsAB, a new broad peak appears at 3850 cm$^{-1}$ (at 18-21 GPa, Supporting Figure S11). According to the results of ab initio calculations, this broad peak may be assigned to $P4/nmm$-$CsH_7$ (Supporting Figure S12). A double peak of molecular hydrogen was detected nearby the vibration frequency of pure $H_2$ (Supporting Figure S11). At the same time, in experiments with LiAB and NaAB, a similar Raman peaks were detected between 2900-3800 cm$^{-1}$ possibly indicating formation of Li and Na polyhydrides (Supporting Figures S7-S8).

Situation for rubidium polyhydrides is similar. Raman spectrum of RbAB/Au after laser heating has a broad peak at 3800-3900 cm$^{-1}$ (Supporting Figures S14-S15), which corresponds to $RbH_{9-x}$ in agreement with the results of direct synthesis of $P6/mmm$-$RbH_9$ from RbH and $H_2$ [11] and ab initio calculations (Supporting Figure S16). In addition, there is a peak of molecular hydrogen (4242 cm$^{-1}$). Thus, for cesium and rubidium polyhydrides, Raman spectroscopy confirms formation of molecular hydrogen, CsH, $CsH_7$ and $RbH_{9-x}$, although the spectra contain other peaks that are difficult to interpret.

*Reflectance and transmittance spectroscopy*

We investigated the relative reflectance $R(\lambda)$ of Cs and CsAB before and after laser heating in the high-pressure DAC X3 at 42 GPa and in DAC Z at 20 GPa in the visible spectral range (400–900 nm, 1.4–3.1 eV). Because the geometry of stressed diamond anvils, Cs particle and $CsNH_2BH_3$ layer is not completely known, it is only possible to determine the relative reflectance or transmittance of samples in comparison with the reflection/transmission from either the rhenium (Re) gasket or an empty place, such as the diamond/$CsNH_2BH_3$ (or AB) boundary. The relative reflectance and transmittance were calculated as $R(\lambda)$ or $T(\lambda) = const \times I_{sample}(\lambda)/I_{ref}(\lambda)$, where $I_{sample}(\lambda)$ and $I_{ref}(\lambda)$ are intensities of light reflected from or transmitted through the sample and the reference, respectively. Reflectance spectra of the compressed technetium hydride $TcH_{1.3}$ were recently measured and processed in a similar manner [41,42].

Surprisingly, transmission measurements of a thin Cs particle at 42 GPa result in a $T(\lambda)$ similar to the literature data at 0 GPa (Supporting Figure S20). After the laser heating of Cs/CsAB, an almost transparent substance (e.g., $CsH_x$) is formed, which makes optical research challenging and the results – qualitative. When using the transmittance of CsAB as a reference, we found good agreement between the calculated transmittance of $CsH_7$ and the experimental data obtained after laser heating of Cs/CsAB (Figure 5d). At the same time, the difference between CsAB and $CsH_x$ is insignificant when using empty DAC as a reference.

Reflectivity of Cs at 20 GPa is also in good agreement with the literature data (Supporting Figure S19). Laser heating of the Cs/CsAB sample at 42 GPa, as expected, leads to a significant drop in the sample reflectivity (compared to Cs), a change in the sign of $dR/d\lambda$ and a monotonic decrease in $R(\lambda)$ in the range of 500-950 nm (Figure 5e). This behavior is in qualitative agreement with the results of DFT calculations for $CsH_7$ (Supporting Figure S23). Therefore, the optical spectroscopy confirms the formation of $CsH_7$ during the thermal decomposition of cesium amidoborane.

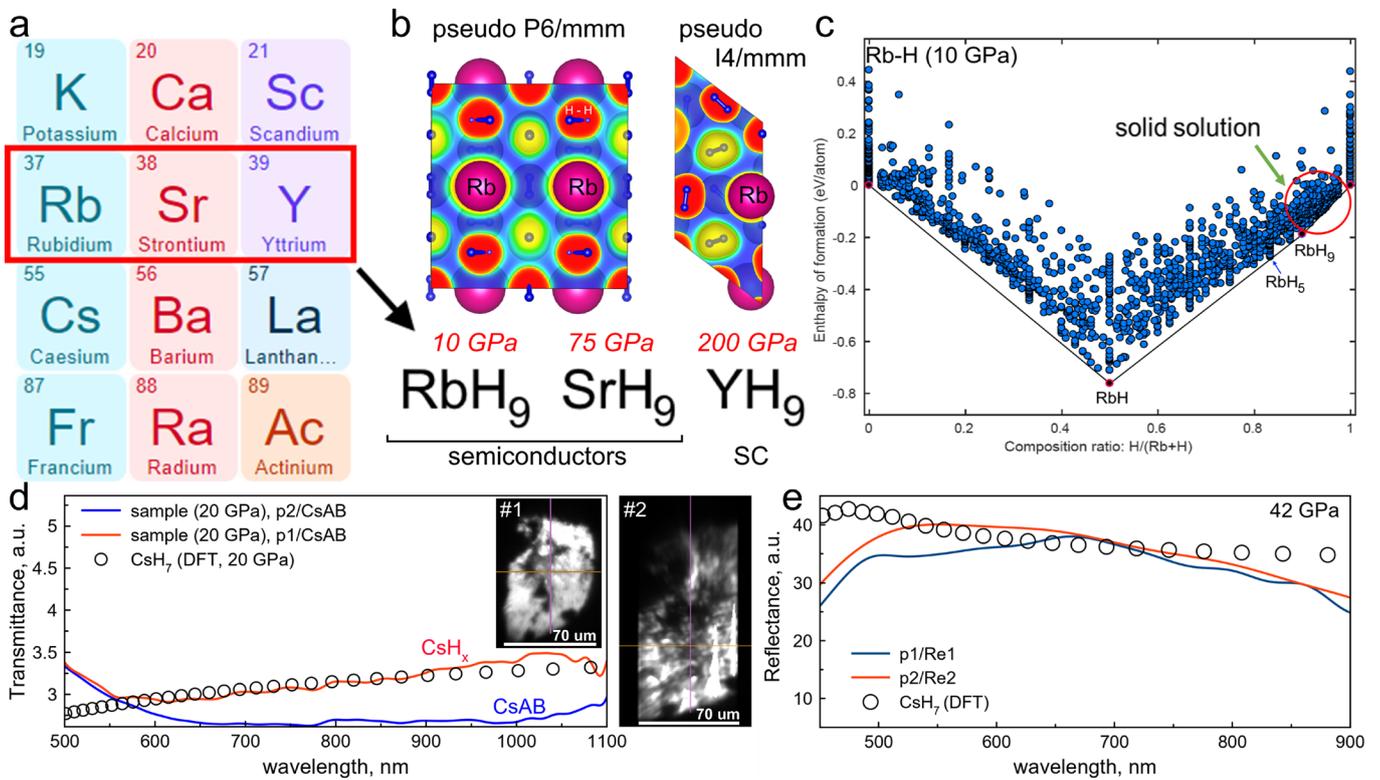

**Figure 5.** Physical properties of cesium and rubidium polyhydrides. (a, b) A series of "isomeric" $MH_9$ polyhydrides formed by neighboring elements: Rb, Sr, Y. The stabilization pressures of the corresponding hydrides are given in red font. "SC" means superconductor. (b) Electron localization functions of both crystal modifications of $RbH_{9-x}$ at 10 GPa. Molecular hydrogen is clearly visible. (c) Thermodynamic convex hull of Rb-H system at 10 GPa and 0 K. The region of solid solutions of $Rb^+$ cations in hydrogen is outlined in red. Stable at this pressure are RbH, $RbH_5$ and $RbH_9$. (d) Experimental ("sample") and calculated ("DFT") relative transmittance of $CsH_7$ at 20 GPa compared to CsAB. Inset: photos of investigated regions, DAC X3. (e) Experimental ("p1,2") and calculated ("DFT") relative reflectance of $CsH_7$ compared to the reflection from a rhenium gasket (Re1,2) at 42 GPa in DAC Z.

## Discussion

*Structural Features of Cs and Rb polyhydrides*

Proposed method for the synthesis of polyhydrides by heating their amidoboranes under pressure opens up prospects for obtaining not only Cs and Rb hydrides, but also polyhydrides of all alkaline and alkaline-earth metals, as well as Y and Zn. Encouraging results from trial experiments with LiAB, NaAB and Ca(AB)$_2$ are given in the Supporting Information. Proposed approach affects not only binary, but also ternary polyhydrides. Indeed, by crystallizing solutions of different amidoboranes from organic solvents, it is easy to obtain mixed amidoboranes. Many of them have been previously described experimentally [43]. Their thermal decomposition under pressure will lead to formation of ternary polyhydrides of Li-Na, Na-Mg, etc.

Discovery of ultrastable cesium and rubidium polyhydrides raises the question of the possibility of further increasing stability of H-rich compounds at ambient conditions. Promising is the design of such compounds using larger ions, atoms and molecules ("pseudo atoms") such as Xe [44], $(CH_3)_3N^+$ and tert-butylammonium cation $(t\text{-}Bu)_3N^+$, methane molecule [45], neopentane $C(CH_3)_4$, various halogen-substituted molecules: $CF_4$, $SF_6$, $CI_4$ and so on.

During this study, we noticed one fascinating analogy between the structure of higher metal polyhydrides and the physics of space plasma. Properties of many polyhydrides of alkali and alkaline earth metals are similar to the behavior of plasma crystals, in which heavy dust grains (say, $Rb^+$, $Sr^{+2}$ etc.) are surrounded by a Debye cloud of much smaller ions (e.g., $H^-$, $H_3^-$) with an opposite charge. In this case, heavy atoms usually form a highly symmetrical sublattice. This is an effect, well-known from the study of plasma crystals in microgravity [46]. Structures of $RbH_{9-x}$, $CsH_{15+x}$ and $SrH_{22}$, are kind of plasma crystals, because a large charged metal ion ($Cs^+$, $Rb^+$) forms a Debye shell of weakly charged light $(H_{2\pm1})^-$ anions around. In the space, these spherical shells are organized into a cubic or hexagonal densest packing, where metal atoms become a kind of raisins in a pie made from hydrogen (Supporting Figure S56).

Tendency to form such solid electrolytes from $Cs^+$ and $Rb^+$ ions dissolved in molecular hydrogen already follows from thermodynamic calculations of Cs-H and Rb-H convex hulls, where the entire region near pure hydrogen is filled with higher polyhydrides even at low pressures (Figures 2e, 5c). These polyhydrides are stabilized by electric field of cesium and rubidium ions separated via a mobile shell of hydrogen. As it has been shown in experiments with $BaH_2$ [30], $SrH_{5.75}$ [7], and by molecular dynamics simulations of $H_3Cl$ [31] and $LaH_{10}$ [47,48], hydrogen in such polyhydrides has very high diffusivity causing significant ionic conductivity and even superionicity, probably exceeding that of $BaH_2$ [30].

## Conclusions

To sum up, using synchrotron X-ray powder diffraction, Raman spectroscopy, optical measurements and first-principles calculations, we demonstrated that Cs and Rb amidoboranes can be used to synthesize corresponding molecular polyhydrides: $P4/nmm$-$CsH_7$, $CsH_{15+x}$ and two crystal modifications of $RbH_{9-x}$. Our proposed approach to polyhydrides from readily available amidoboranes does not require the use of hydrogen loading systems and can be extended to binary and ternary polyhydrides of other alkali (Li and Na) and alkaline earth (Ca) metals. We observed amorphization of Cs and Rb amidoboranes above 20-30 GPa. As a result, they practically do not contribute to the diffraction patterns of Cs and Rb polyhydrides.

All synthesized Cs and Rb polyhydrides are ultrastable molecular-ionic semiconductors with a bandgap of 1-3 eV. Compared to many other hydrides, Cs and Rb require external pressure of only 10–20 GPa, instead of 100–200 GPa for La, Y, Ba, Ce, Th and other metal polyhydrides. During decompression, $CsH_7$ may remain stable until the pressure in DAC is almost completely released. Exploring the thermodynamics of hydride synthesis, we found that cesium and rubidium tend to form nonstoichiometric compounds with excess of hydrogen under pressure. Large $Rb^+$ and $Cs^+$ ions locate in the disordered lattice of molecular hydrogen. Adding or removing one $H_2$ molecule from such a hydrogen "pie" has virtually no effect on the enthalpy of formation of hydrogen-rich compounds. Considering the molecular-ionic nature of the Cs, Rb hydrides, high diffusivity of hydrogen already at room temperature, as well as the low symmetry of the hydrogen sublattice, all the resulting compounds will likely exhibit pronounced ionic conductivity.

Our work demonstrates that cesium and rubidium have prospects for increasing the capacity of hydrogen batteries by stabilizing the shell of molecular hydrogen, and open a way for further reducing the synthesis pressure of polyhydrides.

## Acknowledgments

D. S. and D. Z. thank National Natural Science Foundation of China (NSFC, grant No. 1231101238) and Beijing Natural Science Foundation (grant No. IS23017) for support of this research. The high-pressure experiments were supported by the Russian Science Foundation (Project No. 22-12-00163). We also thank the Center for Energy Science and Technology, the Center for Photonic Science and Engineering and the Center for Engineering Physics of Skoltech for the opportunity to use the center's equipment during the implementation of this project. All authors thank the staff scientists of the SPring-8 and Elettra (proposals No. 20215034 and No. 20220288) synchrotron radiation facilities for their help with X-ray diffraction measurements, especially Dr. Saori Kawaguchi (SPring-8) and Dr. Boby Joseph (Elettra). This work was started at Jilin University in 2019 with the support of Prof. Xiaoli Huang. We express our gratitude to Dr. Mikhail Kuzovnikov (University of Edinburgh) and Dr. Vadim Efimchenko (IPSS RAS, Chernogolovka) for providing CsH samples and valuable discussions. We thank Dr. Pavel Zinin (STC UI RAS, Moscow) for assistance in laser heating.

## Contributions

D.Z., D.S., M.G., F.G.A., I.T., Y.N. and K.S. performed experiments. D.Z., M.G. and D.S. prepared theoretical part of the paper. D.Z., D.S., and A.R.O. analyzed the data and wrote the paper. All the authors discussed the results and offered useful inputs.

## Data availability

Authors declare that the main data supporting our findings of this study are contained within the paper and Supporting Information. All relevant data are available from the corresponding authors upon request.

## Code availability

Quantum ESPRESSO code is free for academic use and available after registration on https://www.quantum-espresso.org/. Vienna ab-initio Simulation Package (VASP) code is available after registration on https://www.vasp.at/. USPEX code is free for academic use and available after registration on https://uspex-team.org/en/.

# SUPPORTING INFORMATION

## Raisins in a Hydrogen Pie: Ultrastable Cesium and Rubidium Polyhydrides

Di Zhou[1,*], Dmitrii Semenok[1,*], Michele Galasso[2], Frederico Gil Alabarse[3], Denis Sannikov[6], Ivan A. Troyan[4], Yuki Nakamoto[5], Katsuya Shimizu[5], and Artem R. Oganov[6,*]

[1] Center for High Pressure Science & Technology Advanced Research, Bldg. #8E, ZPark, 10 Xibeiwang East Rd, Haidian District, Beijing, 100193, China

[2] Institute of Solid State Physics, University of Latvia, 8 Kengaraga str., LV-1063 Riga, Latvia

[3] Elettra Sincrotrone Trieste, Trieste 34149, Italy

[4] Shubnikov Institute of Crystallography, Federal Scientific Research Center Crystallography and Photonics, Russian Academy of Sciences, 59 Leninsky Prospekt, Moscow 119333, Russia

[5] KYOKUGEN, Graduate School of Engineering Science, Osaka University, Machikaneyamacho 1-3, Toyonaka, Osaka, 560-8531, Japan

[6] Skolkovo Institute of Science and Technology, 121205, Bolshoy Boulevard 30, bld. 1, Moscow, Russia

Corresponding authors: Di Zhou (di.zhou@hpstar.ac.cn), Dmitrii Semenok (dmitrii.semenok@hpstar.ac.cn), Artem R. Oganov (a.oganov@skoltech.ru)

# Content

# 1. Methods

*Experiment*

For the high-pressure synthesis, symmetrical diamond anvil cells (Mao type) together with rhenium gaskets were used in experiments. Thickness of gaskets was measured by the optical interference method and was found to be about 30 microns at a pressure of 20 GPa. The pressure gradient between center and edge in DACs was about 5 GPa. Detailed information about the loaded DACs can be found in Table S1.

**Table S1**. List of high-pressure DACs used in this experiment, samples loaded into them and products obtained.

| DACs (culet diam.) | Experiment (date) | Sample | Comments and products |
| --- | --- | --- | --- |
| DAC X1 (culet 250 μm) | SPring-8 (Feb. 2020) | Cs/CsAB | Initial pressure is 40.5 GPa. Products: $CsH + CsH_7 + CsH_{15+x}$ |
| DAC X2 (culet 150 μm) | Elettra (Apr. 2022) | CsAB/Au | Initial pressure is 22 GPa. Products: $CsH + CsH_{15+x}$ |
| DAC X3 | - | Cs/CsAB/Au | Laser heated at 18 GPa. Raman/optical measurements |
| DAC X4 | - | Cs/CsAB/Au | Laser heated at 40, decompressed to 32) Raman measurements. |
| DAC Z (culet is 100 μm) | - | Cs/CsAB | Laser heated at 42 GPa |
| DAC Y (culet is 200 μm) | Elettra (Sept. 2022) | Rb/RbAB/Au | Initial pressure is 12.5 GPa. Products: $RbH + RbH_{9-x}$ |

Cs (99.99 %, 25 mg/ampoule) and Rb (99.99 %, 15 mg/ampoule) were stored in sealed quartz ampoules, which were opened before synthesis in an argon glove box. Synthesis of amidoboranes was carried out by mixing solid sublimated ammonia borane and the metals, or by adding excess of the metals to a saturated solution of ammonia borane in tetrahydrofuran at ambient temperature in an Ar glove box. Release of hydrogen gas served as a marker of the reaction. Subsequent separation of the resulting solution from the unreacted metal and evaporation of tetrahydrofuran led to crystallization of the desired Cs and Rb amidoboranes. When DACs with $CsH_x$ or $RbH_x$ are opened at ambient conditions, Cs/Rb hydrides instantly react with oxygen and moisture of the air. The "broken glass" morphology of amorphous CsAB at 64 GPa is shown in Figure S1.

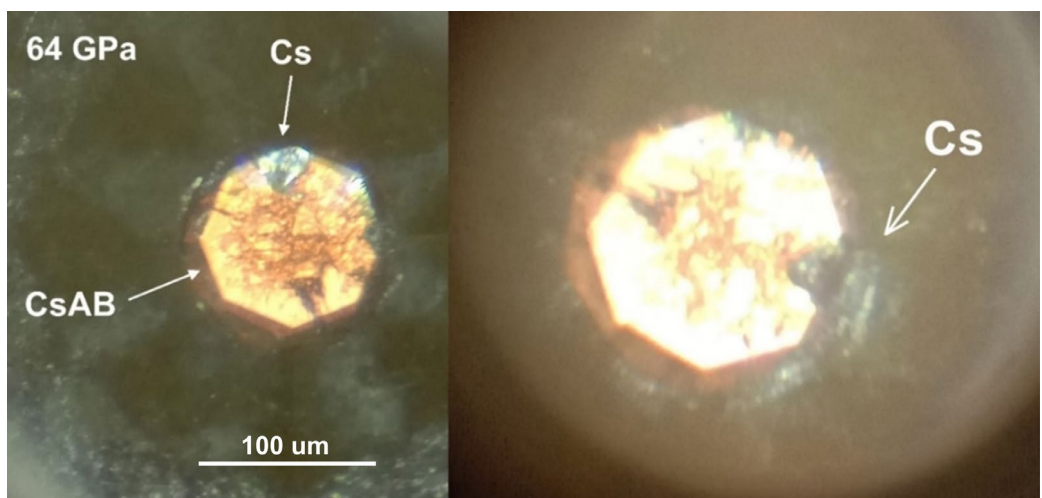

**Figure S1.** Unsuccessful loading of Cs into the CsAB medium and the gasket chamber expansion at 64 GPa. Due to the high surface tension of Cs metal, its droplet is displaced towards the gasket during loading.

It is important to note that the formation of $CsH_{13}$ and $CsH_{17}$ under high-pressure high-temperature conditions is most likely based on the stoichiometry of the decomposition reaction of cesium amidoboranes:

$$3\ CsNH_2BH_3 \rightarrow 2CsH + CsH_{13} + c\text{-}BN. \quad (S1)$$

$$4\ CsNH_2BH_3 \rightarrow 3CsH + CsH_{17} + c\text{-}BN. \quad (S2)$$

Crystal structure of Cs and Rb polyhydrides was determined using the synchrotron powder X-ray diffraction on the BL10XU synchrotron beamline (SPring-8, Japan, proposal No. 2021B1436) using a wavelength of 0.413 Å and on Xpress beamline (Elettra Sincrotrone Trieste, Italy, proposals No. 20215034 and No. 20220288) using a wavelength of 0.4956 Å and a PILATUS3 S 6M (DECTRIS) detector at ambient temperature. The experimental XRD images were integrated and analyzed using the Dioptas 0.5 [1]. To fit the diffraction patterns and obtain the cell parameter, we analyzed the data using Materials Studio and Jana2006 software [2], employing the Le Bail method [3].

Pressure in diamond anvil cells was determined via the Raman signal of diamond at room temperature [4] using LabRAM HR Evolution (HORIBA) and DXR3xi Raman Imaging Microscope (Thermo Scientific). In both cases, 532 nm excitation laser light was used. Laser heating was carried out by 0.1-0.2 second pulses of IR laser (~1 μm), which led to the local temperature rise above 1000 K.

Before the reflectance/transmittance experiments, a series of calibrations were performed to determine the apparent spectral function of the incandescent lamp after passing the light through the optical system (Supporting Figures S18). Then, the relative reflectance and transmittance of the empty diamond anvil cell, rhenium gasket, and Cs particle in CsAB media was measured (Supporting Figures S19-S22). Complex interference pattern was smoothed by the Fourier filter in the Origin [5] to suppress interference between two diamond planes, and multiplied by an arbitrary constant to compare results with the DFT calculations.

*Theory*

Structural searches for stable compounds in the Cs–H and Rb-H systems were performed using the USPEX code [6-8], for pressure of 30 GPa (Cs-H system), and 10 GPa (Rb-H system) with a variable-composition evolutionary search from 1 to 48 atoms of each type (Cs/Rb, H). The first generation of the search (80 structures) was created using a random symmetric generator, all subsequent generations (80 structures) contained 20% of random structures and 80% of those created using the heredity, soft mutation, and transmutation operators. For the Cs-H system, convex hulls were also built at 5, 10 and 20 GPa by optimizing the structures found by USPEX at 30 GPa.

The results contain the files extended_convex_hull and extended_convex_hull_POSCARS, which were postprocessed using the Python scripts change_pressure.py, split_CIFs.py etc. [9]. The script split_CIFs.py converts the set of POSCARS recorded in the extended_convex_hull_POSCARS file into a set of CIF files, simultaneously symmetrizing the unit cells and sorting the files by ascending fitness (the distance from the convex hull). The CIF files created in such a way can be analyzed using Dioptas [1] and JANA2006 [2].

To calculate the equations of state (EoS) of Cs and Rb polyhydrides, we performed structure relaxations of phases at various pressures using the density functional theory (DFT) [10,11] within the generalized gradient approximation (Perdew–Burke–Ernzerhof functional) [12] and the projector augmented wave method [13,14] as implemented in the VASP code [15-17]. A pseudopotential with the number of valence electrons ZAL = 9 was used for calculations of properties of rubidium polyhydrides. The plane wave kinetic energy cutoff was set to 600 eV, and the Brillouin zone was sampled using the Γ-centered k-points meshes with a resolution of $2\pi \times 0.05$ Å$^{-1}$. We calculated the phonon densities of states and phonon band diagrams using the finite displacement method (VASP and PHONOPY) [18,19]. VESTA software was used for visualization [20].

Calculations of optical properties was performed using the VASP code according to the methodology of Gajdoš et al. [21] with LOPTICS = .TRUE. The number of frequency grid points was 20000 (NEDOS), the total number of Kohn-Sham orbitals in the calculation was 120 (NBANDS), cutoff energy was 600 eV. Local field effects were included on the Hartree level only (LRPA=.TRUE., CSHIFT = 0.1).

## 2. Optical images

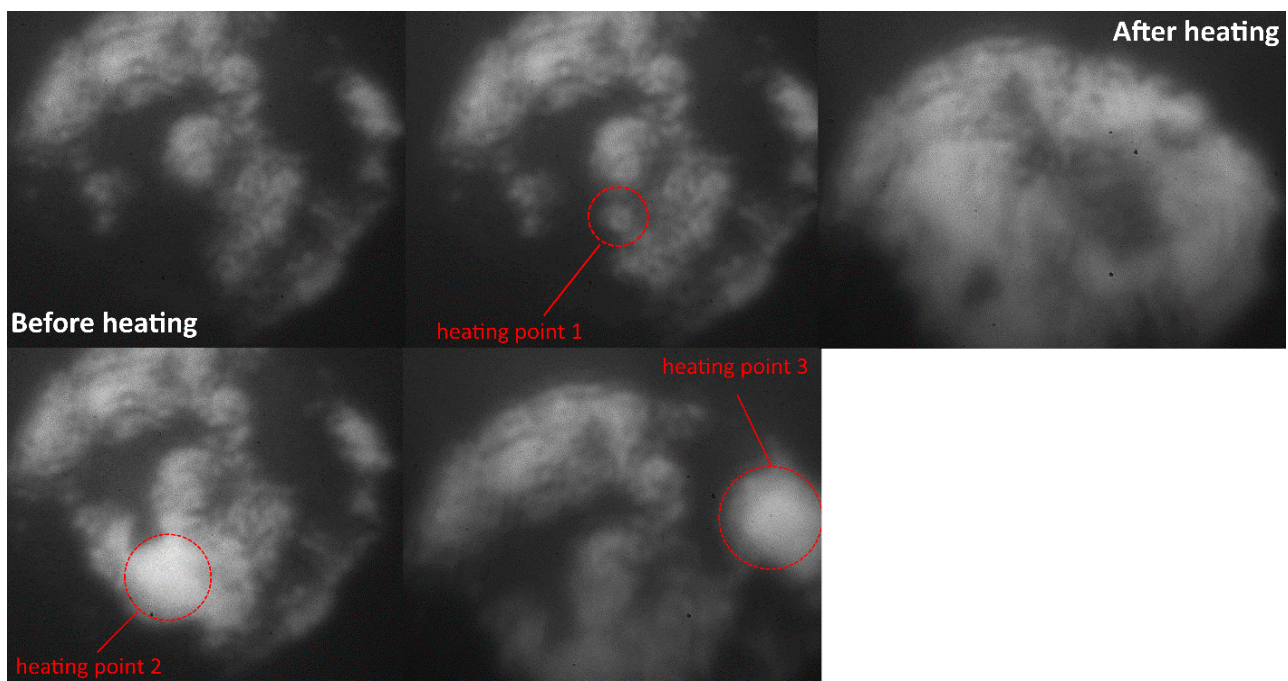

**Figure S2.** Laser heating (pulsed of ~ 0.2 s) of Cs/CsAB mixture in DAC X3 at 20 GPa. Photos were taken with a black/white camera. After heating, opaque Cs particles disappear and windows of transparency appears instead. The sample is used for reflectance and transmission spectroscopy studies.

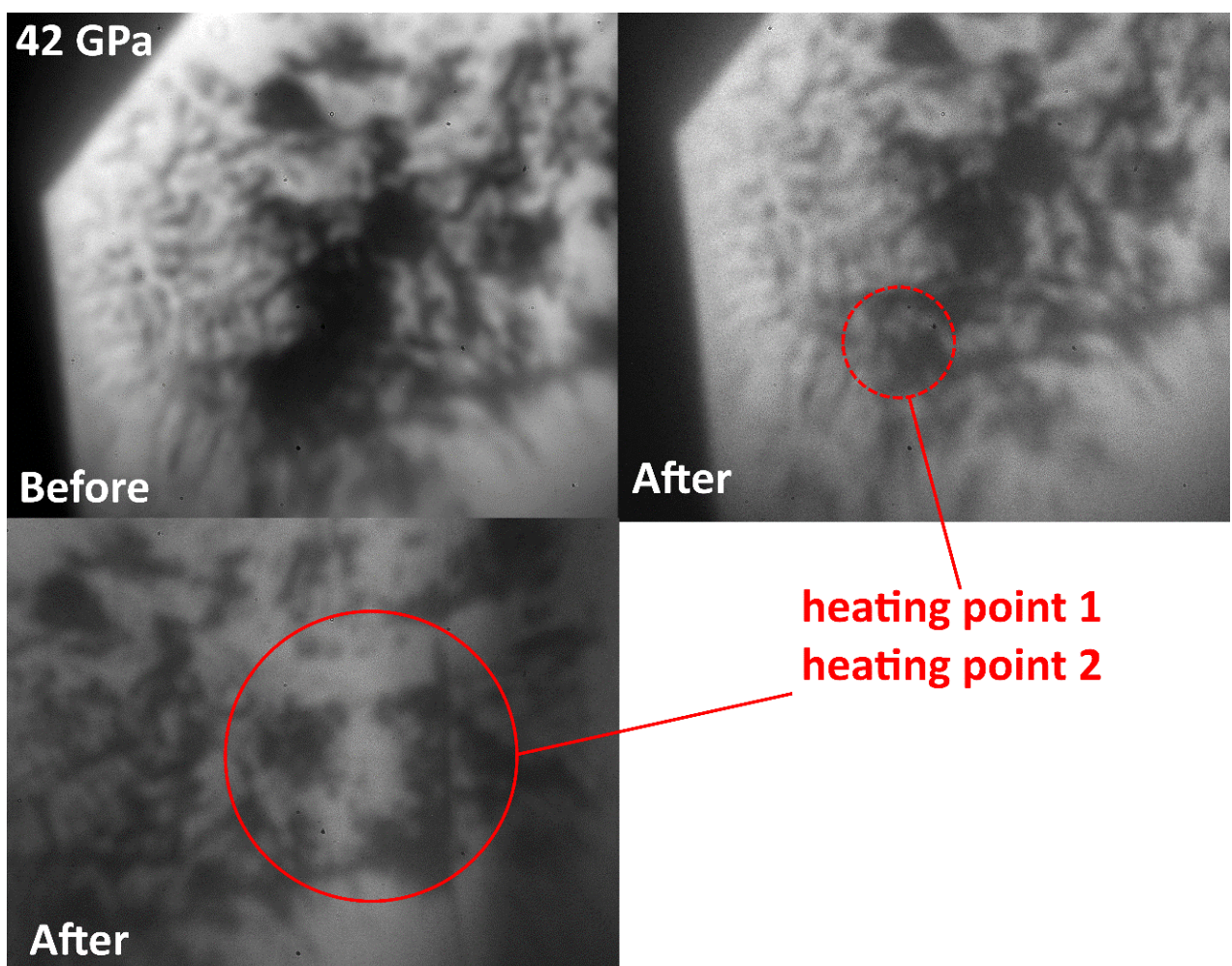

**Figure S3.** Laser heating (pulsed of ~ 0.2 s) of Cs/CsAB mixture in DAC Z at 42 GPa. The sample is used for reflectance and transmission spectroscopy studies.

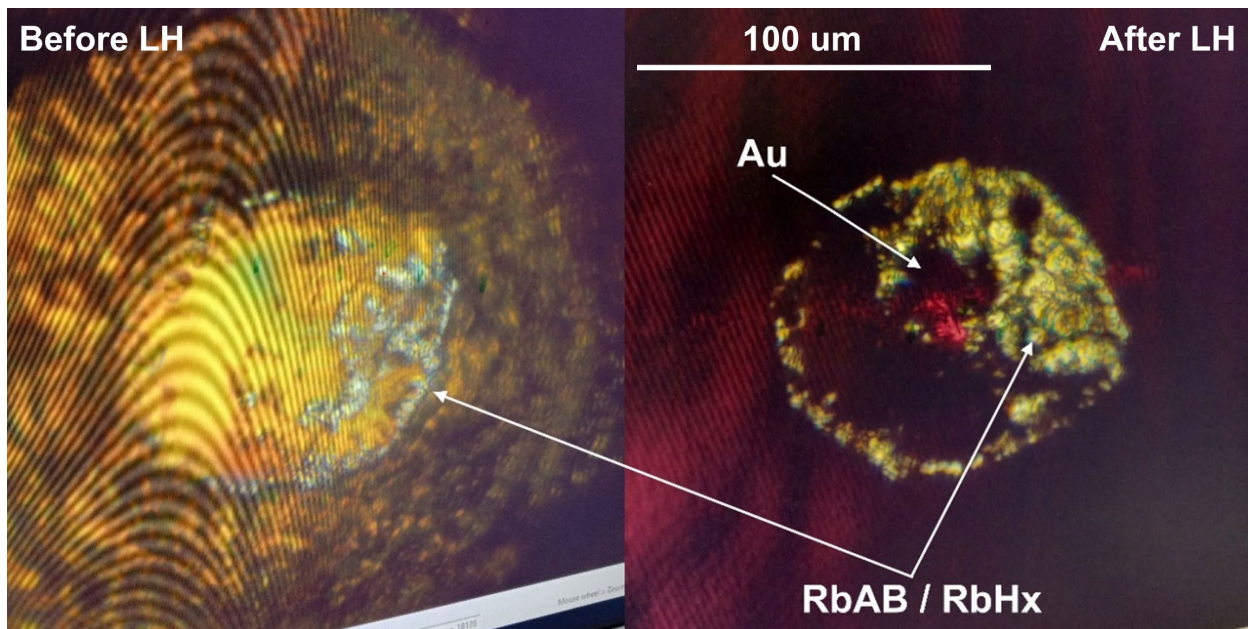

**Figure S4.** Laser heating of excess of Rb metal ("yellow mirror", left panel) with RbAB and Au target. Golden piece remains dark after laser heating at 12 GPa in DAC Y (right panel).

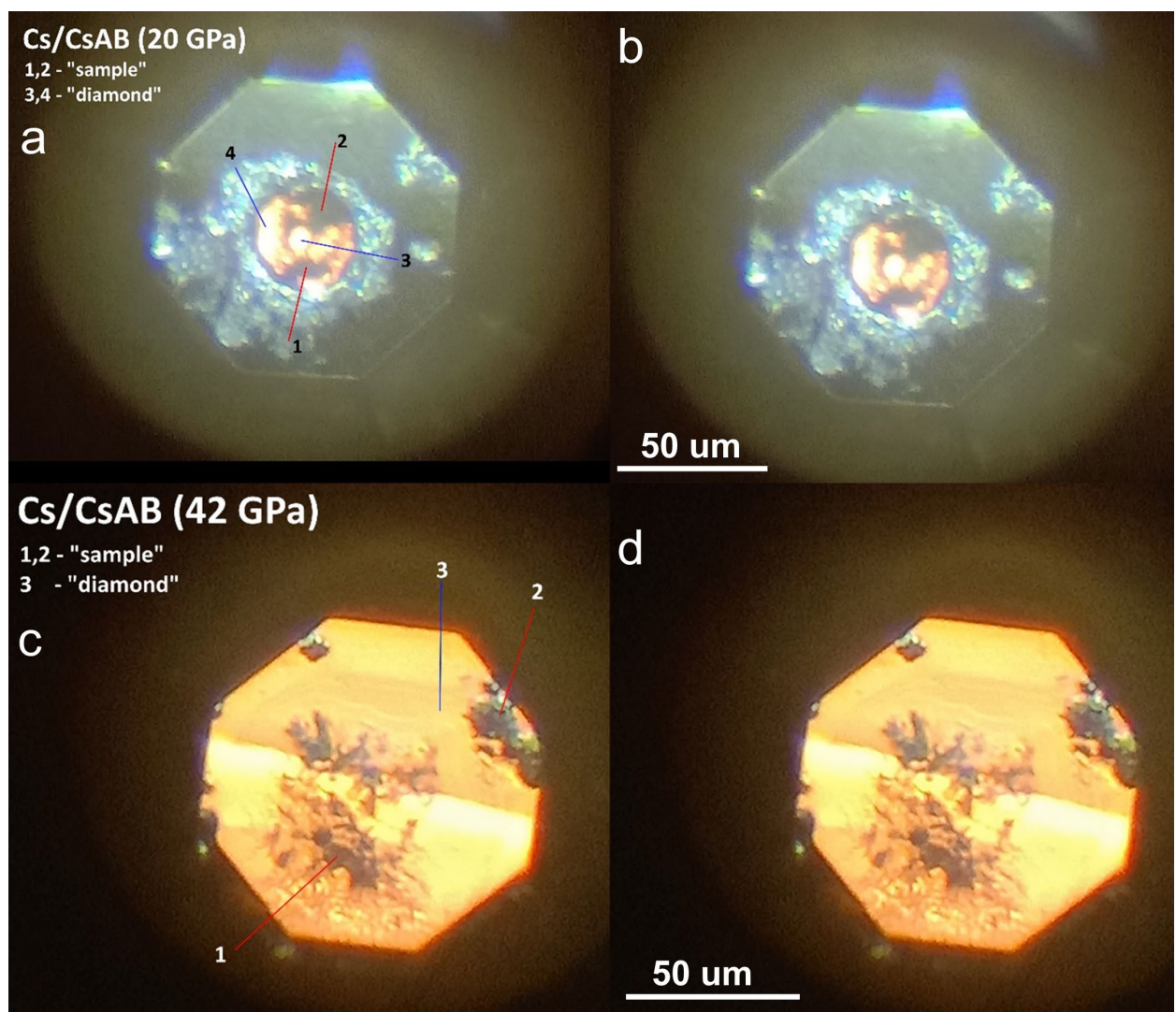

**Figure S5.** Optical photographs of samples in the DACs chambers used for reflectance and transmission spectroscopy studies. (a, b) Cs/CsAB sample after laser heating in DAC X3 at 20 GPa. (c, d) Cs/CsAB sample after laser heating in DAC Z at 42 GPa. Thickness of Re gasket in DAC X3 is 33 μm, and about 4 μm in DAC Z according the optical interference method. DAC Z was used to investigate the transmission spectrum of metallic cesium.

# 3. Thermal decomposition of Na, Li and Ca amidoboranes

Experiments with the thermal decomposition of sodium, lithium, calcium amidoboranes and ultraviolet irradiation of ammonia borane preceded the study of cesium and rubidium polyhydrides. The thermal decomposition results were studied using Raman spectroscopy only. The emergence of many new Raman signals has been observed (see below). In turn, the use of an ultraviolet He-Cd (325 nm, Kimmon Koha KP2014C, tube current 90 mA, tube voltage 3.33 kV) laser at a power of 0.14 mW for several hours does not lead to the formation of molecular hydrogen from ammonia borane, although it does lead to a change in a molecular structure of AB (Figure S6).

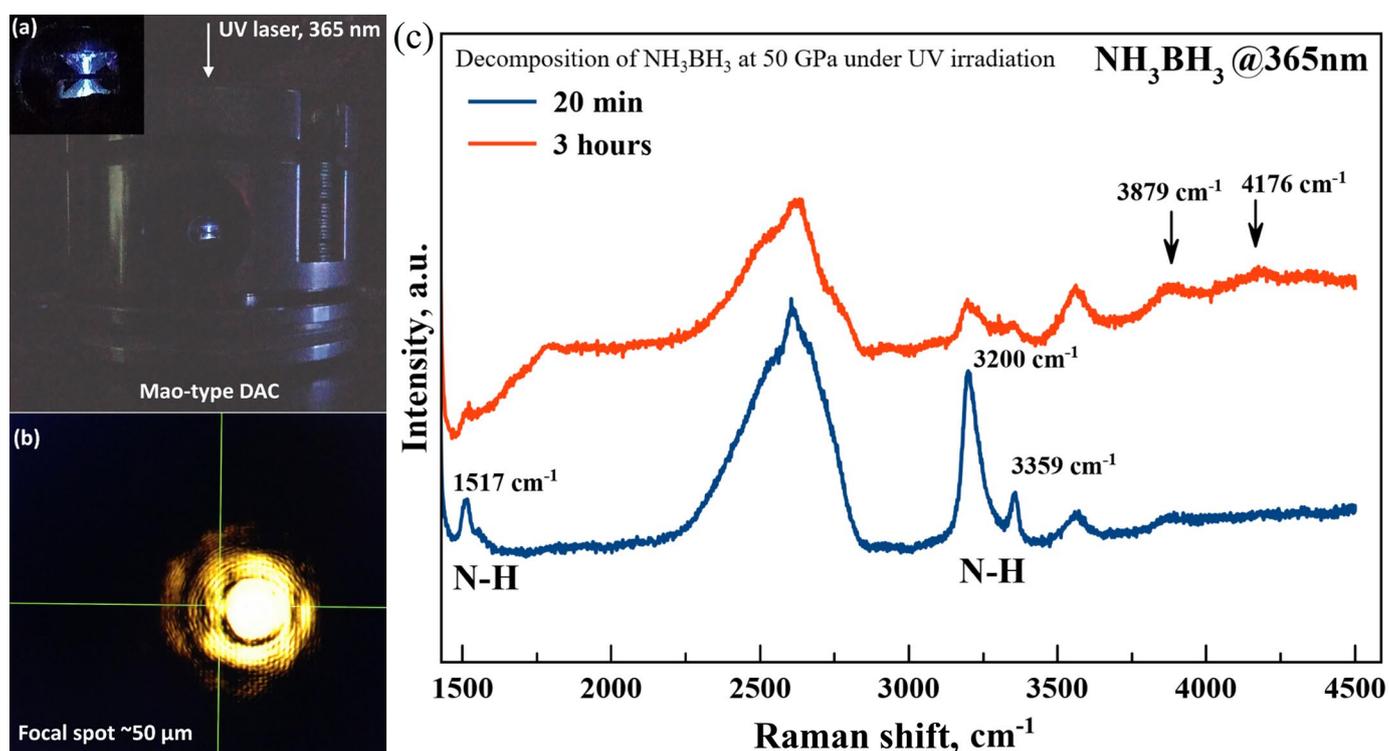

**Figure S6.** An attempt to decompose ammonia borane through ultraviolet laser irradiation (λ = 365 nm, power is 0.14 W) in a DAC at 50 GPa. (a) View of the DAC and fluorescence of diamonds under ultraviolet irradiation. (b) Photograph of culet. (c) Raman spectra of ammonia borane after 20 minutes (blue) and after 3 hours of irradiation (red). One can see a change in the intensity of the N-H Raman signals and the appearance of new weak signals at 3879 cm$^{-1}$ and 4176 cm$^{-1}$. Formation of molecular hydrogen was not observed.

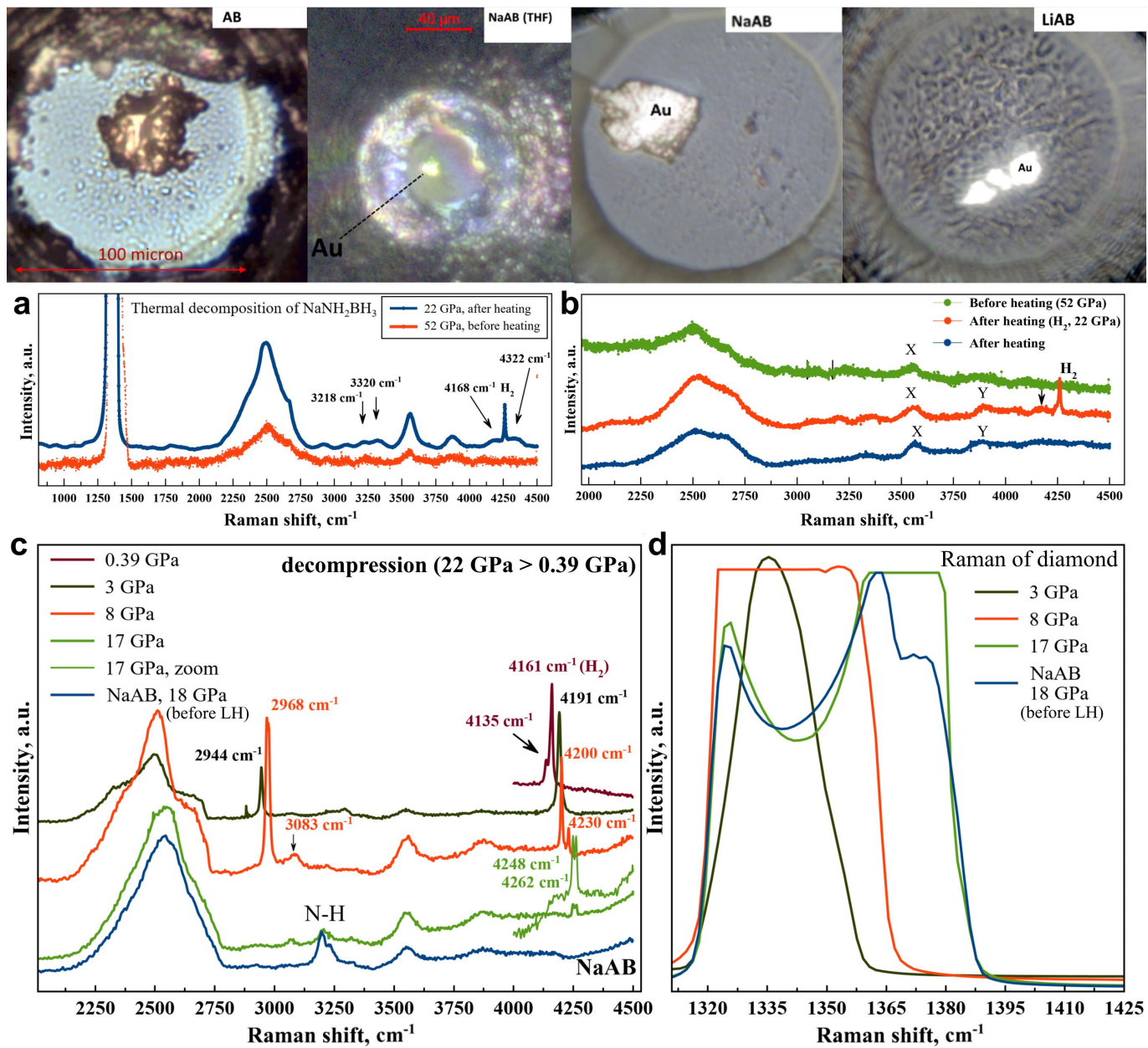

**Figure S7.** Experiments on the thermal decomposition of lithium and sodium amidoboranes at high pressure. The top series of photographs shows the position of gold targets in AB, NaAB (from THF), NaAB and LiAB media. (a) Raman spectra of NaAB before (52 GPa) and after (22 GPa) laser heating. (b) Comparison of Raman spectra collected at different places of the NaAB (THF) sample before and after laser heating. The peak designated as "Y" turns out to be not associated with molecular hydrogen. (c) Results of decompression of another NaAB sample heated at 18 GPa. Formation of molecular hydrogen is visible; the corresponding double peak is at 4248 cm$^{-1}$ and 4262 cm$^{-1}$ at 17 GPa. Appearance of a broad signal at 3800-3900 cm$^{-1}$ and narrow peaks at 2968-2944 cm$^{-1}$ may point to formation of new Na polyhydrides. (d) Raman spectra in the C-C vibration region.

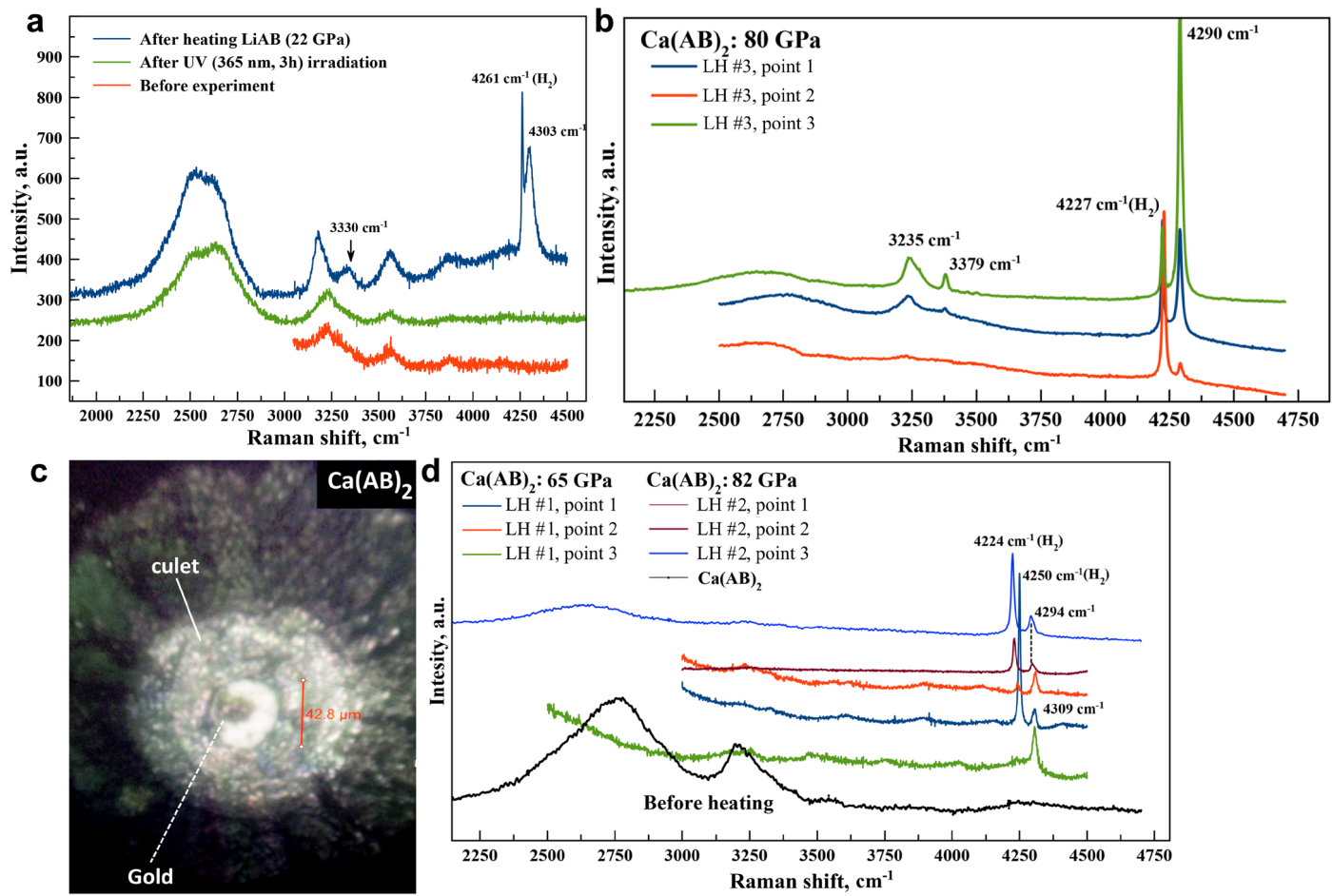

**Figure S8**. Experiments on thermal decomposition of lithium and calcium amidoboranes at high pressure. (a) Raman spectroscopy of thermal decomposition products of LiAB at 22 GPa. We can see two different types of molecular hydrogen in the sample as well as an additional Raman signal at 3330 cm$^{-1}$. Ultraviolet (UV) irradiation does not affect the sample. (b) Raman spectroscopy of calcium amidoborane Ca(AB)$_2$ decomposition products after 3$^{rd}$ laser heating (LH) at 80 GPa. Two peaks of molecular hydrogen can be seen. These peaks change their relative intensity depending on studied position on the sample. (c) Photo of the DAC's culet showing the gold particle used as a heating target. (d) Raman spectroscopy of calcium amidoborane decomposition products after the 1$^{st}$ and 2$^{nd}$ laser heating runs at 65 and 82 GPa. We can see two different types of molecular hydrogen in the system.

In all cases of thermal decomposition of amidoboranes (LiAB, NaAB, Ca(AB)$_2$), the Raman spectra show signs of the formation of new compounds. Considering the powder X-ray diffraction analysis of the decomposition products of RbAB and CsAB, the additional Raman signals should be attributed to the new sodium, lithium, and calcium polyhydrides. However, Raman studies alone are not sufficient to establish the structure of the obtained compounds. The formation of methane CH$_4$ [22] and NH$_3$ [23], as well as their complex compounds with hydrogen [24,25], cannot be completely ruled out. One of the most promising continuations of these studies is the solid state chemical reaction YCl$_3$ + 3 LiNH$_2$BH$_3$ → Y(AB)$_3$ + 3 LiCl, which makes it possible to obtain yttrium superhydrides YH$_6$ and others using only one component of the reaction mixture: Y(AB)$_3$, while LiCl will play a role pressure-transmitting media.

## 4. Raman spectra

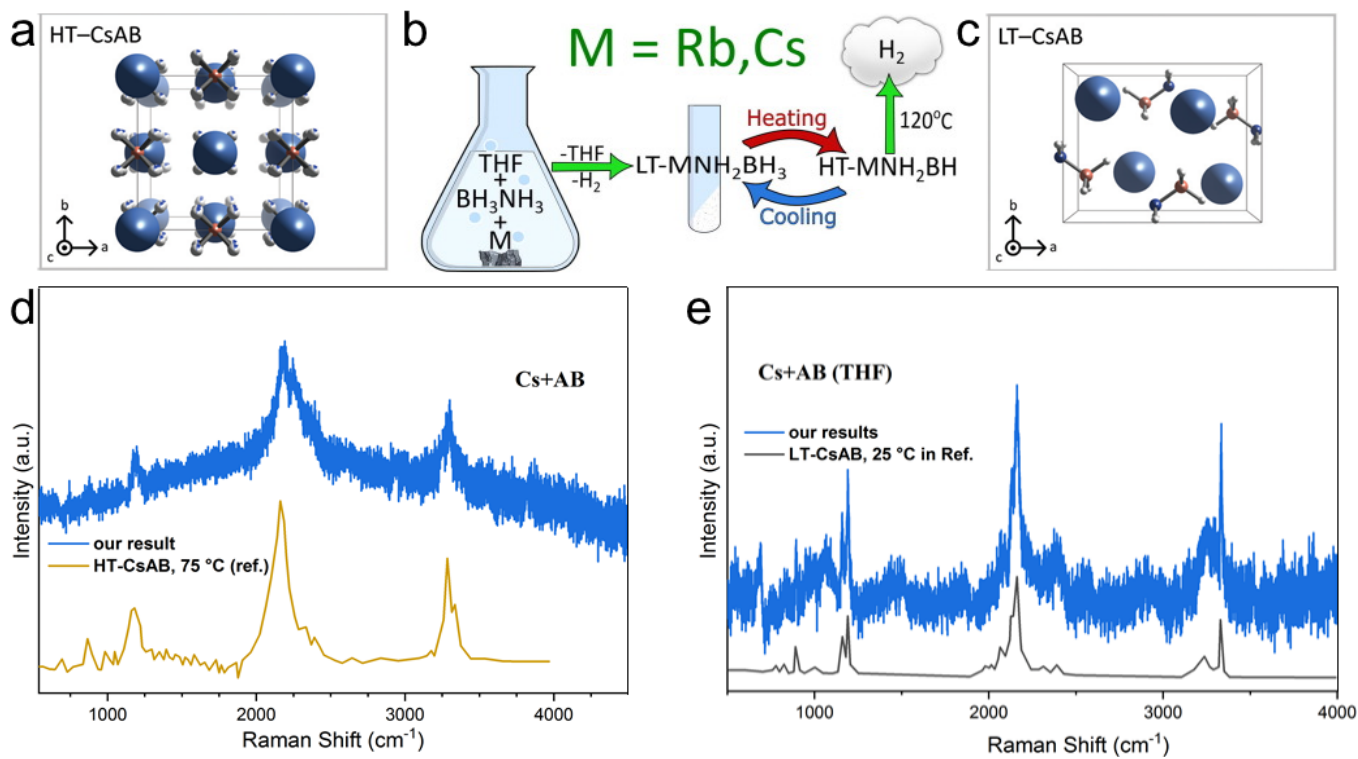

**Figure S9.** Preparation of Cs and Rb amidoboranes. (a) Crystal structure of HT-CsAB. (b) Scheme of CsAB synthesis using THF as a recation media and solvent. (c) Low-temperature crystalline modification of CsAB at 0 GPa. (d) Raman spectra (532 nm) of the result of synthesis (HT-CsAB, blue line) recorded in a diamond anvil cell compared to the literature data (yellow line). (e) The same for synthesized LT-CsAB.

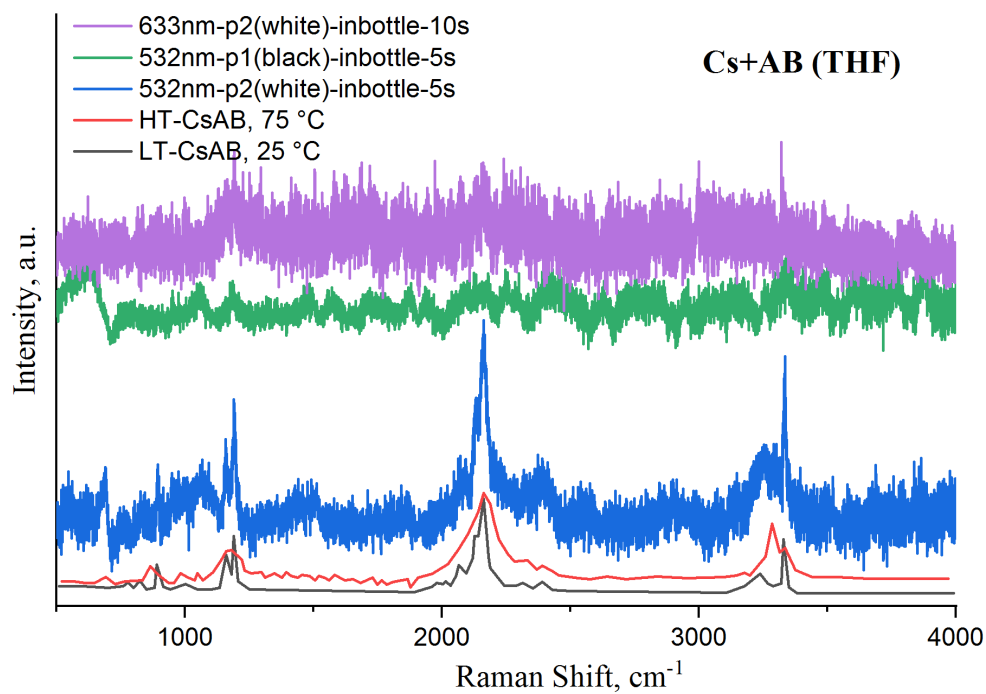

**Figure S10.** Identification of the reaction product of Cs with ammonia borane (AB) in a THF solution at 25 °C using Raman spectroscopy in an argon atmosphere. Raman spectrum of the synthesized product is shown in blue, and the red and black curves correspond to literature data for the high and low temperature modifications of CsAB. It is easy to see that the experimental spectrum corresponds to the low-temperature (LT) modification of CsAB.

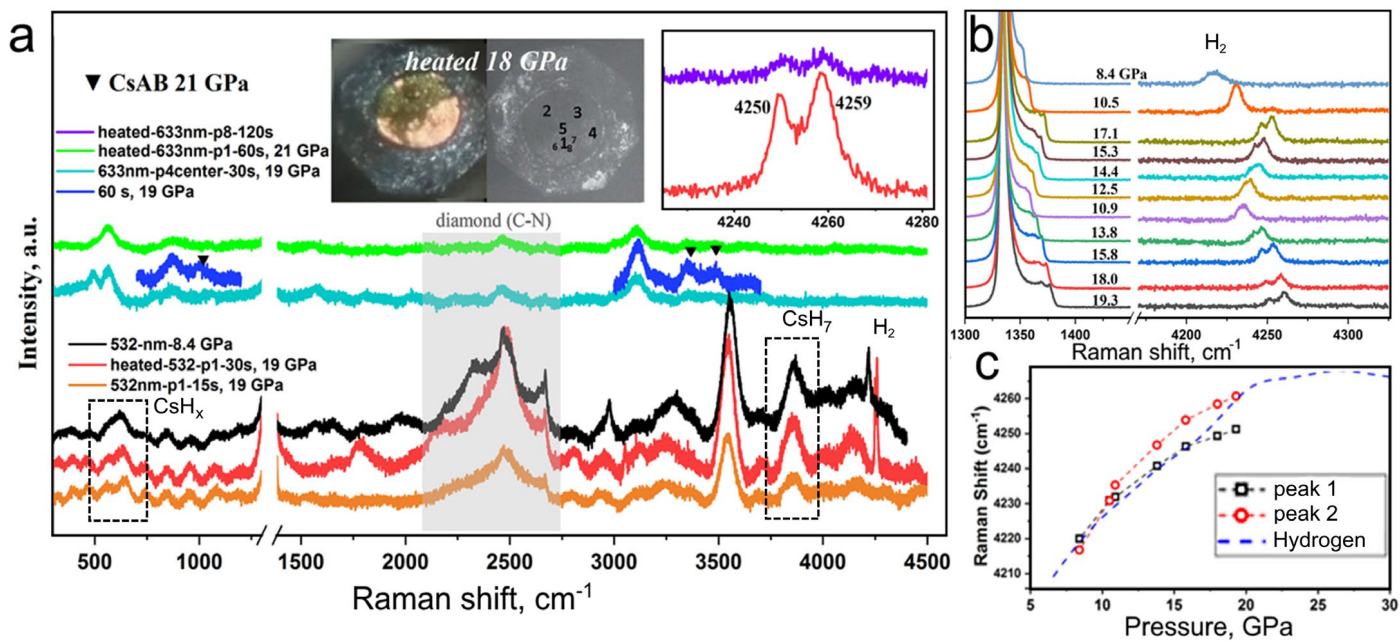

**Figure S11.** Raman spectroscopy of CsAB/Au (DAC X4) decomposition products after laser heating at 18-21 GPa. Difference in pressure is due to its gradient over the sample. (a) Comparison with Raman spectra of unreacted CsAB at 21 GPa (blue curve). (b) Behavior of the hydrogen vibrons during decompression from 19 to 8.4 GPa. (c) Splitting of the hydrogen peak and its behavior in the decompression run. Raman spectra indicate the formation of molecular hydrogen ($H_2$) and tetragonal $CsH_7$ (≈ 3850 cm$^{-1}$). The splitting of the hydrogen peak points to bound $H_2$ molecules. Some broad peaks around 4100-4200 cm$^{-1}$ and 650 cm$^{-1}$ may correspond to $CsH_{15-17}$ and CsH as well.

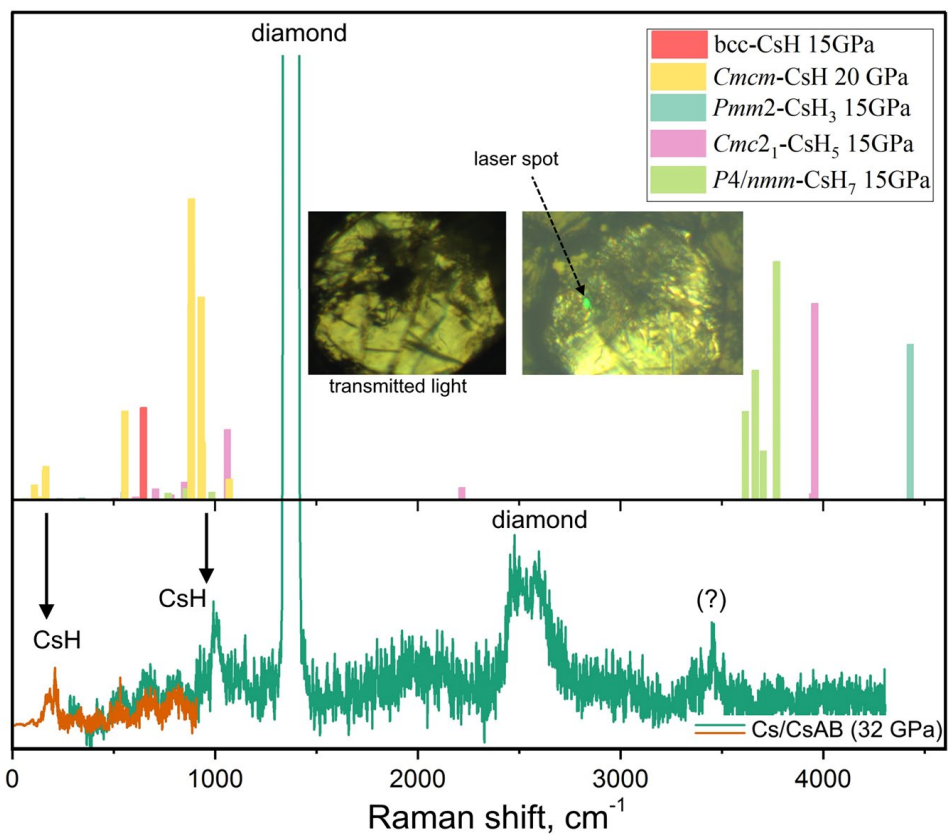

**Figure S12.** Raman spectroscopy of Cs/CsAB/Au sample after laser heating at 32 GPa. The study was carried out in the central part of the DAC using the Raman system of station ID15 (ESRF, France [26]). The upper panel above shows the calculated Raman spectra (VASP code [15-17]) of various cesium hydrides at 15-20 GPa. Despite the small difference in pressure, the experimental Raman signals (bottom panel) at 150-200 cm$^{-1}$ and around 1000 cm$^{-1}$ agree with the theoretical calculations for *Cmcm*-CsH. Insert: microphotographs of a diamond anvil culet in transmitted (left) and reflected light, indicating the point at which this Raman spectrum was obtained.

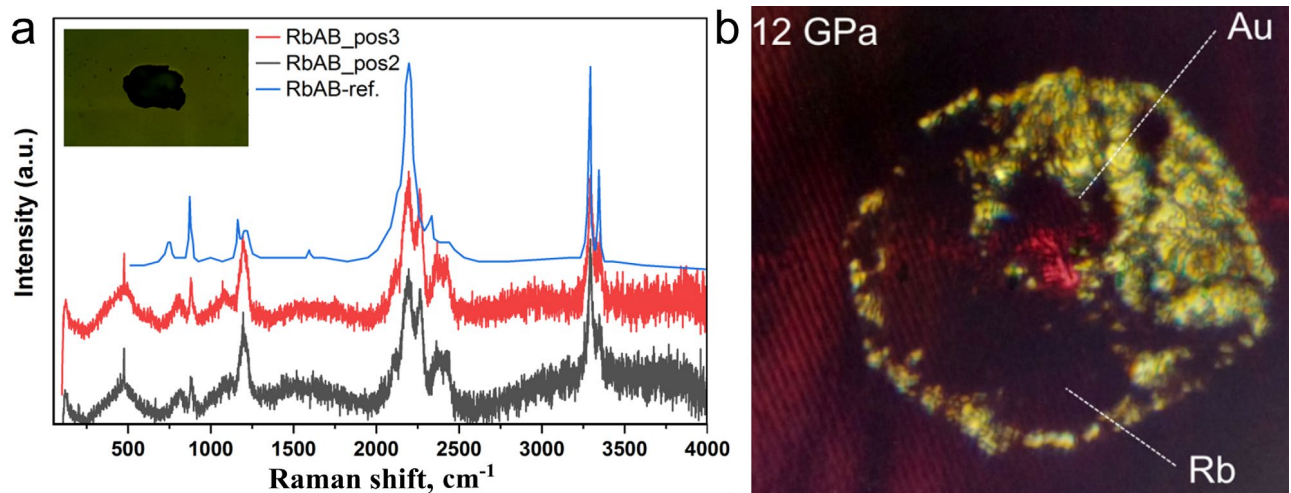

**Figure S13.** Raman spectroscopy of rubidium amidoborane LT-RbAB obtained from Rb and a solution of AB in THF. (a) Literature data is shown as a blue curve (RbAB-ref [27]), results of synthesis are shown in black and red curves. (b) The resulting RbAB was used to load the DAC Y along with the Rb metal and gold target. Photo of the sample at 12 GPa after laser heating. This DAC was further decompressed to 3-4 GPa.

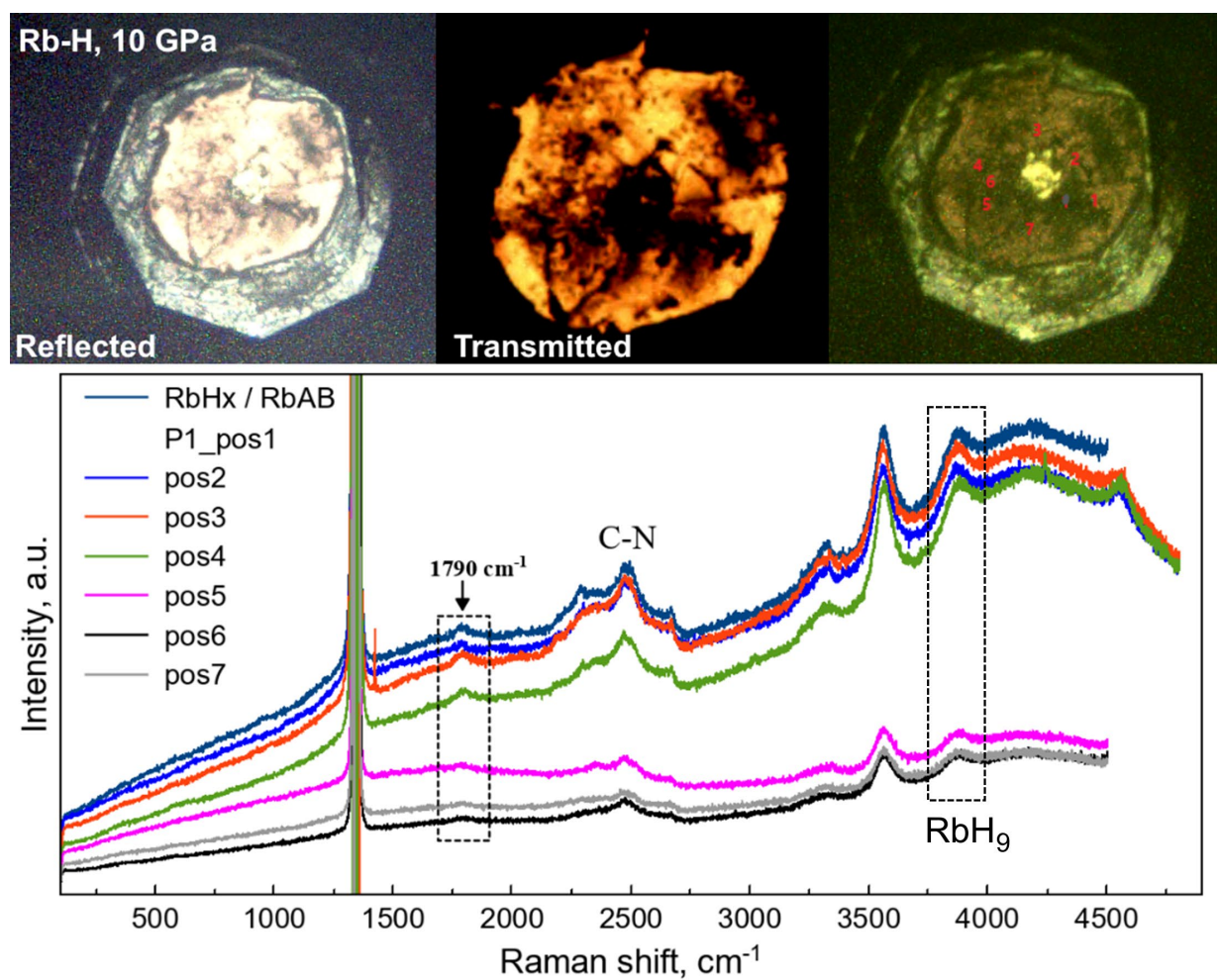

**Figure S14.** Raman study of formation of rubidium polyhydrides. Upper panel: photographs of the RbH$_x$/RbAB sample after laser heating at 10-12 GPa in reflected, transmitted light. Numbers mark the points (#1-7) where the Raman spectra were taken. It is important to note that there are virtually no Raman signals from the parent RbAB salt, which suggests amorphization upon compression. The Raman signal around 3800 cm$^{-1}$ corresponds to RbH$_{9-x}$, see also [28].

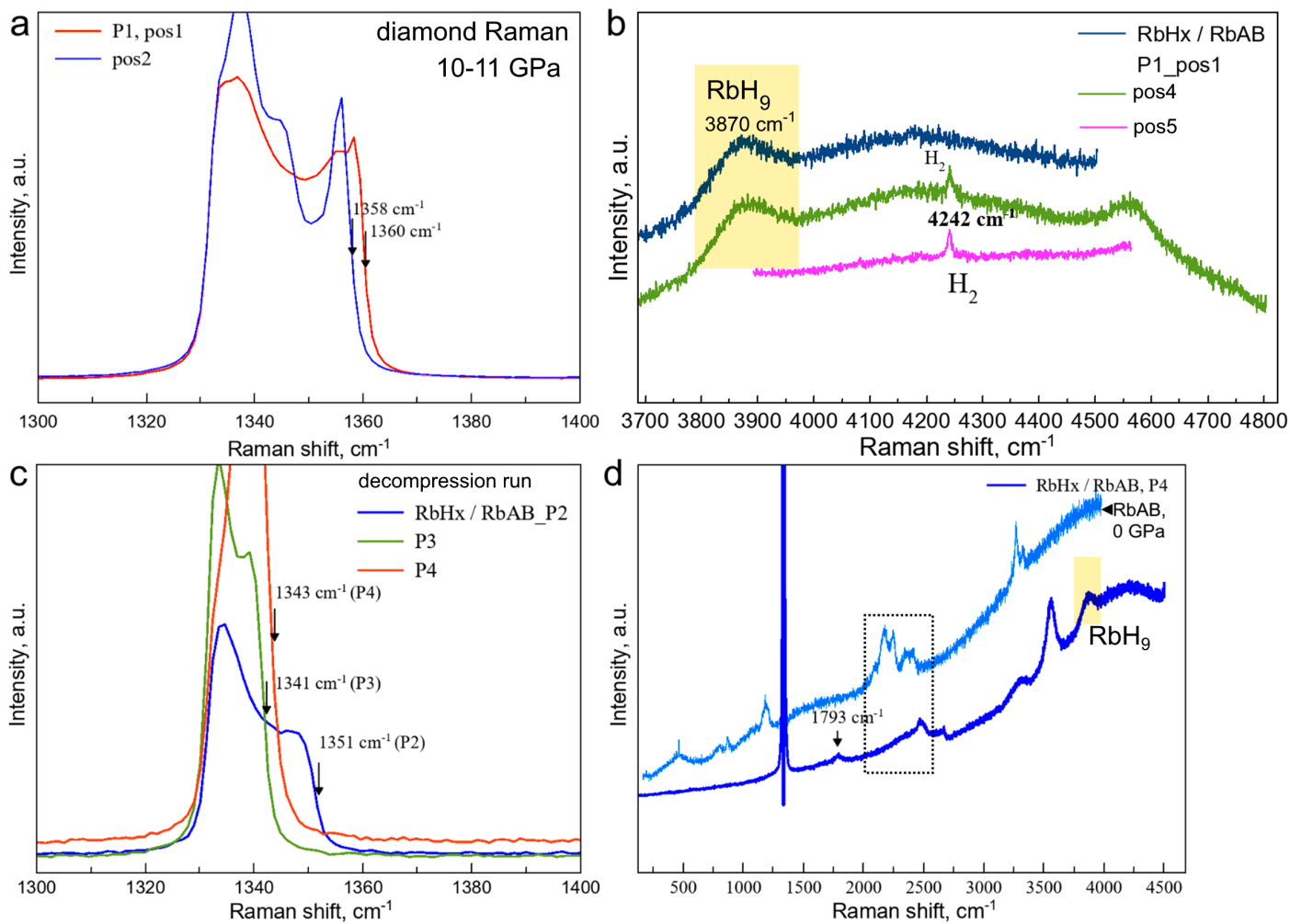

**Figure 15.** Raman spectra of RbH$_x$/RbAB/Au sample in DAC Y after laser heating at pressure P$_1$ ≈ 12 GPa, taken in (a) the region of C-C vibrations; (b) in the region of H-H vibrations for points #1, 2, 4. The pressure gradient between points 1 and 2 is ΔP = 1 GPa. (c) Decompression of DAC Y and compression P$_3$ → P$_4$. (d) Raman spectra of the sample RbH$_x$ at pressure P$_4$ = 4 GPa (dark blue) and starting RbAB at 0 GPa (light blue). There is an additional signal from the sample at 1793 cm$^{-1}$. After laser heating, the RbAB Raman peaks disappear. Pressure determination according to the Akahama scale [4]: 10-11 GPa (P$_1$), 7 GPa (P$_2$), 3 GPa (P$_3$), 4 GPa (P$_4$). According to the hydrogen vibron frequency detected at P$_1$ the pressure is 15 GPa ([29]). At 10 GPa this frequency should be 4230-4235 cm$^{-1}$, so H$_2$ vibron has higher frequency than it should be. Determination of pressure was also done by the equation of state of gold [30]: 12.5 GPa (P$_1$), 8.1 GPa (P$_2$), 1.7 GPa (P$_3$), 3.3 GPa (P$_4$).

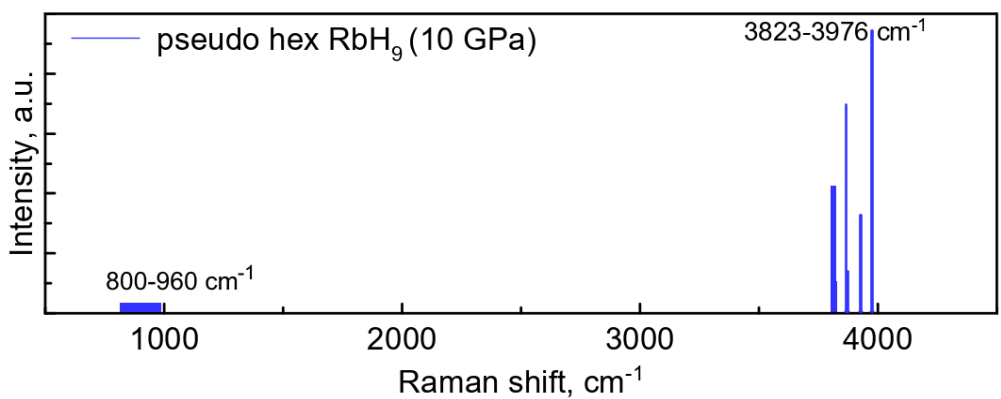

**Figure S16.** Predicted Raman spectrum of pseudo hexagonal RbH$_9$ at 10 GPa. Calculations were made using VASP and Raman off-resonant activity calculator [31]. As can be seen, RbH$_9$ should have a broad Raman peak at the 3823-3976 cm$^{-1}$, which corresponds to the experimental data (Figures S15b, d).

## 5. Reflection and transmission spectroscopy

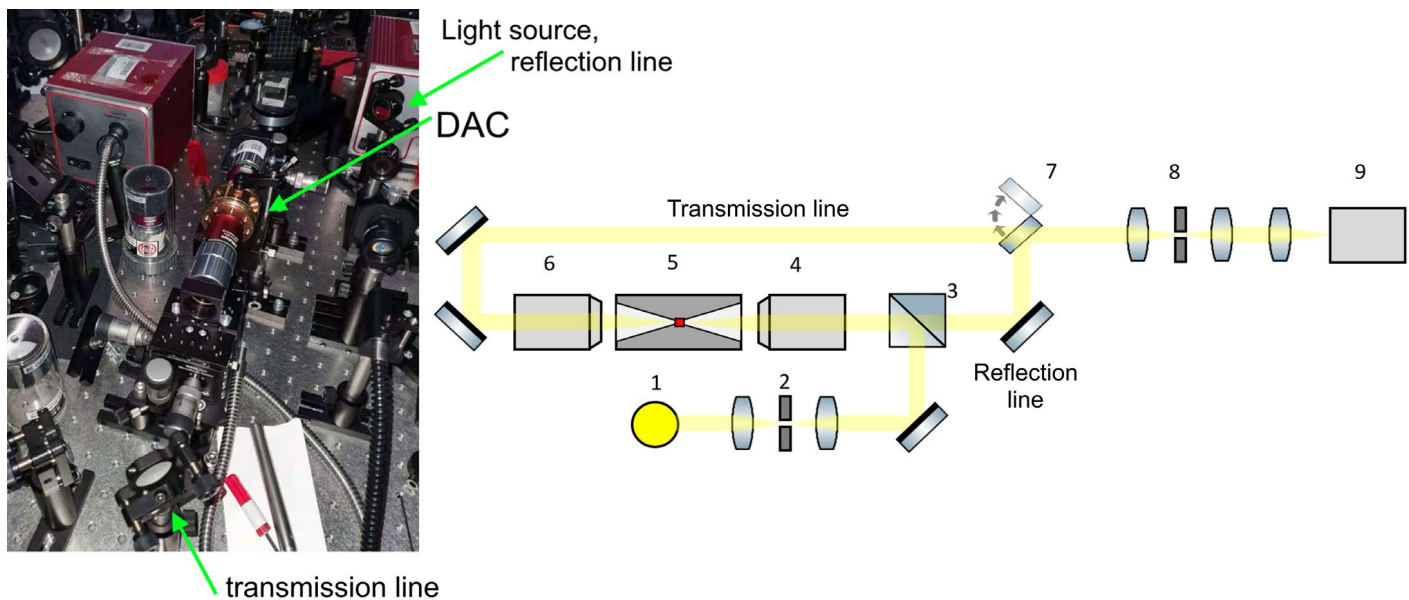

**Figure S17.** Optical scheme for transmission and reflection study of samples in DACs. 1 - Thorlabs OSL2B/OSL2BIR light source, 2 – a pinhole with diameter of 50 μm, 3 – beam splitter BS013, 4 and 6 - Mitutoyo plan apo NIR 20x infinity corrected objectives, 5 – diamond anvil cell, 7 – flip mirror, 8 – a pinhole-type spatial filter and 9 – spectrometer with CCD camera: Action SpectraPro SP-2750 and ProEM HS.

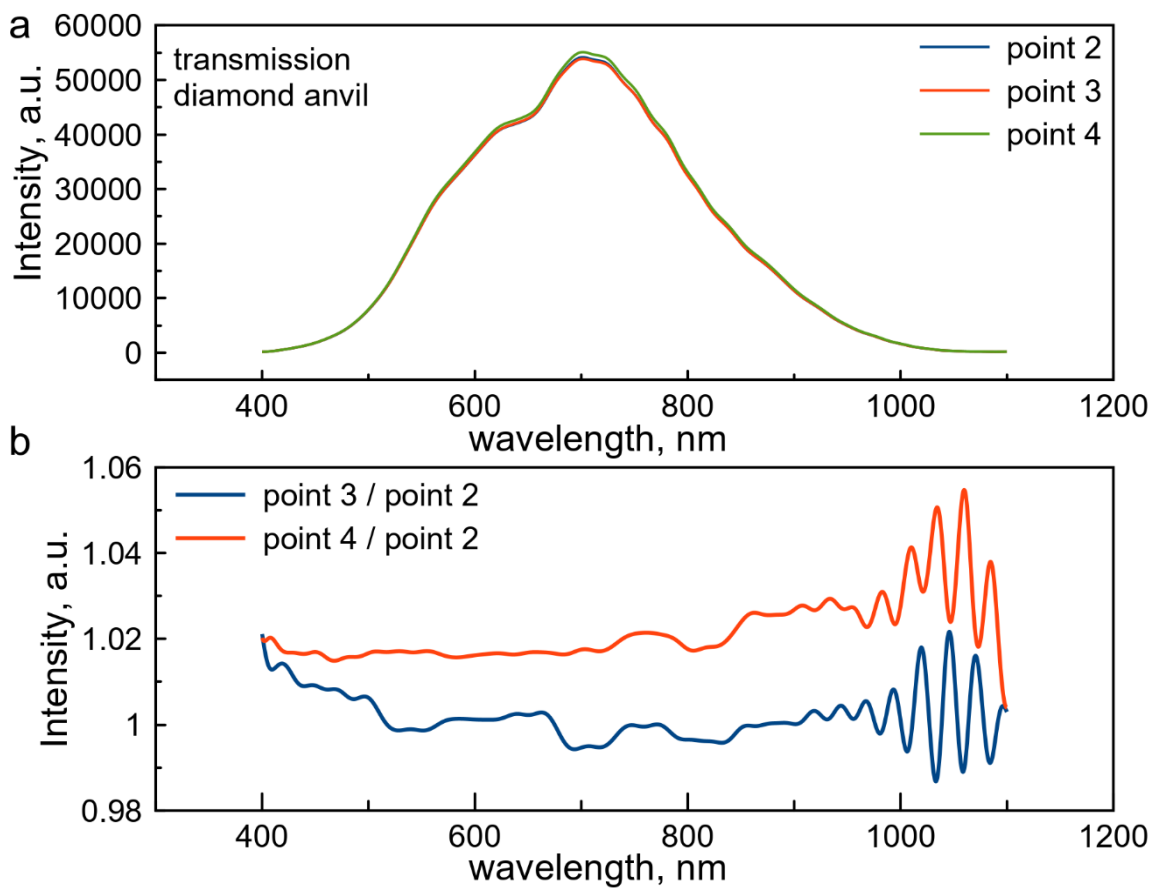

**Figure S18.** Optical measurements in DACs. (a) Fourier-filtered transmission spectra of a diamond anvil without sample measured at different points (2–4) of the culet (see also [32]). (b) Relative transmittance ($I_{sample}/I_{ref}$) of the DAC in points 3,4 with respect to point 2. Above 1000 nm, the signal-to-noise ratio is too low for reliable data interpretation.

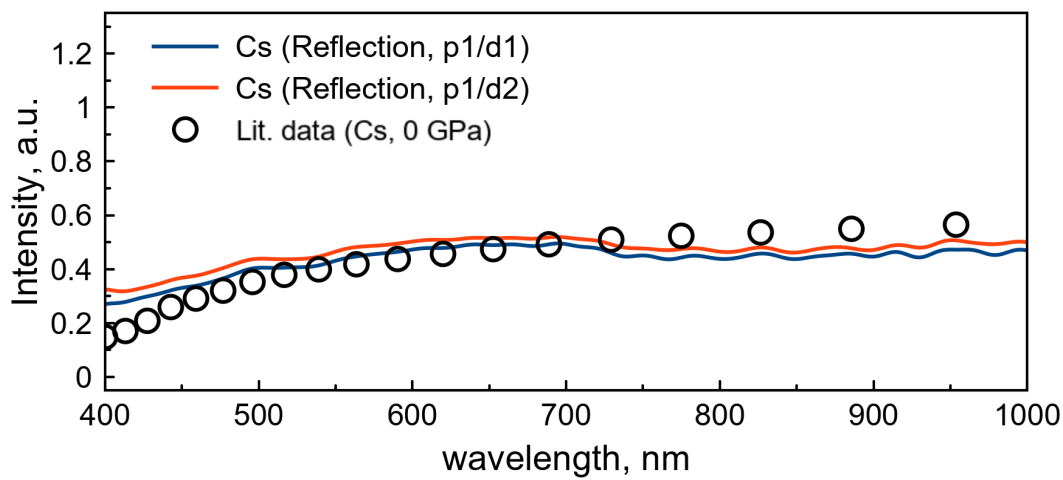

**Figure S19.** Fourier-filtered reflectance spectra of metallic cesium particle in DAC X3 at 20 GPa (p1 – means "point 1") compared with the literature data [33]. The reflection $R(\lambda)$ from two empty places on the diamond culet (d1, d2) was used as the reference.

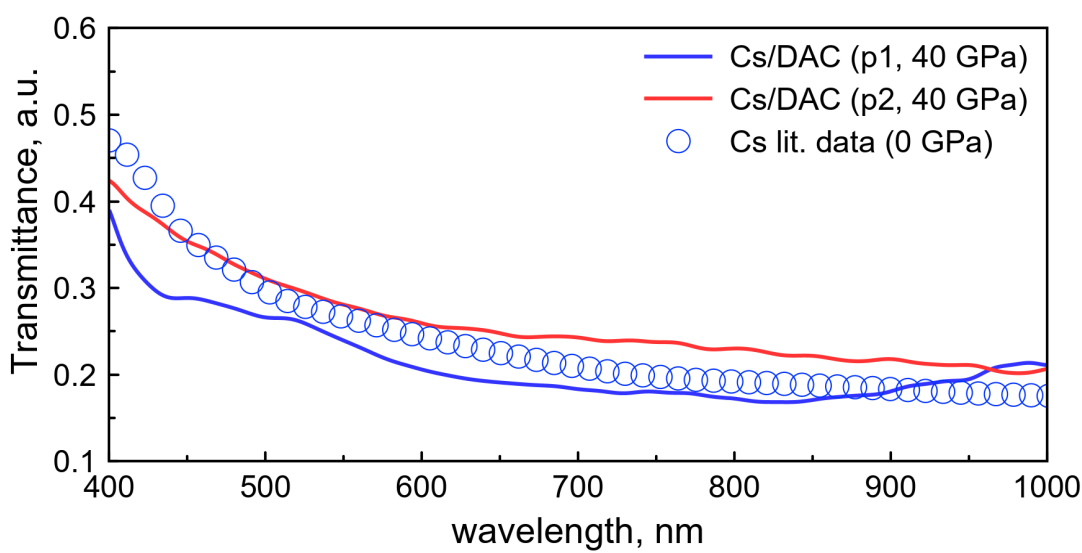

**Figure S20.** Fourier-filtered transmittance spectra of thin metallic cesium thin particle in DAC Z at 40-42 GPa (p1 – means "point 1") compared with the literature data [33] measured in two points (p1-p2). Empty place in DAC was used as the reference (see "diamond" in Figure S5).

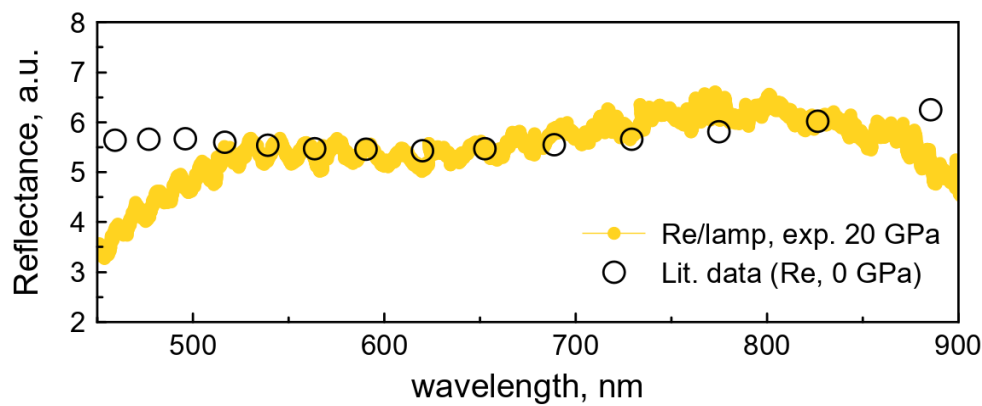

**Figure S21.** Fourier-filtered reflectance spectra of rhenium gasket at about 20 GPa (DAC X3) compared with the literature data [33]. Lamp light was used as the reference. The experiment shows that the lamp can be used in the range of 500-850 nm.

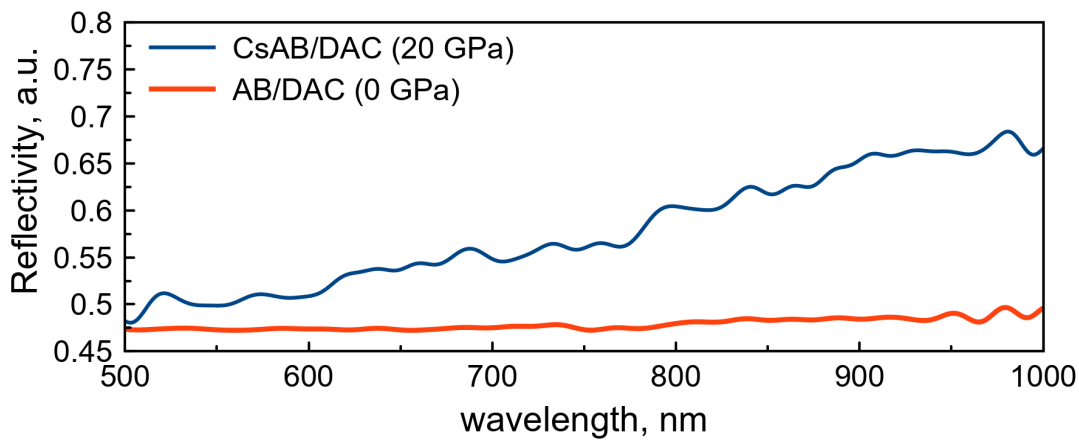

**Figure S22.** Fourier-filtered relative reflectance spectra of a DAC filled with AB compared to the same DAC with Cs amidoborane (CsAB). Transmission of the empty DAC was used as a reference. There is an important difference in CsAB and AB reflectance in the red and near IR range.

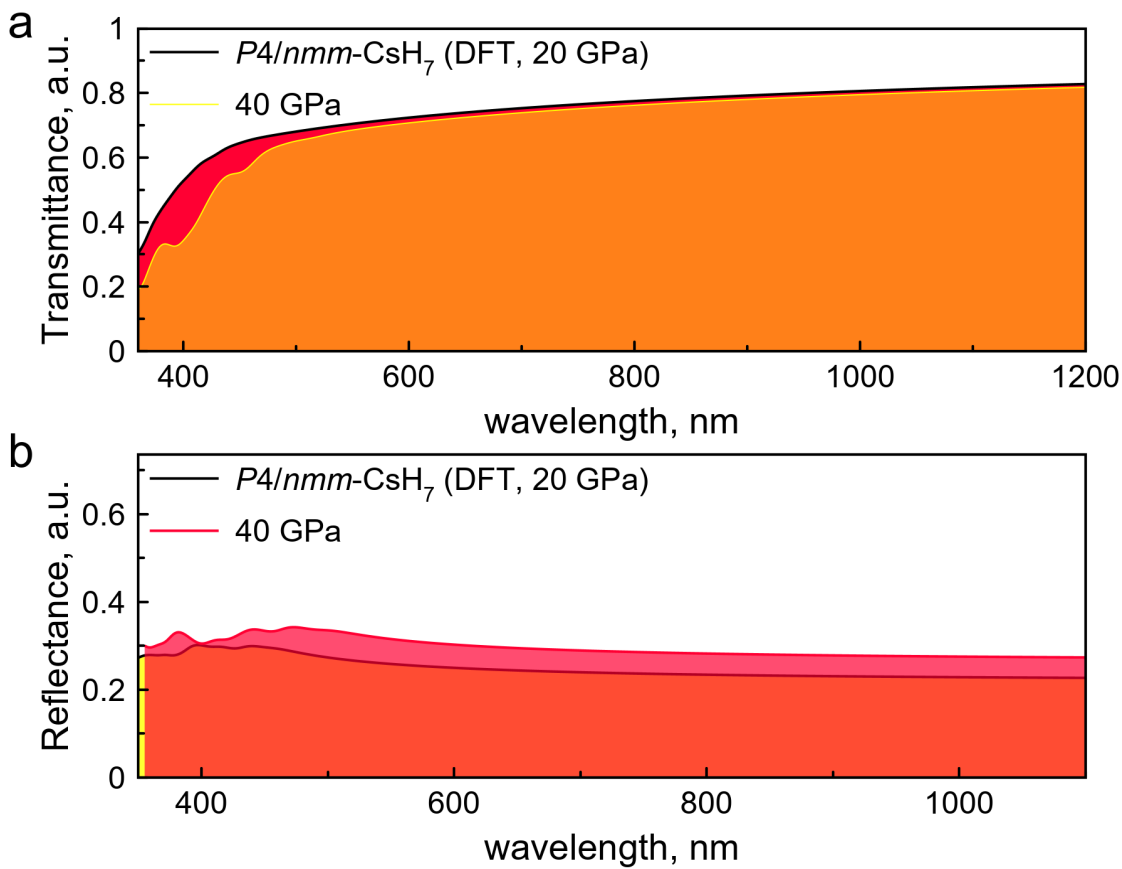

**Figure S23.** Calculated transmittance (a) and reflectance (b) of $P4/nmm$-$CsH_7$ at pressure of 20 and 40 GPa (VASP, PAW PBE). Expected color of $CsH_7$ in transmitted light is light yellow, while monohydride CsH is transparent.

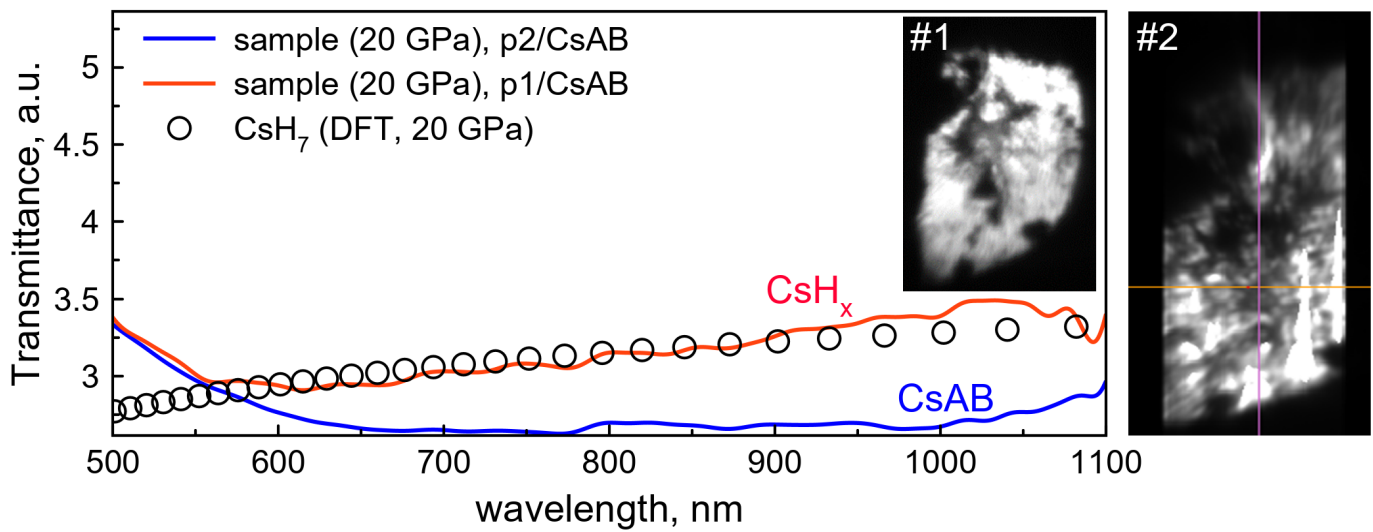

**Figure S24.** Fourier-filtered relative transmittance spectra of DAC X3, loaded with Cs/CsAB, after laser heating at 20 GPa studied in two points (p1, p2). Transmittance of pure CsAB was used as a reference (see Figure S25, grey line). Result of DFT calculations for CsH$_7$ is shown by black circles and was shifted up via multiplication by 4.07 ($T(\lambda)$ → 4.07×$T(\lambda)$). Insets: photographs of the investigated areas. Point #2 contains mostly CsAB, while point #1 (center) may contain CsH$_7$.

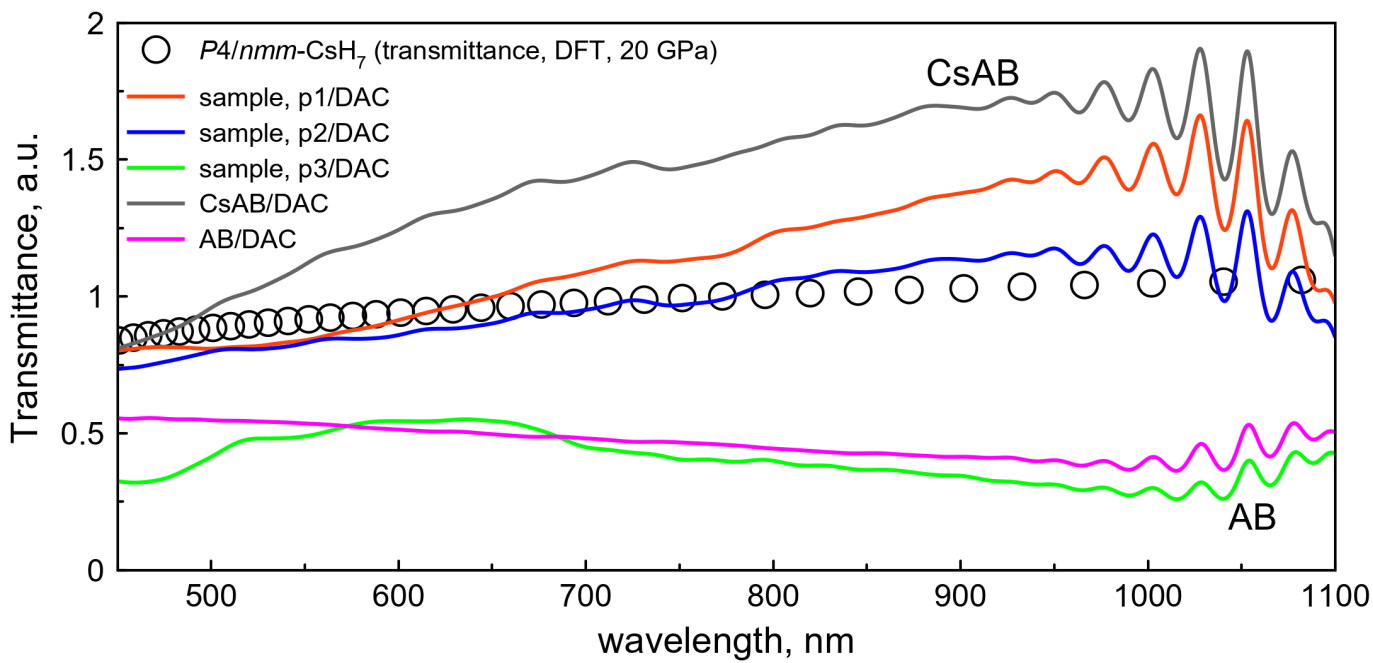

**Figure S25.** Fourier-filtered relative transmittance spectra of DAC X3, loaded with Cs/CsAB, after laser heating at 20 GPa studied in three points (p1, p2, p3). Transmittance of an empty DAC was used as a reference. With such reference CsAB and CsH$_x$ hydrides cannot be well distinguished (see p1, p2 and CsAB/DAC curves). But in the point p3 we have mainly unreacted AB which has a significantly different $T(\lambda)$. Result of DFT calculations for CsH$_7$ is shown by black circles and was shifted up through multiplication by 1.3 (i.e., $T(\lambda)$ → 1.3×$T(\lambda)$).

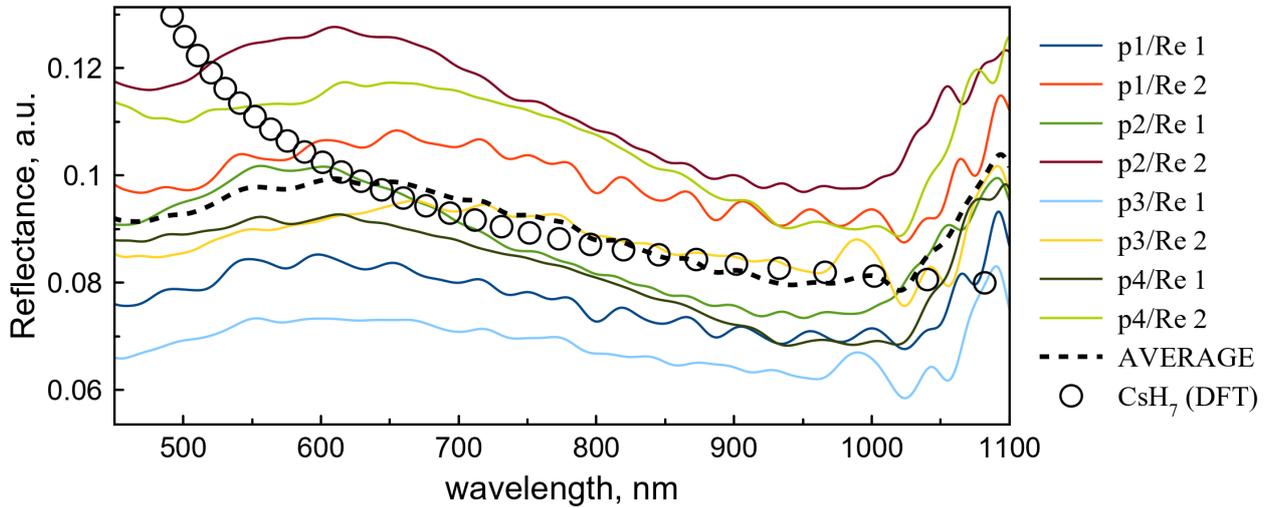

**Figure S26.** Fourier-filtered relative reflectance spectra of DAC X3, loaded with Cs/CsAB sample, after laser heating at 20 GPa studied in four points (p1-4). Reflectance of a Re gasket in two points (Re1, Re2) was used as a reference (see Ref. [32]). In all four investigated points behavior of $R(\lambda)$ is similar. Averaged $R(\lambda)$ corresponds to the predicted $CsH_7$ (black circles), except edges of the wavelength interval ($\approx$ 450 nm and 1100 nm) where signal/noise ratio is small. Result of DFT calculations for $CsH_7$ was shifted down via subtracting a constant of –0.147 (i.e., $R(\lambda) \rightarrow R(\lambda) - 0.147$).

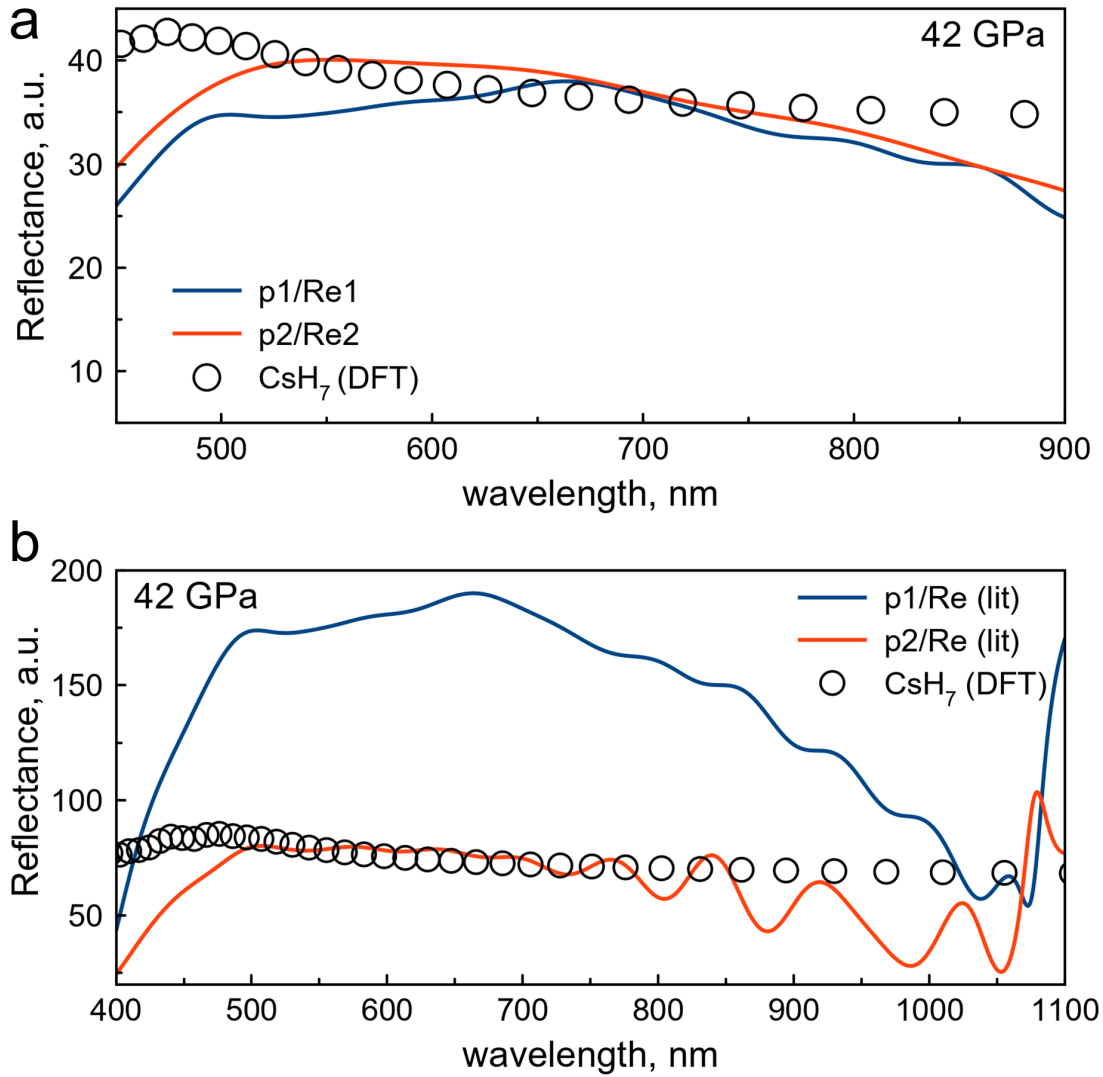

**Figure S27.** Fourier-filtered relative reflectance spectra of DAC Z, loaded with Cs/CsAB, after laser heating at 42 GPa studied in two points (p1-p2). (a) Reflectance of a Re gasket in two points (Re1, Re2) was used as a reference (see Ref. [32]). (b) Literature data on the reflectivity of Re [33] were used as a reference. In general, reflectivity $R(\lambda)$ decreases with increasing wavelength above 500 nm as predicted for $CsH_7$ by ab initio calculations (black circles). Result of DFT calculations for $CsH_7$ was shifted up through multiplication by (a) 125, (b) 250 as, for example, $R(\lambda) \rightarrow 250 \times R(\lambda)$.

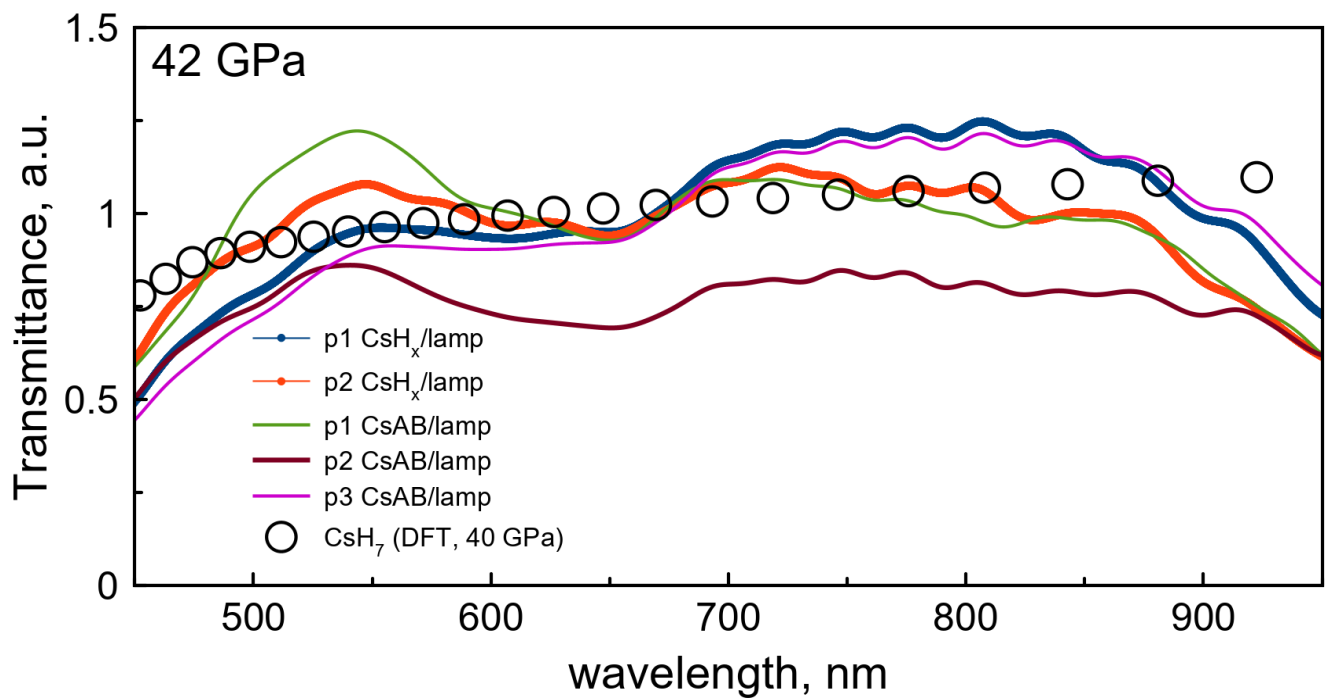

**Figure S28.** Fourier-filtered relative transmittance spectra of a DAC Z with Cs/CsAB/Au before and after laser heating at 42 GPa studied in several points. Lamp light was used as the reference. Points p1-2 correspond reaction products $CsH_x$ after laser heating. Points p1-3 correspond to initial compound, mainly CsAB. Result of DFT calculations for $CsH_7$ (black circles) was shifted up through multiplication by a factor of 1.4 (i.e., $T(\lambda)$ → $1.4 \times T(\lambda)$).

## 6. Powder X-ray diffraction data

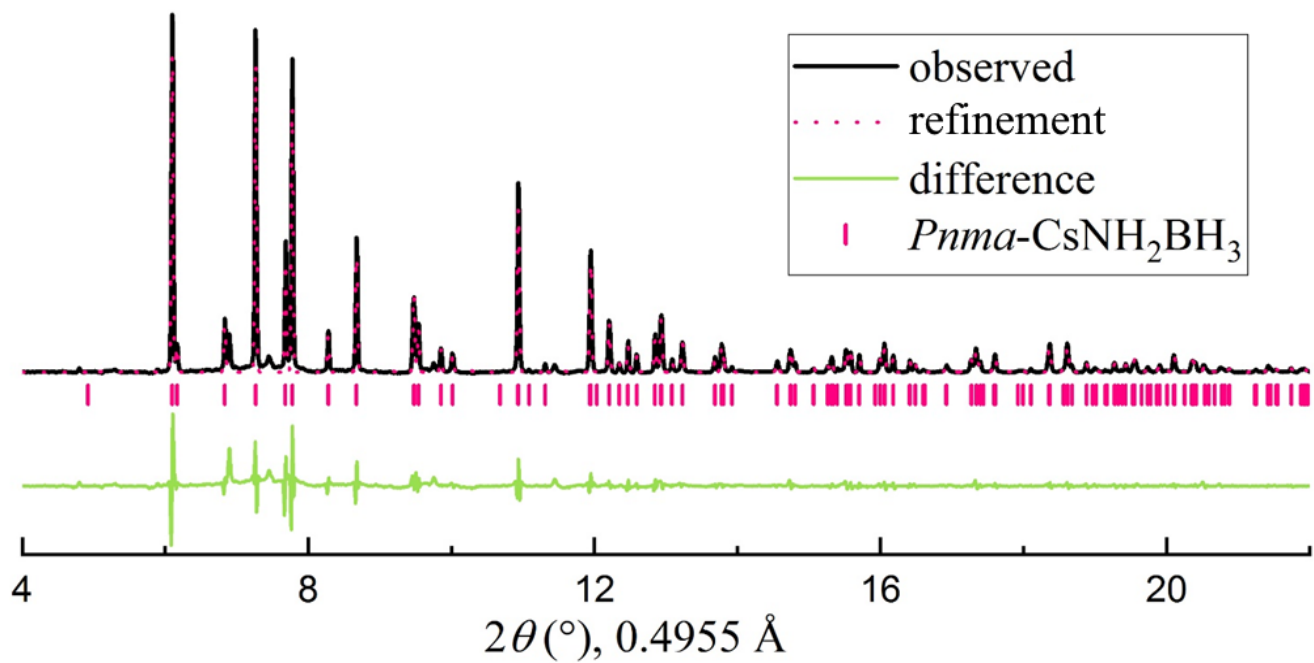

**Figure S29.** Synchrotron XRD pattern (λ = 0.4955 Å) of low-temperature (LT) modification *Pnma*-CsAB [27] at ambient pressure obtained from Cs metal and a solution of AB in THF in an Ar atmosphere. XRD was carried out in a sealed glass vial filled by Ar at the Xpress beamline (Elettra). The black line corresponds to the experimental spectrum, the pink dotted line is the Le Bail refinement, the green line is the difference between the experimental spectrum and the refinement.

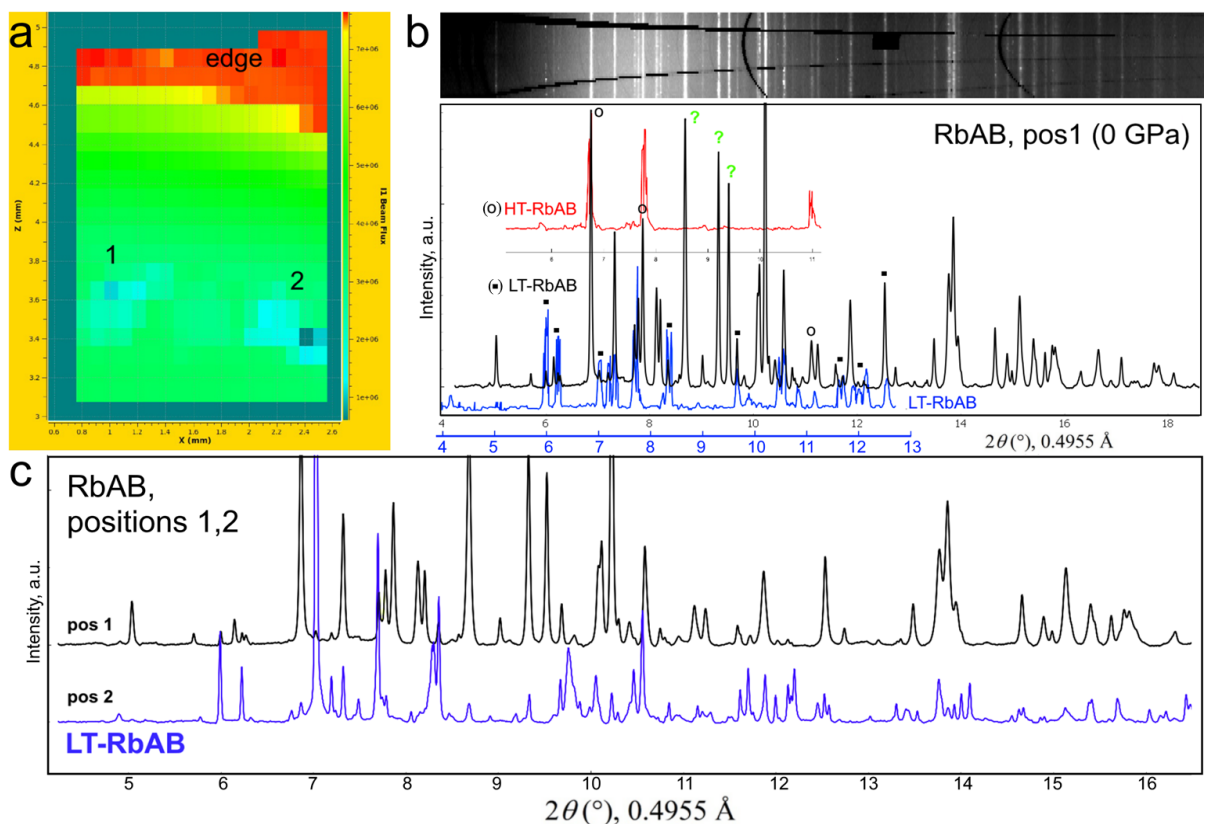

**Figure S30.** Synchrotron XRD patterns (λ = 0.4955 Å) of synthesized RbAB at ambient pressure. The samples were prepared from rubidium metal and a solution of AB in THF under an inert Ar atmosphere. (a) X-ray imaging of two polycrystals (pos 1, pos 2) in a sealed glass vial. (b) Diffraction pattern obtained from the sample in pos.1. The XRD spectrum contains reflections of low-temperature (LT, blue curve) and high-temperature (HT, red curve) modifications of RbAB, as well as an unidentified compound ("?" signs). (c) Diffraction pattern obtained from sample's pos. 2 (blue curve) compared to the pos. 1 (black curve). Sample in position 2 is almost pure LT-RbAB.

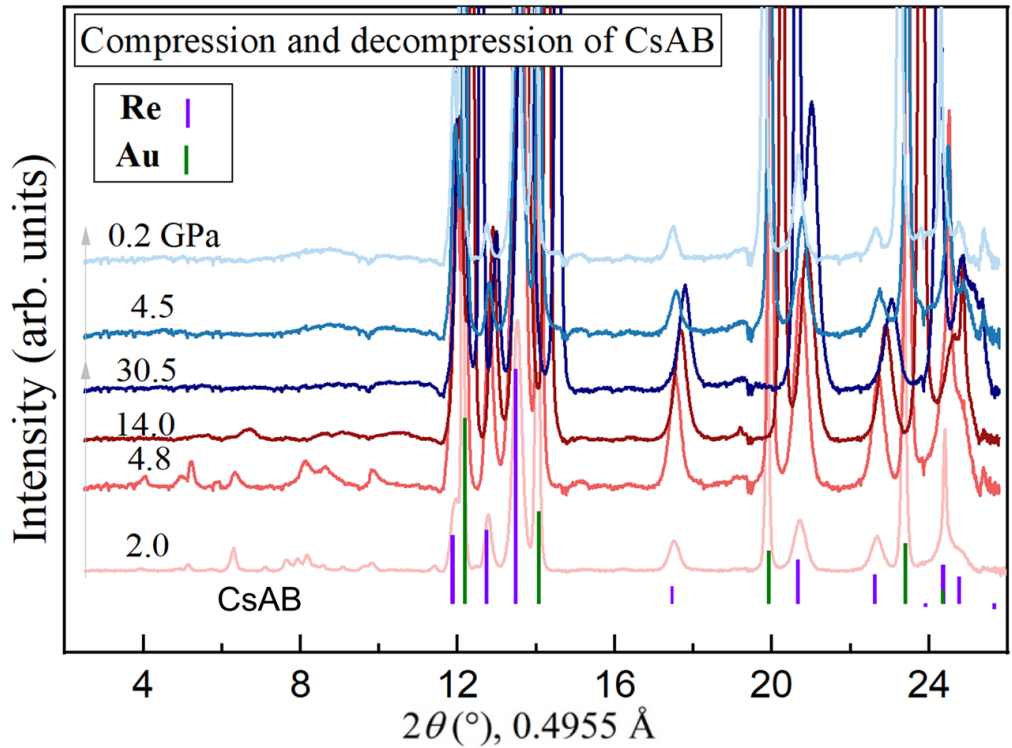

**Figure S31.** A series of XRD patterns (λ = 0.4955 Å) of pure CsAB compressed from 2 GPa to 30.5 GPa in a DAC, and then decompressed back to 0.2 GPa without laser heating. Gold (Au) piece was used as a pressure sensor. Re gasket was used. Experiment was done at the Elettra synchrotron radiation facility (Italy), Xpress beamline. Beam energy was 25 keV, the beam diameter was 80 and 50 μm. Cesium polyhydrides and CsAB correspond to the region 2θ < 12°. It is easy to see that above 14 GPa, almost complete amorphization of CsAB is observed, which turns out to be irreversible as the pressure decreases.

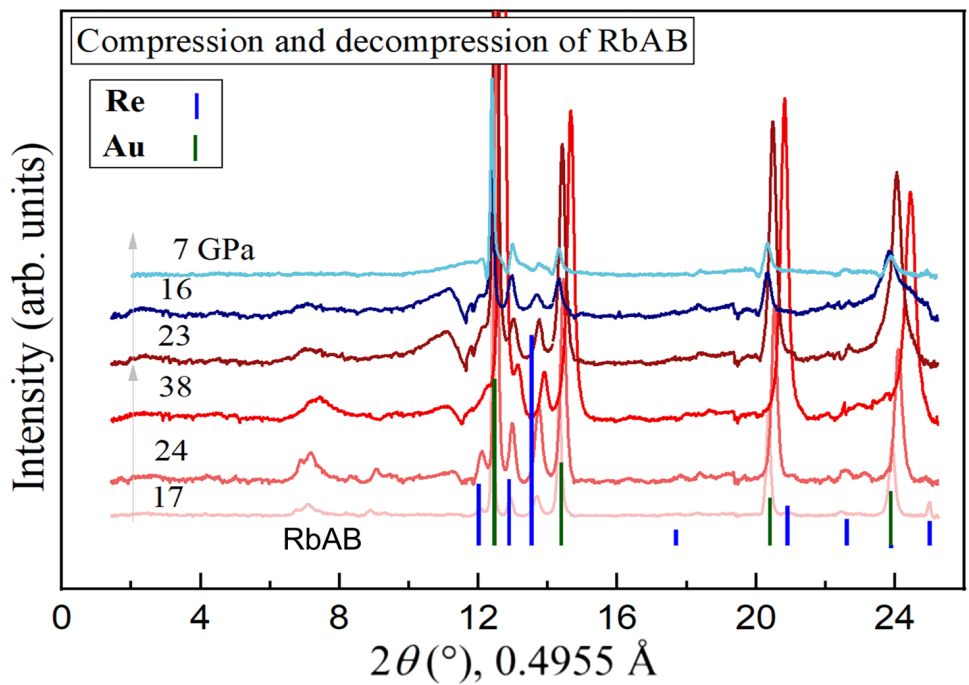

**Figure S32.** A series of XRD patterns (λ = 0.4955 Å) of pure RbAB compressed from 0 GPa to 38 GPa in a DAC, and then decompressed back to 7 GPa. Gold (Au) piece was used as a pressure sensor. Re gasket was used. Experiment was done at the Elettra synchrotron radiation facility (Italy), Xpress beamline. Beam energy was 25 keV, the beam diameter was 80 and 50 μm. Rubidium polyhydrides and RbAB correspond to the region 2θ < 12°. It is easy to see that above 24 GPa, partial amorphization of RbAB is observed, which turns out to be irreversible as the pressure decreases. However, residual RbAB makes a small contribution to the diffraction pattern of Rb polyhydrides, giving a broad peak around 7-7.5°.

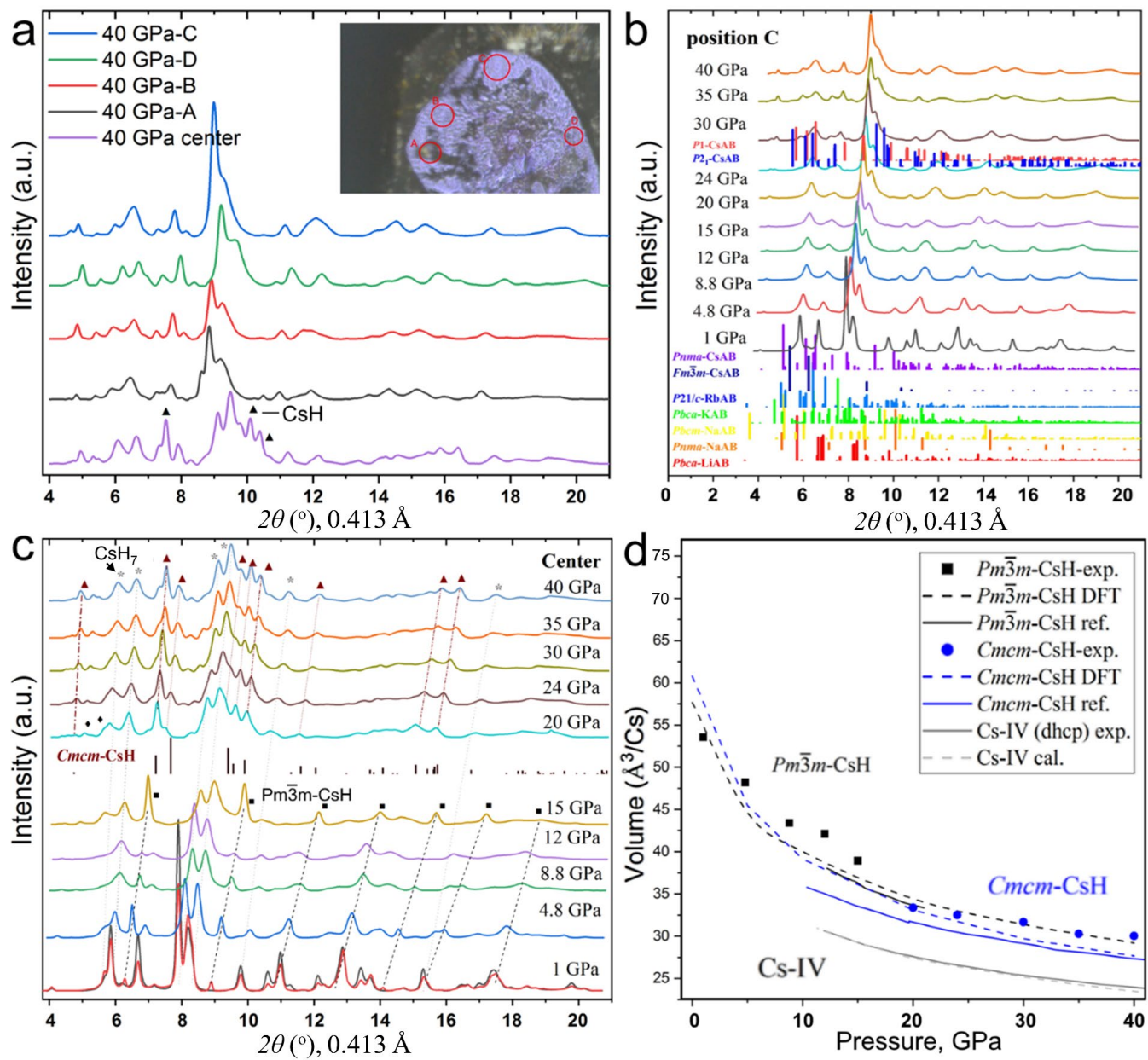

**Figure S33.** X-ray diffraction patterns of Cs/CsAB sample after laser heating in DAC X1 at around 40 GPa and below this pressure. (a) Comparison of XRD patterns measured at different points (A, B, C, D and "center") of the sample. In almost all points there are diffraction peaks from the reaction products: CsH and $CsH_7$. (b) Comparison of the XRD patterns, obtained in our experiment, with previously studied amidoboranes of various alkali metals. None of the previously studied amidoboranes can explain the observed diffraction patterns. (c) Change of XRD patterns in the central part of the sample during decompression of DAC X1 from 40 to 1 GPa. The phase transition of cesium monohydride from $Cmcm$ to $Pm\bar{3}m$ can be seen. Moreover, in panels (b) and (c) there is a specific doublet at ~6° that belongs to $CsH_7$ and can be tracked at low-pressure patterns down to 1 GPa. (d) Comparison of experimental cell parameters of CsH with the results of previous studies. The error in the cell volume reaches 7-8%, which may indicate a non-stoichiometric composition of cesium hydride ($CsH_{1+x}$).

**Table S2.** Experimental unit cell parameters of $Pm\bar{3}m$-CsH (Z=1) and $Cmcm$-CsH (Z = 4), DAC X1.

| Pressure, GPa | V, Å³/Cs | Pressure, GPa | a, Å | b, Å | c, Å | V, Å³/Cs |
|---|---|---|---|---|---|---|
| $Pm\bar{3}m$-CsH | | | | $Cmcm$-CsH | | |
| 1 | 53.56 | 20 | 3.453 | 9.865 | 3.919 | 33.38 |
| 4.8 | 48.20 | 24 | 3.412 | 9.720 | 3.922 | 32.52 |
| 8.8 | 43.41 | 30 | 3.376 | 9.616 | 3.904 | 31.68 |
| 12 | 42.12 | 35 | 3.332 | 9.484 | 3.832 | 30.28 |
| 15 | 38.94 | 40 | 3.318 | 9.445 | 3.832 | 30.02 |

**Table S3.** Experimental unit cell parameters of *P4/nmm*-CsH$_7$ (Z = 2, slightly distorted), obtained from XRD of the center of DAC X1. For DFT calculations we used ENCUT = 300 eV.

| Pressure, GPa | a, Å | c, Å | V, Å$^3$/Cs | V$_{DFT}$, Å$^3$/Cs |
|---|---|---|---|---|
| 1 | 5.753 | 4.447 | 73.6 | 73.0* |
| 4.8 | 5.602 | 4.329 | 67.9 | 66.8* |
| 8.8 | 5.436 | 4.243 | 62.7 | 61.2* |
| 12 | 5.407 | 4.198 | 61.3 | 59.9 |
| 15 | 5.272 | 4.209 | 58.5 | 56.8 |
| 20 | 5.173 | 4.154 | 55.6 | 53.3 |
| 24 | 5.117 | 4.024 | 52.7 | 52.9 |
| 30 | 5.067 | 3.886 | 49.9 | 48.6 |
| 40 | 5.015 | 3.858 | 48.5 | 45.2 |

*For comparison purposes we performed calculations with accounting the van der Waals interactions (IVDW=11, vdw_kernel.bindat)

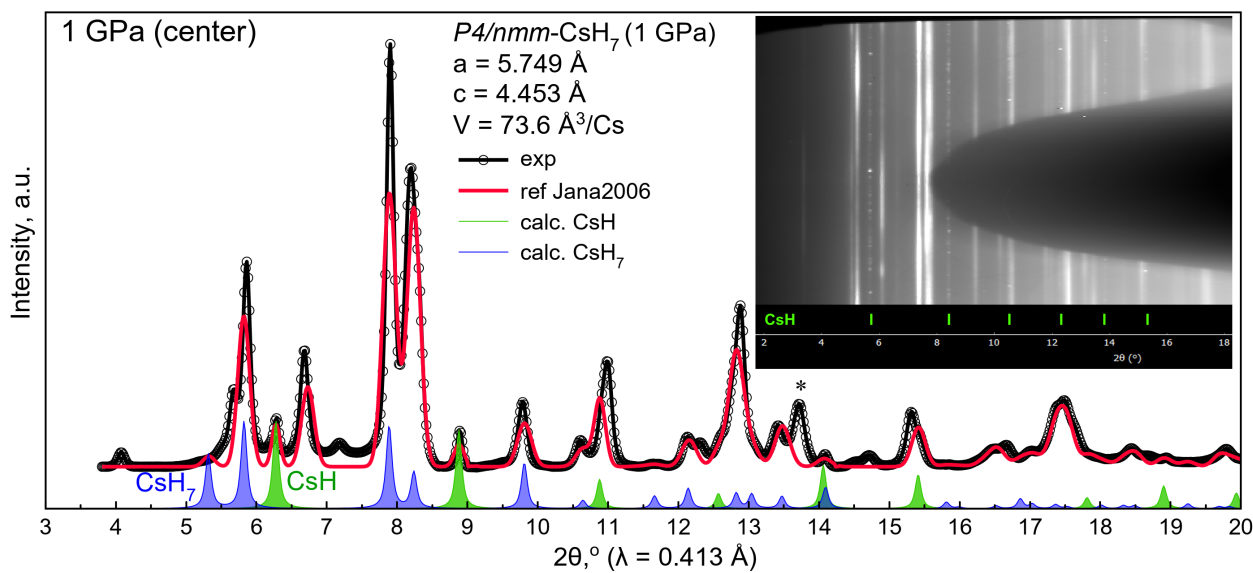

**Figure S34.** Experimental X-ray diffraction and Le Bail refinement of unit cell parameters of *Pm$\bar{3}$m*-CsH and *P4/nmm*-CsH$_7$ at 1 GPa, decompression run. Inset: diffraction image ("cake"), center of the sample. Sample obviously contains sc-CsH, but the refinement for CsH$_7$ is not ideal. Despite the presence of the main reflections, a change in the structure of the XRD signal at 5-6°, as well as a high reflection intensity at 6.7°, indicate a possible distortion of the tetragonal structure of CsH$_7$ or its decomposition.

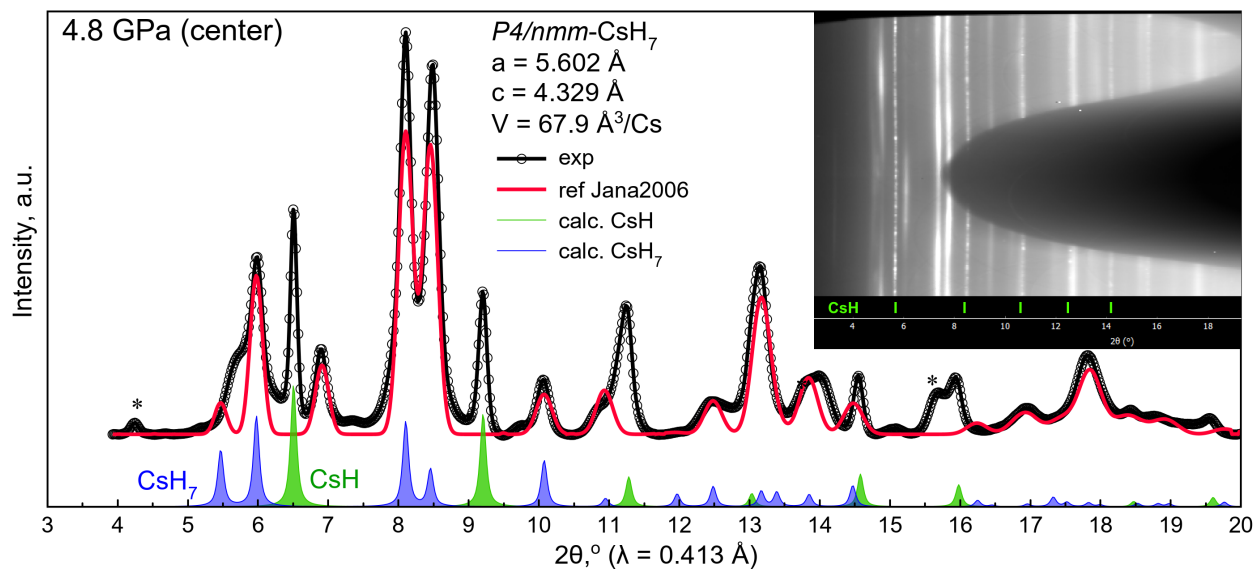

**Figure S35.** Experimental X-ray diffraction and Le Bail refinement of unit cell parameters of *Pm$\bar{3}$m*-CsH and *P4/nmm*-CsH$_7$ at 4.8 GPa, decompression run. Inset: diffraction image ("cake"), center of the sample. Sample contains sc-CsH, but the refinement for CsH$_7$ is not ideal which may indicate a distortion of the tetragonal structure of CsH$_7$.

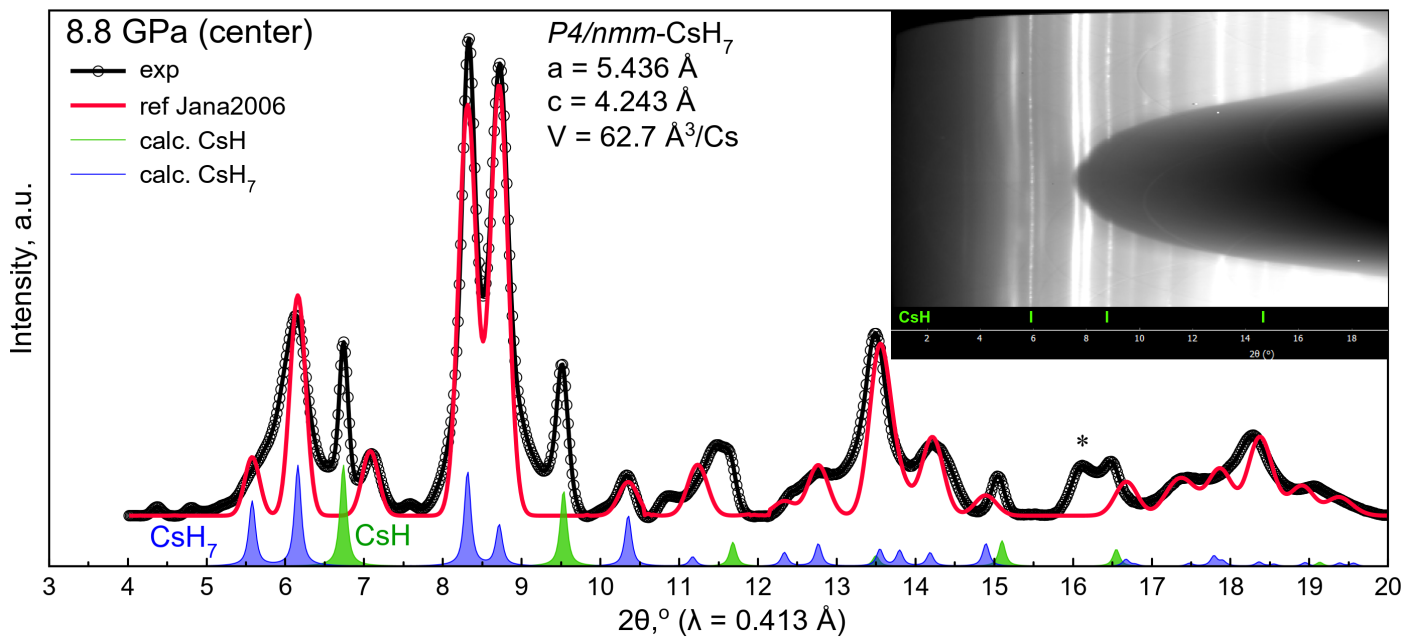

**Figure S36.** Experimental X-ray diffraction and Le Bail refinement of unit cell parameters of $Pm\bar{3}m$-CsH and $P4/nmm$-CsH$_7$ at 8.8 GPa, decompression run. Inset: diffraction image ("cake"), center of the sample. Sample contains sc-CsH, but the refinement for CsH$_7$ is not ideal which may indicate a distortion of the tetragonal structure of CsH$_7$.

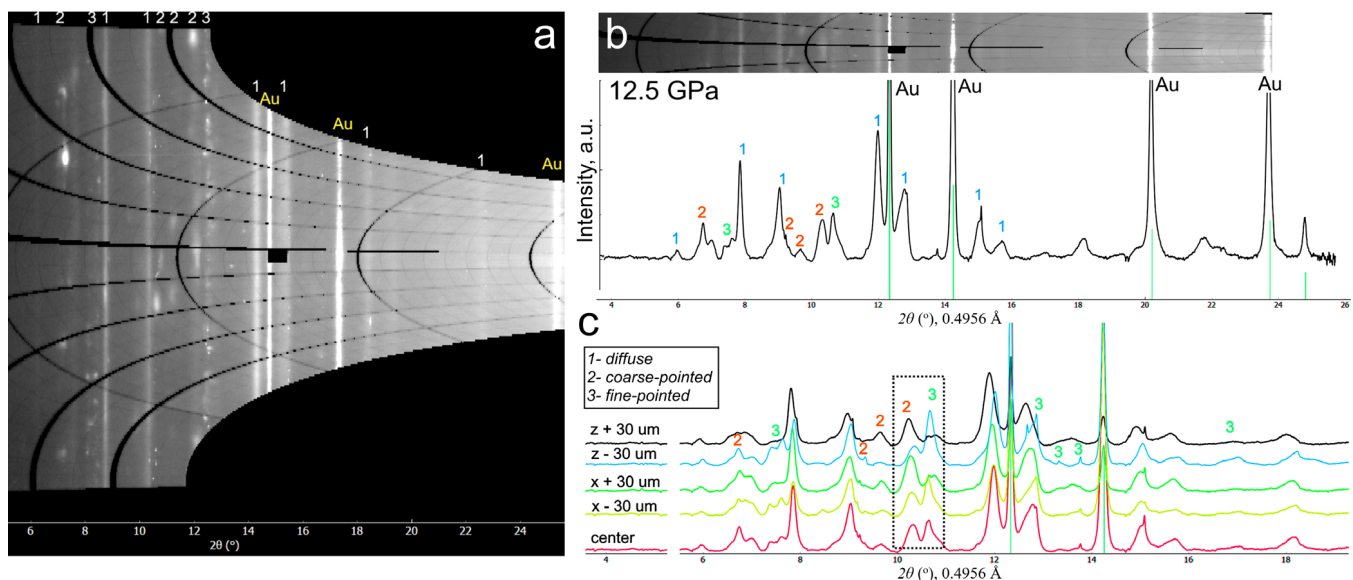

**Figure S37.** X-ray diffraction patterns ($\lambda = 0.4956$ Å, Elettra) in different points of Rb/RbAB/Au sample after laser heating in DAC Y at around 12.5 GPa. (a) XRD image ("cake") that allows one to qualitatively distinguish three (1-3) different phases in the mixture. (b) XRD pattern measured in the center of sample. (c) XRD pattern measured on the sample edges with a shift from the center of ±30 μm. As shown in the main text of the article, 1 – is pseudo tetragonal RbH$_{9-x}$, 2 – is residual RbAB, 3 – is pseudo hexagonal RbH$_{9-x}$.

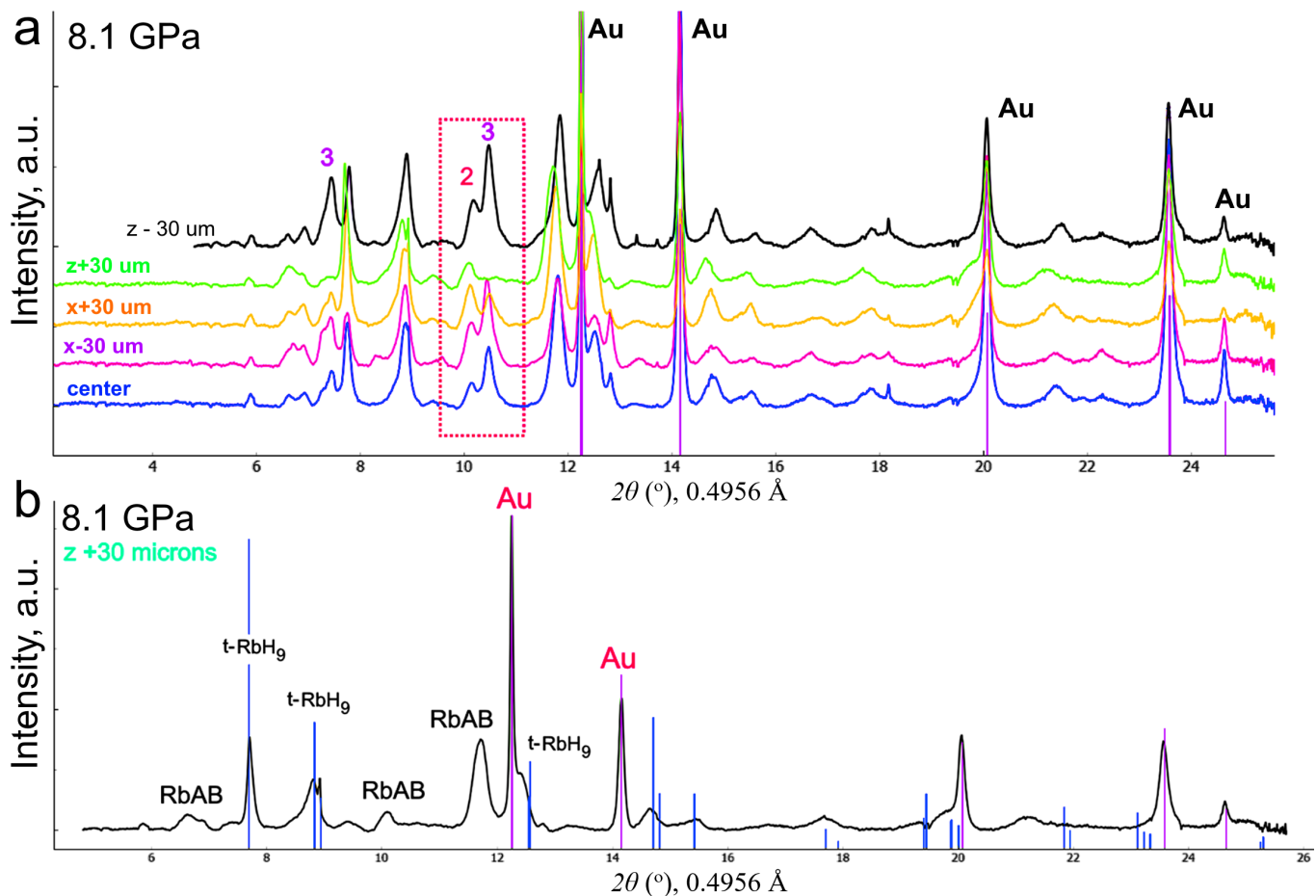

**Figure S38.** X-ray diffraction patterns (λ = 0.4956 Å, Elettra) in different points of Rb/RbAB/Au sample after laser heating in DAC Y decompressed to 8.1 GPa. (a) A series of XRD patterns measured over the sample with a shift from the center of ±30 μm. As shown in the main text of the article, 2 – is residual RbAB, 3 – is pseudo hexagonal $RbH_{9-x}$. (b) XRD pattern measured in the z+30 μm point of the sample. The main phase at this point is pseudo tetragonal $t\text{-}RbH_{9-x}$ (mixed with starting RbAB), the expected reflections of which are shown in blue lines (Dioptas 0.5 software [1]).

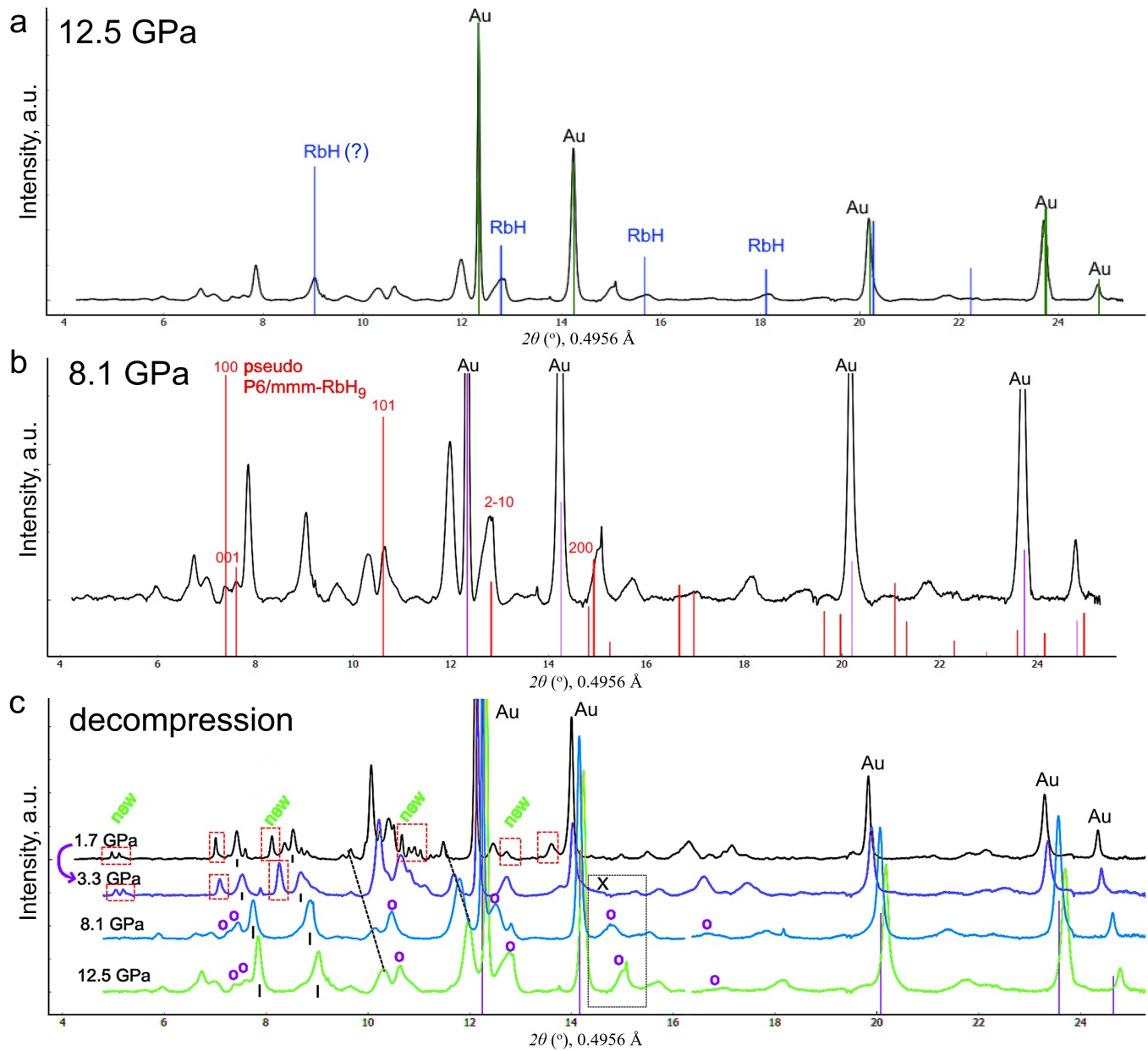

**Figure S39.** X-ray diffraction patterns (λ = 0.4956 Å, Elettra) in different points of Rb/RbAB/Au sample after laser heating in DAC Y. (a) X-ray diffraction pattern at 12.5 GPa which indicates possibility of formation of rubidium monohydride (several XRD peaks coincide with pseudo tetragonal $t$-RbH$_{9-x}$). (b) XRD pattern at 8.1 GPa with approximate peak indexing performed using pseudo hexagonal phase RbH$_{9-x}$ (Dioptas 0.5 software [1]). (c) A series of XRD patterns measured during decompression of DAC Y from 12.5 GPa down to 1.7 GPa. The pressure was initially reduced to 1.7 GPa and then increased again to 3.3 GPa (see arrow in panel (c)). One can see the formation of new compounds at 1.7 GPa ("new", red rectangles), which persist as the pressure increases. Marks "o" correspond to pseudo hexagonal RbH$_{9-x}$, whereas dashes "|" correspond to pseudo tetragonal RbH$_{9-x}$.

**Table S4.** Experimental and calculated unit cell parameters of $Pm\bar{3}m$-RbH (Z = 1) and pseudo $P6/mmm$-RbH$_{9-x}$ (Z = 1).

| | $Pm\bar{3}m$-RbH | | | pseudo $P6/mmm$-RbH$_{9-x}$ | | | |
|---|---|---|---|---|---|---|---|
| Pressure, GPa | a, Å | V(exp), Å$^3$/Rb | V(theory), Å$^3$/Rb | a, Å | c, Å | V(exp), Å$^3$/Rb | V(theory), Å$^3$/Rb |
| 12.5 | 3.148 | 31.20 | 31.9 | 4.44 | 3.74 | 63.88 | 65.67 (69.75)* |
| 8.1 | 3.215 | 33.23 | 34.74 | 4.41 | 3.86 | 65.01 | 72.39 (76.81)* |
| 3.3 | 3.444 | 40.88 | 40.36 | decomposed | | | 85.03 |
| 1.7 | 3.505 | 43.06 | 43.17 | decomposed | | | 91.7 |

*Calculations were done with the Van der Waals correction to interatomic interactions (IVDW=11, vdw_kernel.bindat).

**Table S5.** Experimental and calculated unit cell parameters of possible pseudo $I4/mmm$-RbH$_{9-x}$ ($Z = 2$). A large discrepancy between the calculated and experimental unit cell volume at low pressure may indicate loss of molecular hydrogen during decompression. At 3.3 GPa, the refinement quality is already significantly lower.

| | pseudo $I4/mmm$-RbH$_{9-x}$ | | | |
|---|---|---|---|---|
| **Pressure, GPa** | **a, Å** | **c, Å** | **V(exp), Å³/Rb** | **V(theory), Å³/Rb** |
| 12.5 | 4.43 | 6.29 | 61.7 | 67.76 (69.9)* |
| 8.1 | 4.52 | 6.26 | 64.1 | 73.94 (76.81)* |
| 3.3 | ≈4.87 | ≈6.91 | ≈81.9 | 86.86 |

*Calculations were done with the Van der Waals correction to interatomic interactions (IVDW=11, vdw_kernel.bindat).

# 7. Theoretical calculations

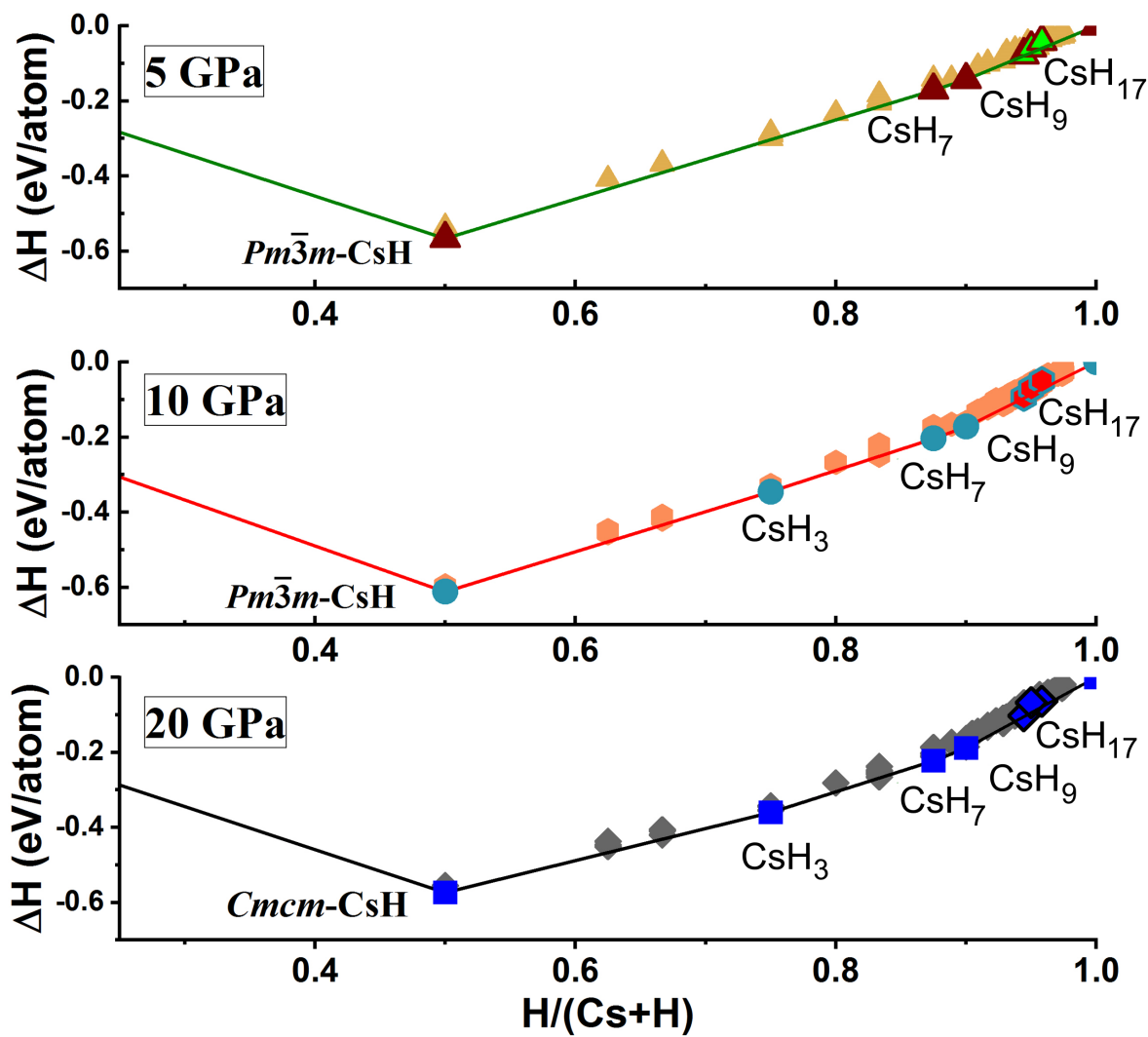

**Figure S40.** Thermodynamically stable phases in the Cs-H system at pressures of 5, 10 and 20 GPa. Calculations were carried out considering ZPE using USPEX [6-8] and VASP [15-17] codes.

**Table S6.** Part of results of an evolutionary search for the most stable structures in the Rb-H system at 10 GPa (27 generations, 80 structures in each). PBE pseudopotentials, USPEX [6-8] and VASP [15-17] codes were used.

| Identification number | Rb | H | Enthalpy, eV/atom | Volume, Å³/atom | Fitness, eV/block | Symmetry group | Comments |
|---|---|---|---|---|---|---|---|
| 102 | 0.0 | 16.0 | -2.9567 | 5.0671 | 0.0 | 7 | Molecular hydrogen ($H_2$) |
| 194 | 6.0 | 0.0 | 2.3551 | 39.2729 | 0.0 | 1 | Rubidium (Rb) |
| **980** | **1.0** | **9.0** | **-2.6138** | **6.9979** | **0.0** | **1** | **Pseudo tetragonal $RbH_9$** |
| **2449** | **3.0** | **3.0** | **-1.0623** | **16.6662** | **0.0** | **221** | **sc-RbH** |
| 1316 | 7.0 | 0.0 | 2.3561 | 38.6299 | 0.001 | 2 | Rb |
| 171 | 0.0 | 8.0 | -2.9555 | 5.0581 | 0.0013 | 1 | $H_2$ |
| 1565 | 2.0 | 0.0 | 2.3566 | 38.5236 | 0.0015 | 139 | Rb |

| 57 | 0.0 | 24.0 | -2.9547 | 5.0719 | 0.0021 | 1 | $H_2$ |
| 142 | 14.0 | 0.0 | 2.3572 | 38.9963 | 0.0021 | 1 | Rb |
| **1** | **4.0** | **36.0** | **-2.6115** | **6.8578** | **0.0023** | **66** | **Pseudo hexagonal $RbH_9$** |
| 2241 | 4.0 | 0.0 | 2.3579 | 38.6486 | 0.0028 | 2 | Rb |
| 778 | 2.0 | 0.0 | 2.3579 | 39.115 | 0.0029 | 1 | Rb |
| 2664 | 8.0 | 0.0 | 2.3584 | 37.8916 | 0.0033 | 225 | Rb |
| 1553 | 5.0 | 0.0 | 2.3585 | 38.0571 | 0.0034 | 1 | Rb |
| 1108 | 8.0 | 0.0 | 2.3587 | 38.6925 | 0.0036 | 2 | Rb |
| 2122 | 6.0 | 0.0 | 2.3592 | 38.3738 | 0.0041 | 1 | Rb |
| 195 | 0.0 | 6.0 | -2.9523 | 5.1079 | 0.0045 | 5 | $H_2$ |
| 174 | 0.0 | 8.0 | -2.9518 | 5.1233 | 0.0049 | 4 | $H_2$ |
| 3036 | 6.0 | 0.0 | 2.3602 | 38.3594 | 0.0051 | 58 | Rb |
| 1417 | 10.0 | 0.0 | 2.3606 | 38.1673 | 0.0056 | 1 | Rb |
| 23 | 0.0 | 32.0 | -2.9503 | 5.106 | 0.0065 | 3 | $H_2$ |
| 1790 | 6.0 | 0.0 | 2.3621 | 38.2211 | 0.007 | 1 | Rb |
| **1158** | **1.0** | **17.0** | **-2.7588** | **6.1544** | **0.0074** | **1** | **$RbH_{17}$** |
| **2619** | **1.0** | **19.0** | **-2.7778** | **6.031** | **0.0074** | **1** | **$RbH_{19}$** |
| **235** | **4.0** | **20.0** | **-2.3472** | **8.3137** | **0.008** | **36** | **$RbH_5$** |
| 213 | 2.0 | 0.0 | 2.3631 | 39.1415 | 0.0081 | 229 | Rb |

**Table S7**. Crystal structures (POSCARS) of the most important rubidium hydrides found by USPEX code at 10 GPa.

| Hydrogen | Rubidium |
|---|---|
| EA102  8.415 2.438 3.952 90.35 89.95 90.04 Sym.group:  7<br>1.0<br>  8.414646   0.005475   -0.024468<br>  -0.003270   2.437968   -0.005095<br>  0.015295   -0.015849   3.951953<br>  H<br>  16<br>Direct<br>  0.290526   0.348730   0.289403<br>  0.259009   0.032228   0.805214<br>  0.962023   0.532448   0.917835<br>  0.199936   0.852355   0.890475<br>  0.335387   0.530236   0.407659<br>  0.504582   0.357087   0.807198<br>  0.036874   0.031578   0.306998<br>  0.455136   0.550628   0.908764<br>  0.790699   0.347839   0.297674<br>  0.758528   0.047539   0.816151<br>  0.591106   0.857706   0.395812<br>  0.700607   0.853692   0.892889<br>  0.836265   0.539169   0.408613<br>  0.997803   0.345829   0.788887<br>  0.535920   0.062083   0.320907 | EA194  6.608 6.544 6.618 80.78 60.38 99.36 Sym.group:  1<br>1.0<br>  -0.072264   -2.002011   6.297061<br>  -6.020172   2.537594   -0.379164<br>  -3.513667   -5.347421   1.692222<br>  Rb<br>  6<br>Direct<br>  0.254867   0.749624   0.747010<br>  0.747408   0.253157   0.252456<br>  0.081912   0.251977   0.585137<br>  0.920393   0.750802   0.414454<br>  0.414087   0.248951   0.920637<br>  0.588054   0.753803   0.079009 |

| Pseudo hexagonal RbH₉ | sc-RbH |
|---|---|
| EA980   4.555   4.571   4.591  118.47  63.64  85.48  Sym.group: 1 | EA2449  4.552  5.570  4.553  89.91  119.99  90.08  Sym.group: 221 |
| 1.0 | 1.0 |
|   4.549334   -0.189192   0.110854 |   4.549053   -0.057274   0.166865 |
|   0.547643    4.537667   0.070703 |   0.062295    5.569551   0.003667 |
|   1.855332   -2.481724   3.387929 |  -2.418738    0.031879   3.857527 |
|  Rb  H |  Rb  H |
|   1   9 |   3   3 |
| Direct | Direct |
|   0.341768   0.129124   0.677775 |   0.458536   0.338173   0.768744 |
|   0.004748   0.887874   0.093643 |   0.792070   0.004702   0.435306 |
|   0.112431   0.604106   0.514775 |   0.125335   0.671278   0.102240 |
|   0.122311   0.864299   0.165327 |   0.126235   0.171844   0.101920 |
|   0.702629   0.601666   0.927550 |   0.792316   0.504855   0.435603 |
|   0.287697   0.442294   0.350239 |   0.459858   0.838648   0.768484 |
|   0.726080   0.552672   0.493857 |   |
|   0.698689   0.441220   0.943105 |   |
|   0.725468   0.006861   0.945939 |   |
|   0.715820   0.401516   0.329733 |   |
| Pseudo tetragonal RbH₉ | RbH₅ (see also Kuzovnikov et al. [28]) |
| EA1   7.569   7.902   4.587   89.66   90.06   90.04   Sym.group: 66 | EA235   3.771   8.812   6.004   89.97   89.99   89.98   Sym.group: 36 |
| 1.0 | 1.0 |
|   7.568477   0.045572   0.049503 |   3.770829   -0.019174   0.001409 |
|  -0.053365   7.901633  -0.053180 |   0.047894    8.812252   0.012039 |
|  -0.035308   0.058006   4.586117 |  -0.001039   -0.004664   6.004393 |
|  Rb  H |  Rb  H |
|   4  36 |   4  20 |
| Direct | Direct |
|   0.000998   0.767223   0.914865 |   0.501142   0.414081   0.648563 |
|   0.500728   0.768199   0.913042 |   0.495946   0.584075   0.149455 |
|   0.999671   0.267186   0.411693 |   0.998117   0.914240   0.649693 |
|   0.500058   0.268156   0.413691 |   0.998102   0.083972   0.148941 |
|   0.751365   0.970134   0.197365 |   0.498868   0.289868   0.105146 |
|   0.250313   0.015714   0.159606 |   0.499187   0.707994   0.605562 |
|   0.749681   0.472829   0.124859 |   0.991949   0.790683   0.108562 |
|   0.249341   0.518914   0.082022 |   0.003369   0.208014   0.605242 |
|   0.543972   0.413346   0.915771 |   0.498972   0.268765   0.229738 |
|   0.249815   0.515590   0.250179 |   0.498607   0.729456   0.729981 |
|   0.750499   0.558812   0.204115 |   0.994594   0.768741   0.232867 |
|   0.546299   0.124793   0.912696 |   0.001108   0.229171   0.729769 |
|   0.045025   0.445746   0.880389 |   0.500165   0.125994   0.832193 |
|   0.250955   0.018476   0.740145 |   0.494258   0.873852   0.333306 |
|   0.750111   0.065809   0.698895 |   0.998058   0.626686   0.834106 |
|   0.749353   0.475851   0.703189 |   0.999869   0.371242   0.331133 |
|   0.750352   0.979803   0.619775 |   0.500252   0.123094   0.693675 |
|   0.750293   0.562759   0.626412 |   0.494412   0.875385   0.195206 |
|   0.455043   0.589504   0.446133 |   0.999031   0.623385   0.696207 |
|   0.250662   0.021706   0.571952 |   0.999302   0.374536   0.192573 |
|   0.750638   0.767773   0.413686 |   0.496060   0.873118   0.973668 |
|   0.953469   0.909413   0.416878 |   0.498804   0.124037   0.473801 |
|   0.251412   0.808374   0.367718 |   0.000329   0.373825   0.972927 |
|   0.251351   0.727103   0.461208 |   0.998640   0.623851   0.474067 |
|   0.751049   0.056857   0.119956 |   |
|   0.749966   0.267802   0.913486 |   |
|   0.249032   0.308524   0.868634 |   |
|   0.249856   0.226647   0.960440 |   |
|   0.955346   0.622670   0.413072 |   |
|   0.547224   0.912539   0.414880 |   |
|   0.955129   0.122692   0.910126 |   |
|   0.250313   0.517391   0.664015 |   |
|   0.952828   0.410718   0.913464 |   |

Row above tables: 0.092736   0.848981   0.399189

|  |  |  |
|---|---|---|
| 0.454204 | 0.089479 | 0.946044 |
| 0.045479 | 0.944699 | 0.383368 |
| 0.547093 | 0.624397 | 0.411807 |
| 0.047364 | 0.588401 | 0.449666 |
| 0.455154 | 0.946758 | 0.378880 |
| 0.047280 | 0.088257 | 0.945013 |
| 0.451806 | 0.447216 | 0.879106 |

| **RbH$_{17}$** | **RbH$_{19}$** |
|---|---|
| EA1158 6.081 4.637 4.642 114.82 96.98 68.91 Sym.group: 1<br>1.0<br>  6.069008   0.309859  -0.231807<br>  1.451278   4.402391   0.114629<br> -0.301550 -2.061061   4.148370<br> Rb  H<br>  1  17<br>Direct<br>  0.969790   0.557097   0.919346<br>  0.414313   0.063006   0.744231<br>  0.488362   0.008314   0.154090<br>  0.347458   0.682437   0.266829<br>  0.393622   0.494295   0.206209<br>  0.164518   0.181051   0.264086<br>  0.474085   0.623113   0.797245<br>  0.555923   0.479511   0.660938<br>  0.063744   0.907025   0.622518<br>  0.410992   0.135723   0.627303<br>  0.746448   0.560240   0.402689<br>  0.167677   0.859269   0.724526<br>  0.053754   0.158910   0.306022<br>  0.824731   0.109931   0.442566<br>  0.758615   0.730691   0.442539<br>  0.591130   0.055962   0.256932<br>  0.748830   0.191654   0.017975<br>  0.749133   0.151752   0.841215 | EA2619 6.196 4.665 4.688 91.25 70.33 72.60 Sym.group: 1<br>1.0<br>  4.642878   0.310835   4.090852<br>  1.485362   4.420956   0.090683<br> -1.740157   0.387687   4.336303<br> Rb  H<br>  1  19<br>Direct<br>  0.337768   0.290500   0.449274<br>  0.431766   0.779247   0.717165<br>  0.178590   0.641878   0.022797<br>  0.940245   0.201342   0.951972<br>  0.451487   0.762489   0.103680<br>  0.605485   0.187383   0.819114<br>  0.817784   0.281919   0.938013<br>  0.544581   0.671463   0.590177<br>  0.842124   0.384039   0.478225<br>  0.944755   0.296335   0.332112<br>  0.167224   0.483132   0.027928<br>  0.562534   0.665251   0.132548<br>  0.498410   0.127980   0.870593<br>  0.771064   0.847804   0.342820<br>  0.846727   0.698268   0.732035<br>  0.211935   0.046595   0.011935<br>  0.835926   0.954305   0.270801<br>  0.891759   0.725037   0.854706<br>  0.122228   0.889300   0.573533<br>  0.137584   0.897781   0.407455 |
| **Rb-sublattice of pseudo tetragonal RbH$_9$ (DFT)** | **Rb-sublattice of pseudo tetragonal RbH$_9$ (Refined, 12.5 GPa)** |
| Rb-sublattice (I4/mmm)<br>1.0<br>    4.4299998283    0.0000000000    0.0000000000<br>    0.0000000000    4.4299998283    0.0000000000<br>    0.0000000000    0.0000000000    6.2899999619<br> Rb<br>  2<br>Direct<br>   0.000000000   0.000000000   0.000000000<br>   0.500000000   0.500000000   0.500000000 | Rb-sublattice (Fm-3m)<br>1.0<br>    6.2733101845    0.0000000000    0.0000000000<br>    0.0000000000    6.2733101845    0.0000000000<br>    0.0000000000    0.0000000000    6.2733101845<br> Rb<br>  4<br>Direct<br>   0.000000000   0.000000000   0.000000000<br>   0.000000000   0.500000000   0.500000000<br>   0.500000000   0.000000000   0.500000000<br>   0.500000000   0.500000000   0.000000000 |

**Table S8.** Part of results of an evolutionary search for the most stable structures in the Cs-H system at 30 GPa (37 generations, 80 structures in each). PBE pseudopotentials, USPEX [6-8] and VASP [15-17] codes were used.

| Identification number | Cs | H | Enthalpy, eV/atom | Volume, Å$^3$/atom | Fitness, eV/block | Symmetry group | Comments |
|---|---|---|---|---|---|---|---|
| **115** | 2.0 | 14.0 | -1.5832 | 6.2246 | 0.0 | 129 | Tetragonal CsH$_7$ |

| **141** | **2.0** | **2.0** | **1.3991** | **15.6933** | **0.0** | **221** | **CsH** |
|---|---|---|---|---|---|---|---|
| 1044 | 3.0 | 9.0 | -0.6422 | 8.9769 | 0.0 | 38 | $CsH_3$ |
| 1519 | 48.0 | 0.0 | 6.1806 | 25.5348 | 0.0 | 194 | Cs |
| 2122 | 4.0 | 36.0 | -1.7628 | 5.7009 | 0.0 | 66 | $CsH_9$ |
| 3897 | 0.0 | 16.0 | -2.4301 | 3.5663 | 0.0 | 7 | hydrogen |
| 4274 | 3.0 | 5.0 | 0.3568 | 11.8975 | 0.0 | 71 | |
| 3486 | 1.0 | 3.0 | -0.6411 | 8.9967 | 0.001 | 38 | |
| 3702 | 2.0 | 6.0 | -0.6411 | 8.9372 | 0.0011 | 15 | |
| 3826 | 0.0 | 8.0 | -2.4285 | 3.5694 | 0.0016 | 2 | |
| 2283 | 2.0 | 18.0 | -1.7609 | 5.6859 | 0.0019 | 40 | $CsH_9$ |
| 4226 | 0.0 | 16.0 | -2.4281 | 3.5874 | 0.0019 | 1 | |
| **1261** | **1.0** | **17.0** | **-2.0564** | **4.7692** | **0.003** | **1** | **$P1$-$CsH_{17}$ (Prototype for $CsH_{15-17}$)** |
| 884 | 0.0 | 24.0 | -2.4266 | 3.5726 | 0.0035 | 1 | |
| 2107 | 48.0 | 0.0 | 6.1841 | 25.5613 | 0.0035 | 6 | |
| 3564 | 4.0 | 8.0 | 0.0274 | 10.975 | 0.0036 | 2 | |
| 2657 | 3.0 | 9.0 | -0.6384 | 8.9362 | 0.0038 | 38 | |
| 3443 | 3.0 | 9.0 | -0.6383 | 8.9283 | 0.0038 | 38 | |
| 3852 | 4.0 | 36.0 | -1.7585 | 5.6989 | 0.0043 | 62 | $CsH_9$ |
| 125 | 2.0 | 10.0 | -1.2649 | 7.2058 | 0.0046 | 63 | $CsH_5$ |
| 4286 | 48.0 | 0.0 | 6.1856 | 25.532 | 0.005 | 63 | |
| 2906 | 2.0 | 10.0 | -1.2627 | 7.1016 | 0.0068 | 65 | |
| 3385 | 4.0 | 36.0 | -1.7545 | 5.6854 | 0.0083 | 60 | |
| 3332 | 4.0 | 8.0 | 0.0331 | 10.8524 | 0.0093 | 38 | |
| 3482 | 0.0 | 6.0 | -2.4205 | 3.6104 | 0.0096 | 1 | |
| 2991 | 9.0 | 7.0 | 2.0065 | 16.4477 | 0.0097 | 8 | |
| 3355 | 0.0 | 8.0 | -2.4195 | 3.6157 | 0.0106 | 1 | |
| 3071 | 1.0 | 21.0 | -2.116 | 4.5564 | 0.0108 | 1 | $CsH_{21}$ |
| 1295 | 1.0 | 7.0 | -1.5718 | 6.2283 | 0.0113 | 99 | |
| 1647 | 1.0 | 7.0 | -1.5714 | 6.23 | 0.0118 | 8 | |
| 3783 | 1.0 | 13.0 | -1.9416 | 5.0941 | 0.0119 | 44 | $CsH_{13}$ |
| 3028 | 4.0 | 4.0 | 1.412 | 14.9421 | 0.0129 | 194 | |
| 3249 | 3.0 | 5.0 | 0.3699 | 12.0385 | 0.0132 | 1 | |
| 977 | 4.0 | 36.0 | -1.7495 | 5.6933 | 0.0133 | 1 | |

**Table S9.** Crystal structures (POSCARS) of the most important cesium hydrides found by USPEX code at 30 GPa. Without considering ZPE, simple cubic (sc) CsH turns out to be the most stable monohydride.

| Hydrogen | Cesium |
|---|---|
| EA3897 7.494 2.176 3.500 90.40 90.09 90.25 Sym.group: 7 | EA1519 5.220 17.536 13.390 89.97 90.00 90.00 Sym.group: 194 |
| 1.0 | 1.0 |
|   7.487832  0.278273  0.094329 |   5.219956  -0.020597  -0.005887 |
|   -0.090568  2.173902  0.022633 |   0.069061  17.536100  -0.017214 |
|   -0.047007  -0.062920  3.498765 |   0.015205  0.020111  13.389536 |
| H | Cs |
| 16 | 48 |
| Direct | Direct |
|   0.287728  0.334951  0.288769 |   0.046237  0.239697  0.835882 |
|   0.263209  0.049183  0.810622 |   0.546154  0.129310  0.584413 |
|   0.956663  0.546485  0.915147 |   0.046164  0.072924  0.959725 |
|   0.196617  0.834562  0.891154 |   0.546192  0.129393  0.084219 |
|   0.338451  0.548191  0.414049 |   0.546161  0.129222  0.334538 |
|   0.507483  0.333711  0.793247 |   0.046157  0.239602  0.586097 |
|   0.030667  0.050253  0.308598 |   0.546185  0.129464  0.834119 |
|   0.454738  0.549002  0.911792 |   0.046179  0.073092  0.459344 |
|   0.787760  0.336002  0.288410 |   0.546119  0.462644  0.586188 |
|   0.763139  0.051230  0.809789 |   0.046160  0.073072  0.709582 |
|   0.596742  0.836057  0.391004 |   0.546137  0.296051  0.460444 |
|   0.697107  0.837354  0.893091 |   0.046181  0.072974  0.209485 |
|   0.837814  0.551069  0.413796 |   0.046149  0.406420  0.710853 |
|   0.005176  0.335489  0.785690 |   0.546160  0.462637  0.836418 |
|   0.530504  0.053116  0.313432 |   0.046161  0.239646  0.335703 |
|   0.097278  0.840516  0.392386 |   0.046216  0.406209  0.211428 |
| |   0.546190  0.296109  0.710376 |
| |   0.046138  0.406231  0.461314 |
| |   0.046238  0.239744  0.085488 |
| |   0.546262  0.295954  0.960780 |
| |   0.546178  0.462693  0.335779 |
| |   0.046231  0.406397  0.960958 |
| |   0.546210  0.295883  0.210841 |
| |   0.546217  0.462698  0.086022 |
| |   0.046105  0.739605  0.709240 |
| |   0.546086  0.629461  0.960120 |
| |   0.046187  0.572988  0.335083 |
| |   0.546178  0.629251  0.460560 |
| |   0.546075  0.629390  0.710243 |
| |   0.046251  0.739592  0.459105 |
| |   0.546193  0.629312  0.210451 |
| |   0.046085  0.572947  0.835443 |
| |   0.546117  0.962698  0.958709 |
| |   0.046161  0.572878  0.085405 |
| |   0.546276  0.796026  0.334063 |
| |   0.046108  0.573066  0.585110 |
| |   0.046148  0.906314  0.083707 |
| |   0.546211  0.962746  0.208255 |
| |   0.046212  0.739699  0.208682 |
| |   0.046223  0.906355  0.583600 |
| |   0.546210  0.796156  0.583718 |
| |   0.046099  0.906452  0.833314 |
| |   0.046071  0.739717  0.958819 |
| |   0.546066  0.796063  0.833959 |
| |   0.546157  0.962669  0.708847 |
|  | 0.046275  0.906233  0.333966 |
|  | 0.546136  0.795925  0.084320 |
|  | 0.546252  0.962696  0.458386 |
| **Tetragonal CsH$_7$** | **sc-CsH (the best without ZPE)** |
| EA115  4.903 4.903 4.143 90.00 90.00 90.00 Sym.group: 129 | EA141  4.460 4.459 4.461 60.03 60.04 60.02 Sym.group: 221 |
| 1.0 | 1.0 |
|   4.903047   0.003759  -0.017750 |   4.459072  -0.071472  -0.003845 |
|  -0.003891   4.902549  -0.020357 |   2.245700   1.298281  -3.626540 |
|   0.014714   0.017502   4.143135 |   2.289957   3.828654   0.048428 |
|  Cs  H |  Cs  H |
|  2  14 |  2  2 |
| Direct | Direct |
|   0.499030   0.497126   0.497248 |   0.260748   0.240964   0.253652 |
|   0.999755   0.996885   0.497363 |   0.760698   0.740967   0.753643 |
|   0.498867   0.997290   0.205899 |   0.010533   0.989853   0.005106 |
|   0.997317   0.497225   0.786413 |   0.511179   0.492766   0.502780 |
|   0.998799   0.496810   0.598746 |  |
|   0.499824   0.997252   0.393384 |  |
|   0.998547   0.497176   0.244521 |  |
|   0.498868   0.997314   0.749913 |  |
|   0.801112   0.299133   0.032446 |  |
|   0.195277   0.693197   0.032564 |  |
|   0.195397   0.301760   0.033497 |  |
|   0.798774   0.697165   0.029572 |  |
|   0.694291   0.801315   0.959828 |  |
|   0.300049   0.197382   0.964716 |  |
|   0.300067   0.797625   0.964197 |  |
|   0.697331   0.194593   0.962117 |  |
| **CsH$_3$** | **CsH$_9$** |
| EA1044  4.820 4.819 4.907 100.83 79.19 103.23 Sym.group: 38 | EA2122  6.770 7.572 4.451 88.20 90.01 90.00 Sym.group: 66 |
| 1.0 | 1.0 |
|   4.771006   0.545627  -0.414947 |   6.769761   0.030274   0.014460 |
|  -1.632987   4.533845  -0.000211 |  -0.033758   7.571227  -0.057418 |
|   1.391644  -0.478765   4.681186 |  -0.011267   0.173629   4.447604 |
|  Cs  H |  Cs  H |
|  3  9 |  4  36 |
| Direct | Direct |
|   0.352076   0.100105   0.385501 |   0.000789   0.768289   0.915019 |
|   0.018759   0.432903   0.052721 |   0.501283   0.768370   0.914330 |
|   0.687687   0.768881   0.719486 |   0.000799   0.268324   0.413684 |
|   0.858326   0.273894   0.551786 |   0.501018   0.268369   0.413777 |
|   0.372597   0.120315   0.885314 |   0.751328   0.975606   0.210695 |
|   0.521794   0.602473   0.219046 |   0.251082   0.018240   0.164307 |
|   0.697003   0.778003   0.219590 |   0.750740   0.478477   0.112080 |
|   0.850477   0.931976   0.220220 |   0.250799   0.512700   0.079703 |
|   0.040718   0.455403   0.552899 |   0.551224   0.407747   0.912091 |
|   0.522496   0.270694   0.884679 |   0.250956   0.524188   0.249791 |
|   0.190479   0.604777   0.554235 |   0.751212   0.557035   0.215479 |
|   0.192120   0.938881   0.885917 |   0.550160   0.127875   0.915569 |
|  |   0.050408   0.447229   0.874026 |
|  |   0.251205   0.023954   0.746825 |
|  |   0.751220   0.059929   0.717115 |
|  |   0.750504   0.474609   0.710567 |
|  |   0.750905   0.981030   0.614390 |
|  |   0.751025   0.560914   0.617827 |
|  |   0.450911   0.589943   0.454074 |

| | | |
|---|---|---|
| 0.251257 | 0.012664 | 0.576687 |
| 0.750988 | 0.768436 | 0.414511 |
| 0.948983 | 0.907136 | 0.413689 |
| 0.251214 | 0.813770 | 0.371262 |
| 0.251073 | 0.724503 | 0.455312 |
| 0.751064 | 0.061494 | 0.116802 |
| 0.750836 | 0.268608 | 0.913931 |
| 0.250674 | 0.314569 | 0.869639 |
| 0.251024 | 0.225155 | 0.952986 |
| 0.950634 | 0.628348 | 0.416538 |
| 0.553696 | 0.906836 | 0.412638 |
| 0.952342 | 0.127302 | 0.915351 |
| 0.250967 | 0.518765 | 0.663796 |
| 0.949524 | 0.407074 | 0.911790 |
| 0.448914 | 0.089232 | 0.955791 |
| 0.050163 | 0.946696 | 0.375975 |
| 0.552268 | 0.629081 | 0.416037 |
| 0.051885 | 0.589486 | 0.456393 |
| 0.452506 | 0.946661 | 0.375573 |
| 0.053625 | 0.088990 | 0.956244 |
| 0.450369 | 0.447567 | 0.872952 |

## $P1$-CsH$_{17}$

```
EA1261  5.245  4.274  4.291 113.83 83.25 82.95 Sym.group: 1
1.0
  5.244461   0.093315  -0.046358
  0.449703   4.250501   0.030674
  0.571298  -1.831805   3.838183
 Cs  H
 1  17
Direct
  0.000041   0.564876   0.827073
  0.902463   0.939218   0.501504
  0.031597   0.182068   0.192864
  0.304327   0.571275   0.273134
  0.295927   0.384616   0.235014
  0.885859   0.245500   0.207569
  0.463086   0.709831   0.979694
  0.518526   0.545736   0.815580
  0.300773   0.971390   0.656752
  0.048614   0.935299   0.429170
  0.611784   0.581314   0.401780
  0.300479   0.012510   0.844720
  0.260686   0.997743   0.269178
  0.607260   0.186539   0.447686
  0.669616   0.734022   0.407677
  0.481735   0.116538   0.384282
  0.634700   0.146857   0.008088
  0.649555   0.112528   0.821969
```

## CsH$_5$

```
EA125  3.444  4.698  5.742 89.82 90.00 68.53 Sym.group: 63
1.0
  3.443672  -0.041524   0.003026
  1.771939   4.351175  -0.023574
 -0.004711   0.052239   5.742253
 Cs  H
 2  10
Direct
  0.086013   0.828652   0.669139
  0.915519   0.169998   0.168446
  0.207836   0.585408   0.103547
  0.793944   0.414549   0.601585
  0.208422   0.583910   0.237439
  0.793885   0.413751   0.735472
  0.367146   0.265536   0.836038
  0.635270   0.730775   0.337296
  0.371653   0.257234   0.667898
  0.630273   0.741180   0.169315
  0.635046   0.731744   0.001095
  0.365547   0.268033   0.499843
```

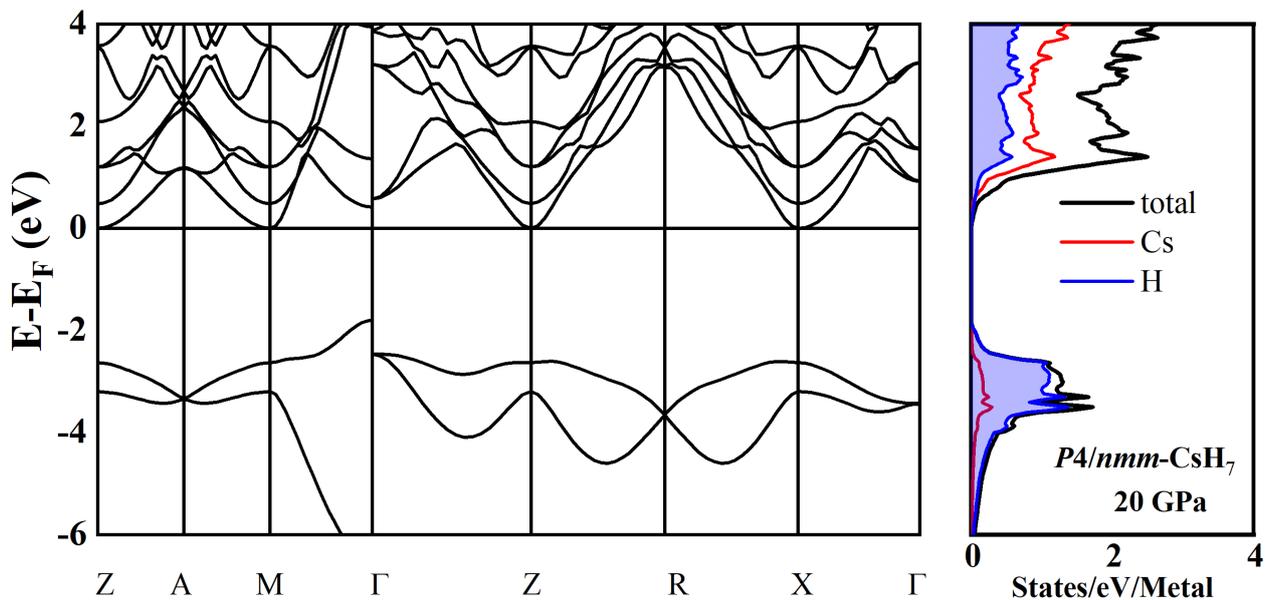

**Figure S41.** Electronic band structure and density of states of *P*4/*nmm*-CsH$_7$ at 20 GPa (0 K). The bandgap is about 1.77 eV (700 nm).

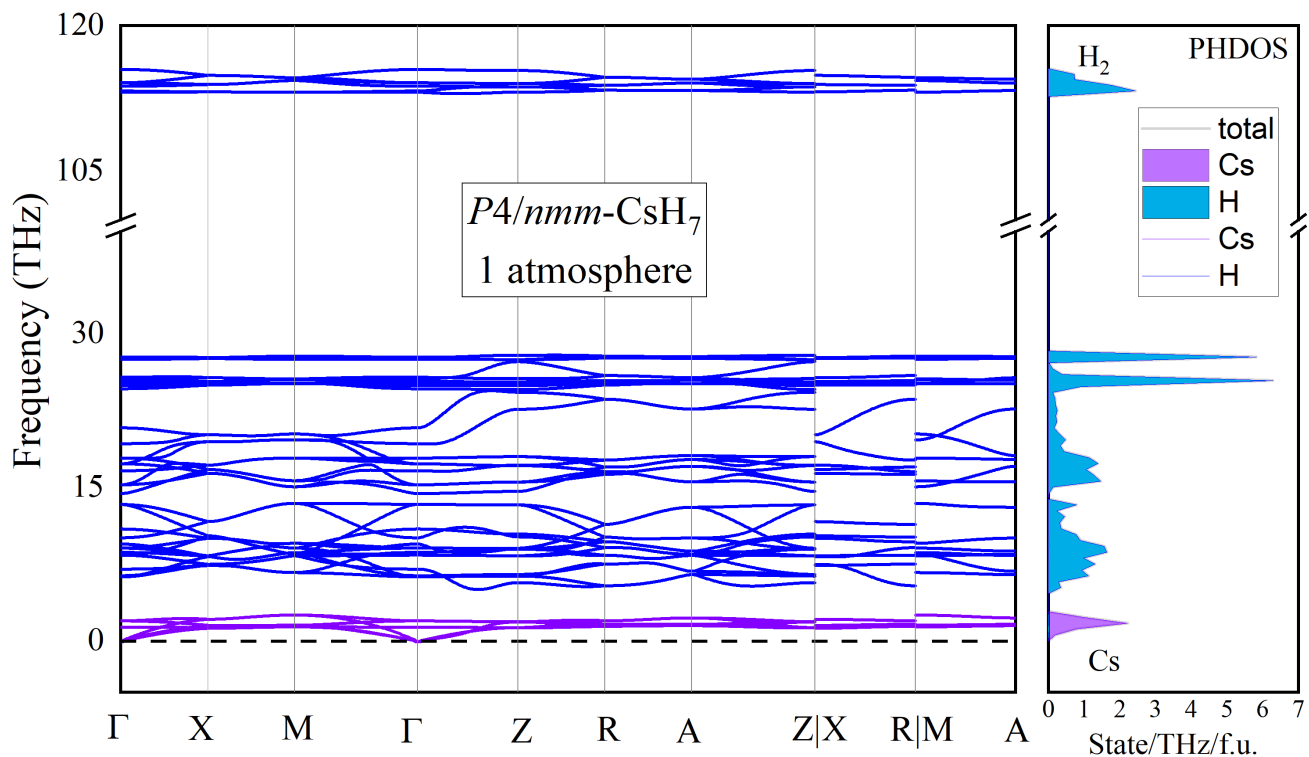

**Figure S42.** Phonon band structure and density of states in *P*4/*nmm*-CsH$_7$ at ambient pressure (0 K) calculated in the harmonic approximation. CsH$_7$ is dynamically stable.

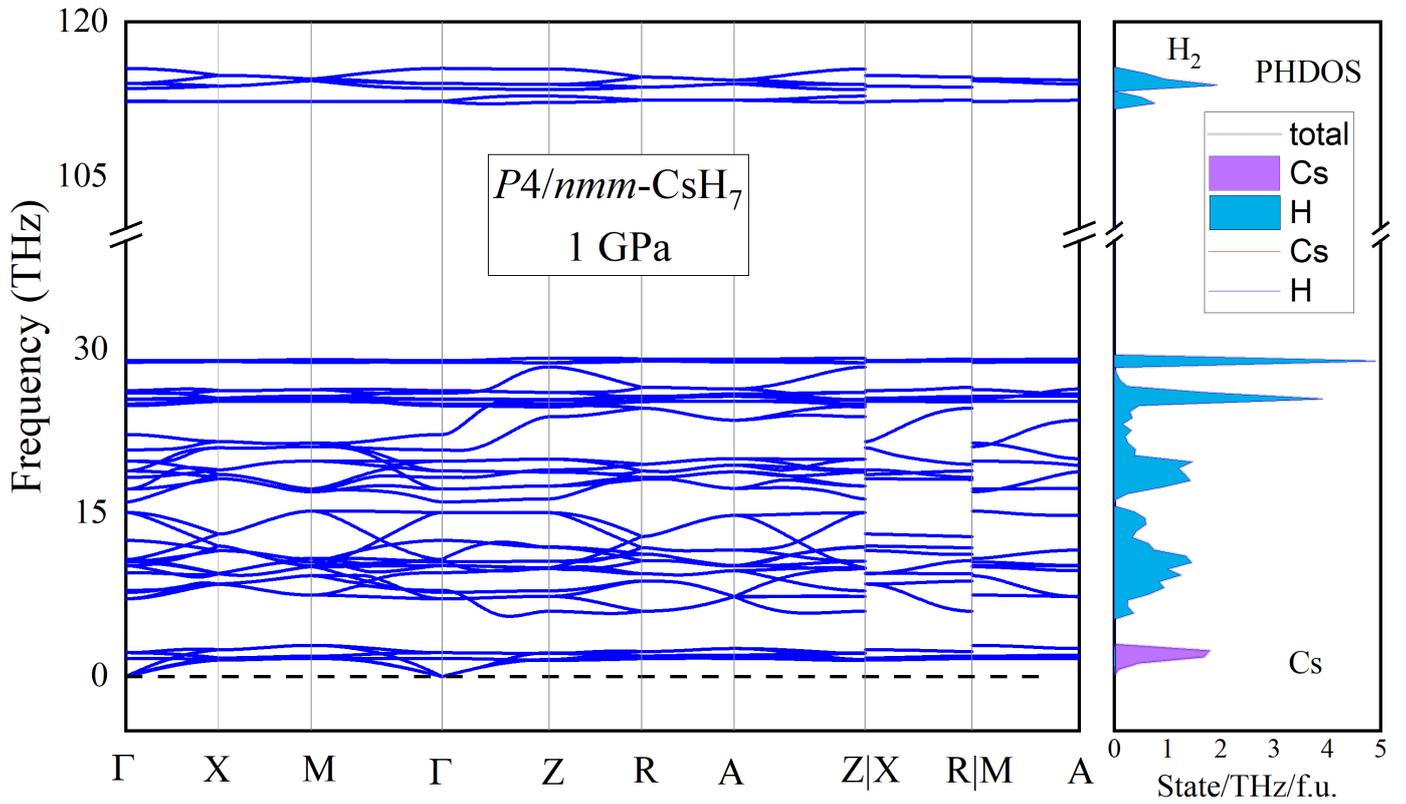

**Figure S43.** Phonon band structure and density of states in *P*4/*nmm*-CsH$_7$ at 1 GPa (0 K) calculated in the harmonic approximation. CsH$_7$ is dynamically stable at 1 GPa. There is a visible splitting of the molecular hydrogen peak.

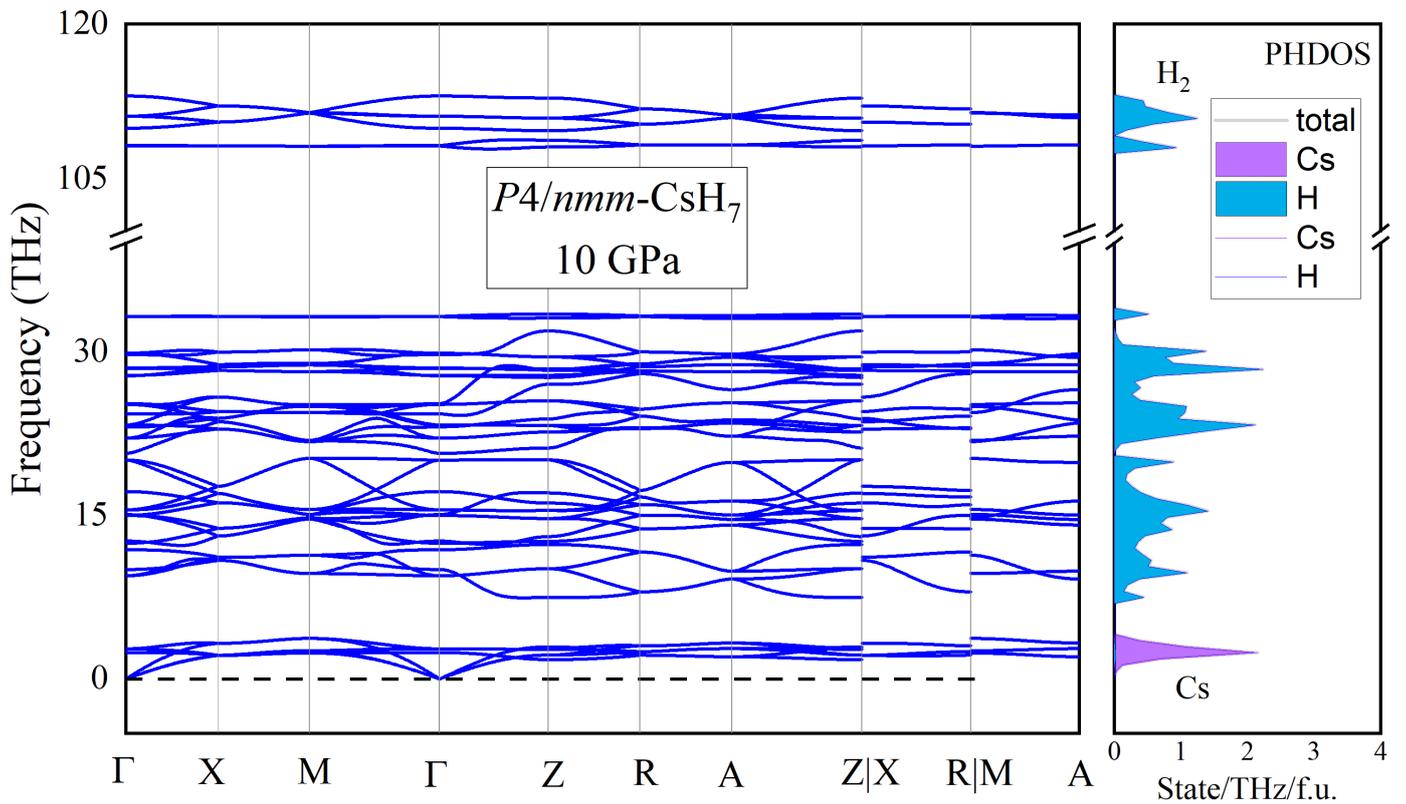

**Figure S44.** Phonon band structure and density of states in *P*4/*nmm*-CsH$_7$ at 10 GPa (0 K) calculated in the harmonic approximation. CsH$_7$ is dynamically stable at 10 GPa. There is a visible splitting of the molecular hydrogen peak.

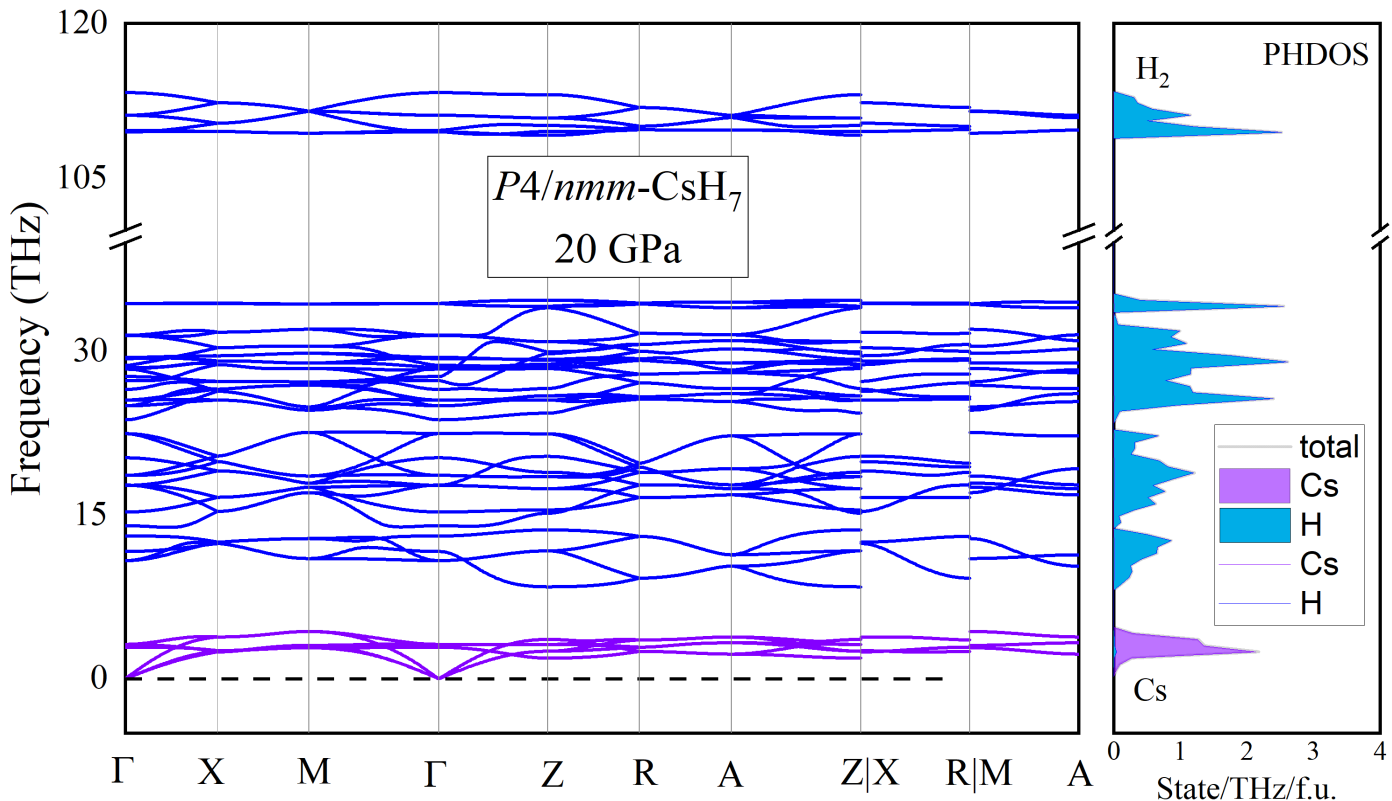

**Figure S45.** Phonon band structure and density of states in $P4/nmm$-CsH$_7$ at 20 GPa (0 K) calculated in the harmonic approximation. CsH$_7$ is dynamically stable at 20 GPa. There is a visible splitting of the molecular hydrogen peak.

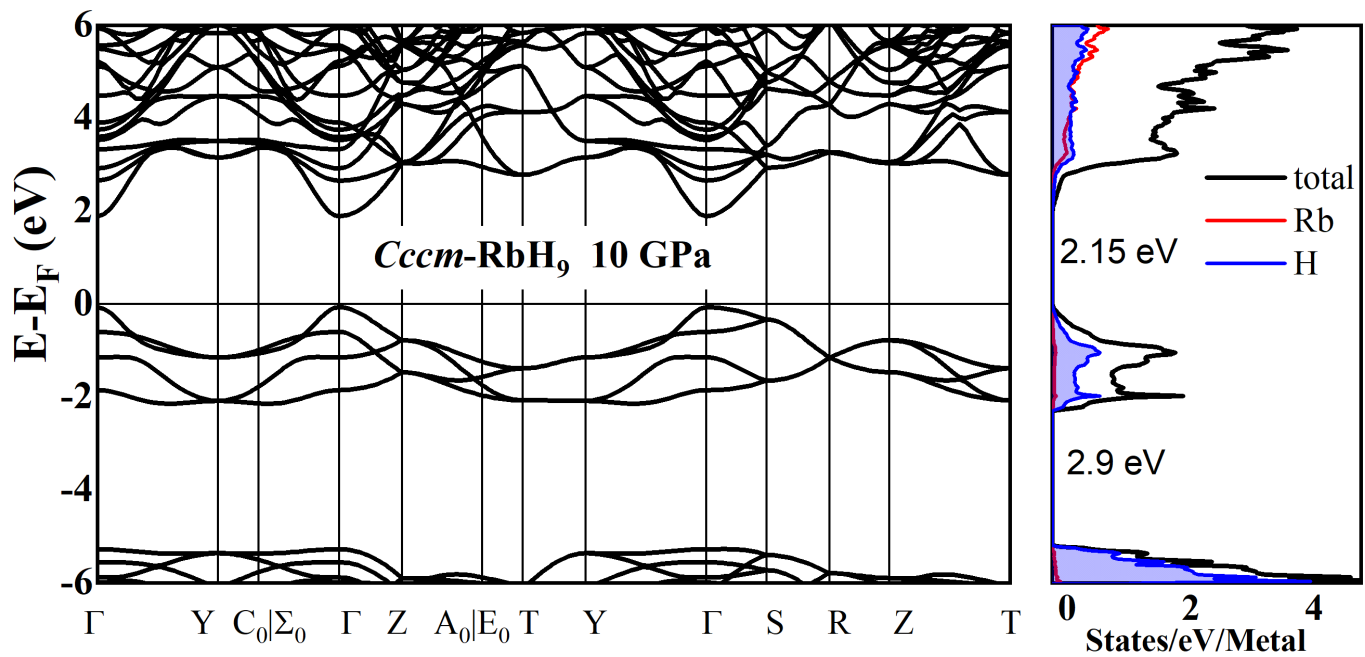

**Figure S46.** Electronic band structure and density of states of pseudo hexagonal RbH$_9$ at 10 GPa (0 K). The bandgap is about 2.15 eV (576 nm). The large difference between total DOS and sum of partial contributions of H and Rb is probably caused by the significant $d$-character of electronic bands above the Fermi energy ($E > E_F$).

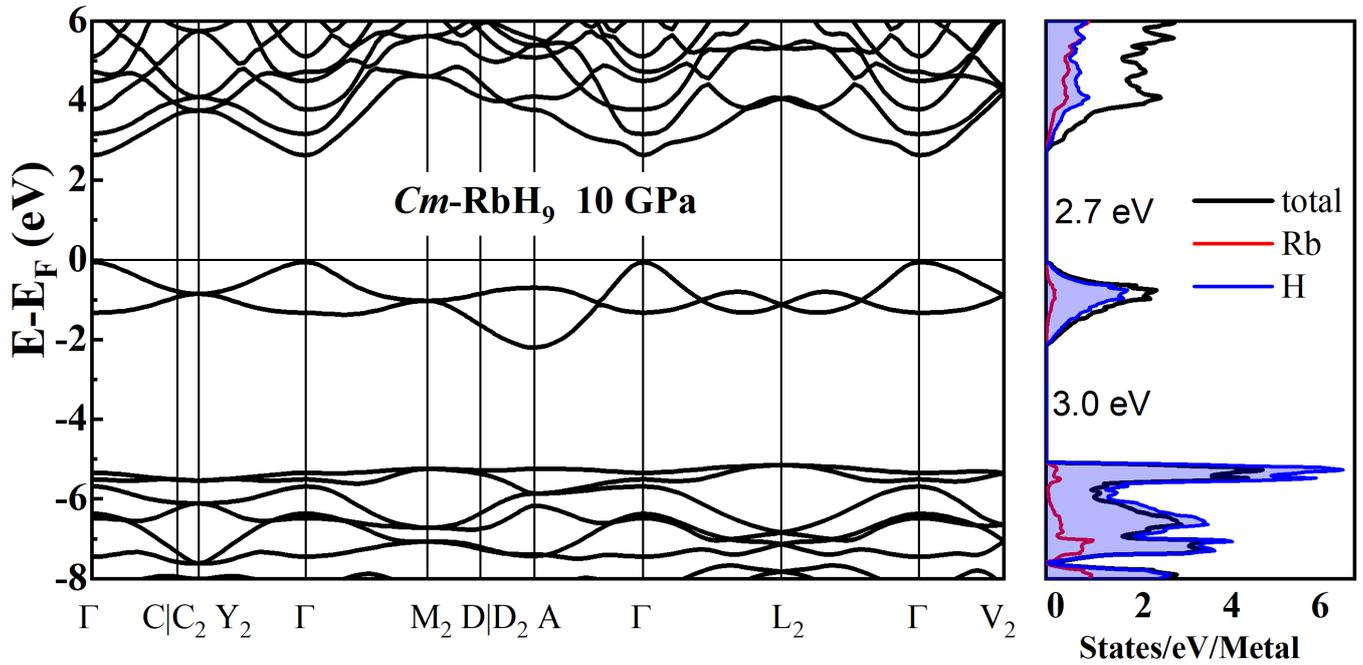

**Figure S47.** Electronic band structure and density of states of pseudo tetragonal RbH$_9$ at 10 GPa (0 K). The bandgap is about 2.7 eV (459 nm). The large difference between total DOS and sum of partial contributions of H and Rb is probably caused by the significant *d*-character of electronic bands above the Fermi energy ($E > E_F$).

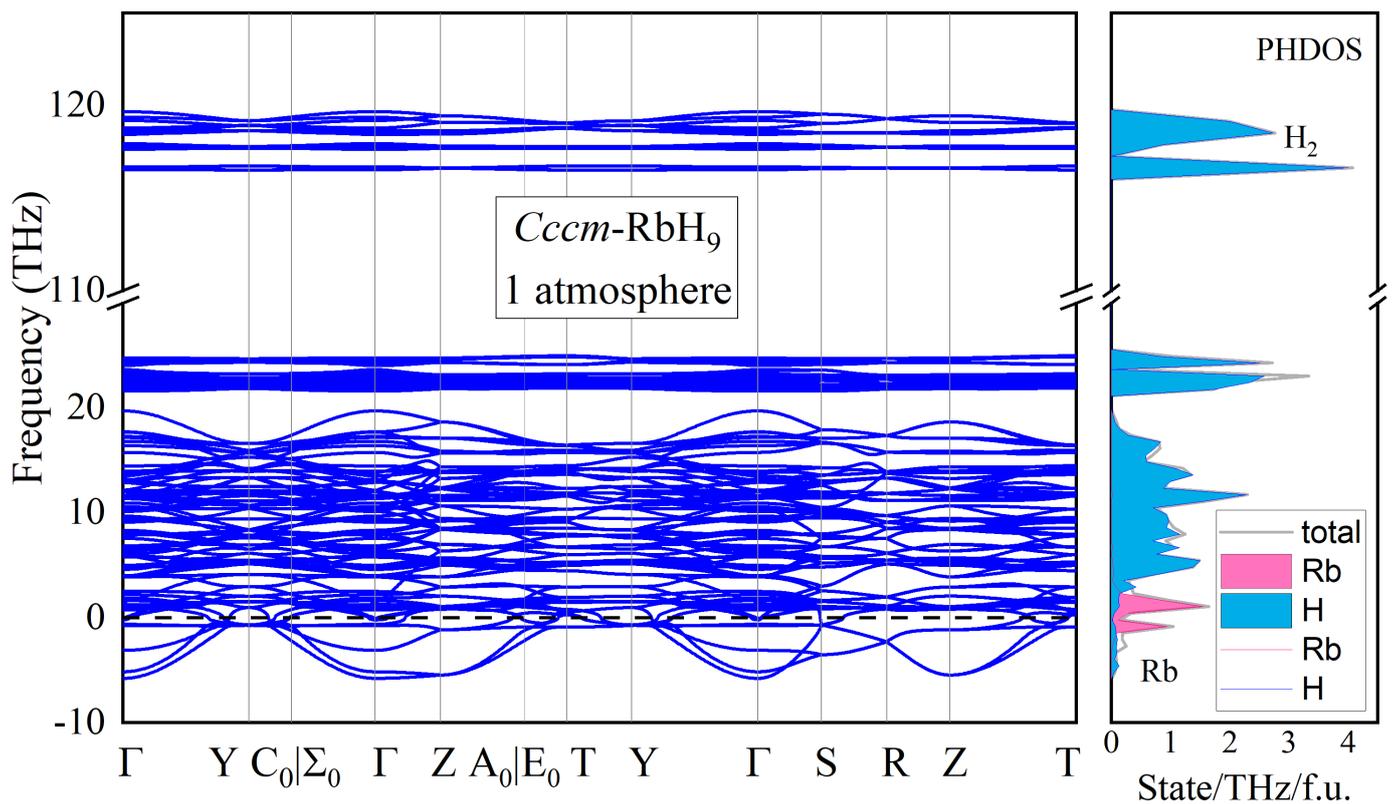

**Figure S48.** Phonon band structure and density of states in pseudo hexagonal RbH$_9$ at ambient pressure (0 K) calculated in the harmonic approximation. The compound is dynamically unstable. The experiment confirms the decomposition of RbH$_9$ below 8 GPa at 300 K.

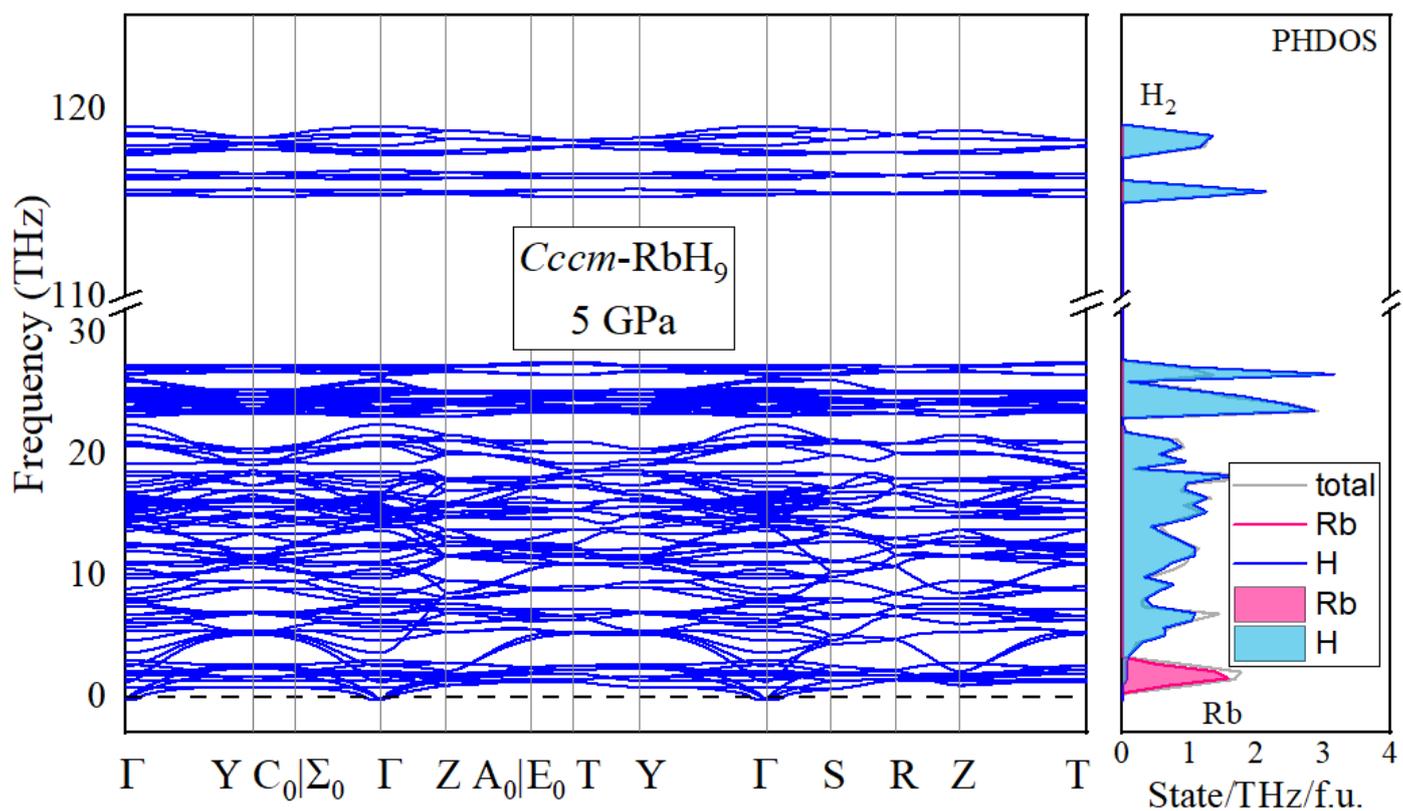

**Figure S49.** Phonon band structure and density of states in pseudo hexagonal RbH$_9$ at 5 GPa (0 K) calculated in the harmonic approximation. The compound is dynamically stable at zero temperature. However, at a finite temperature, the decomposition of RbH$_9$ occurs already below 8 GPa at 300 K.

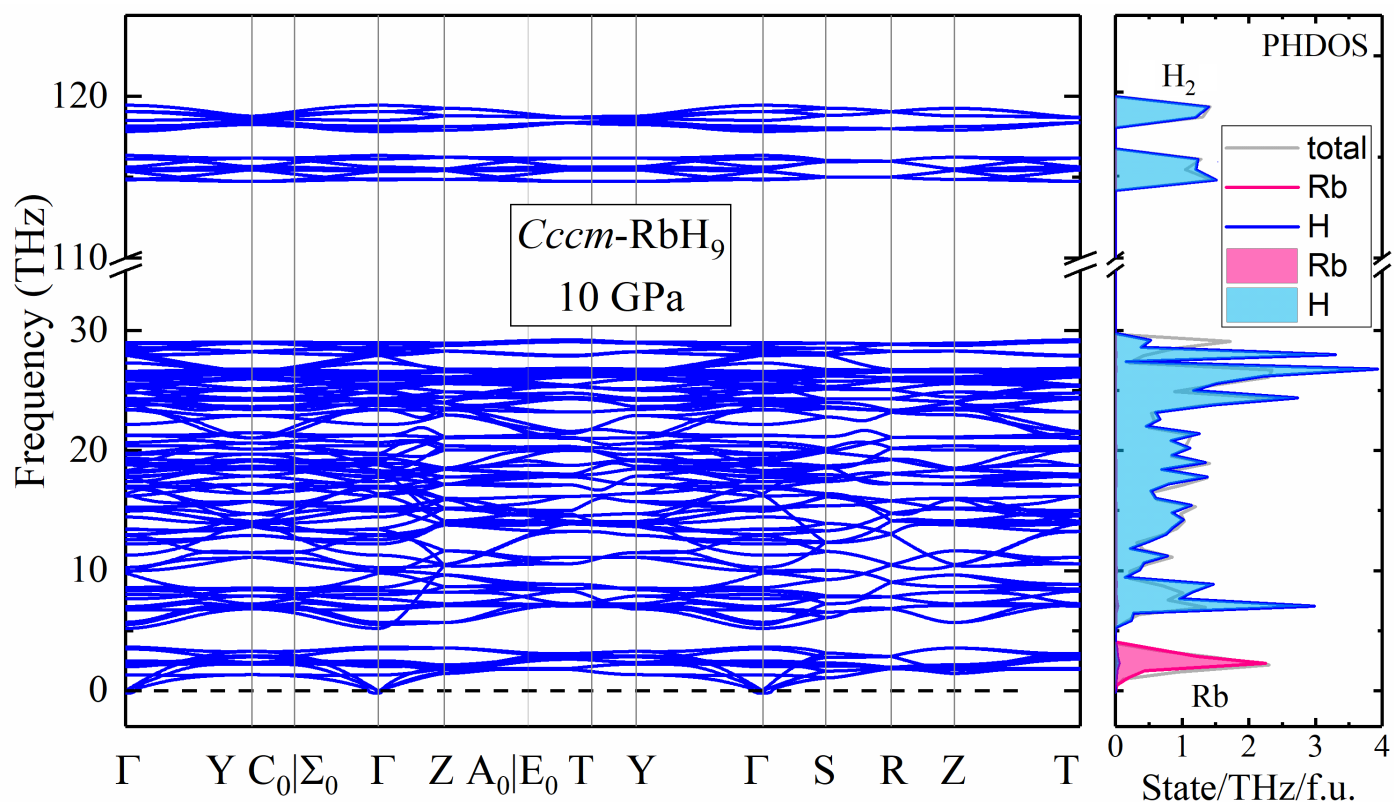

**Figure S50.** Phonon band structure and density of states in pseudo hexagonal RbH$_9$ at 10 GPa (0 K) calculated in the harmonic approximation. RbH$_9$ is dynamically stable at 10 GPa. Optical phonon bands in 115-118 THz range correspond to detected Raman signals at 3800-3900 cm$^{-1}$.

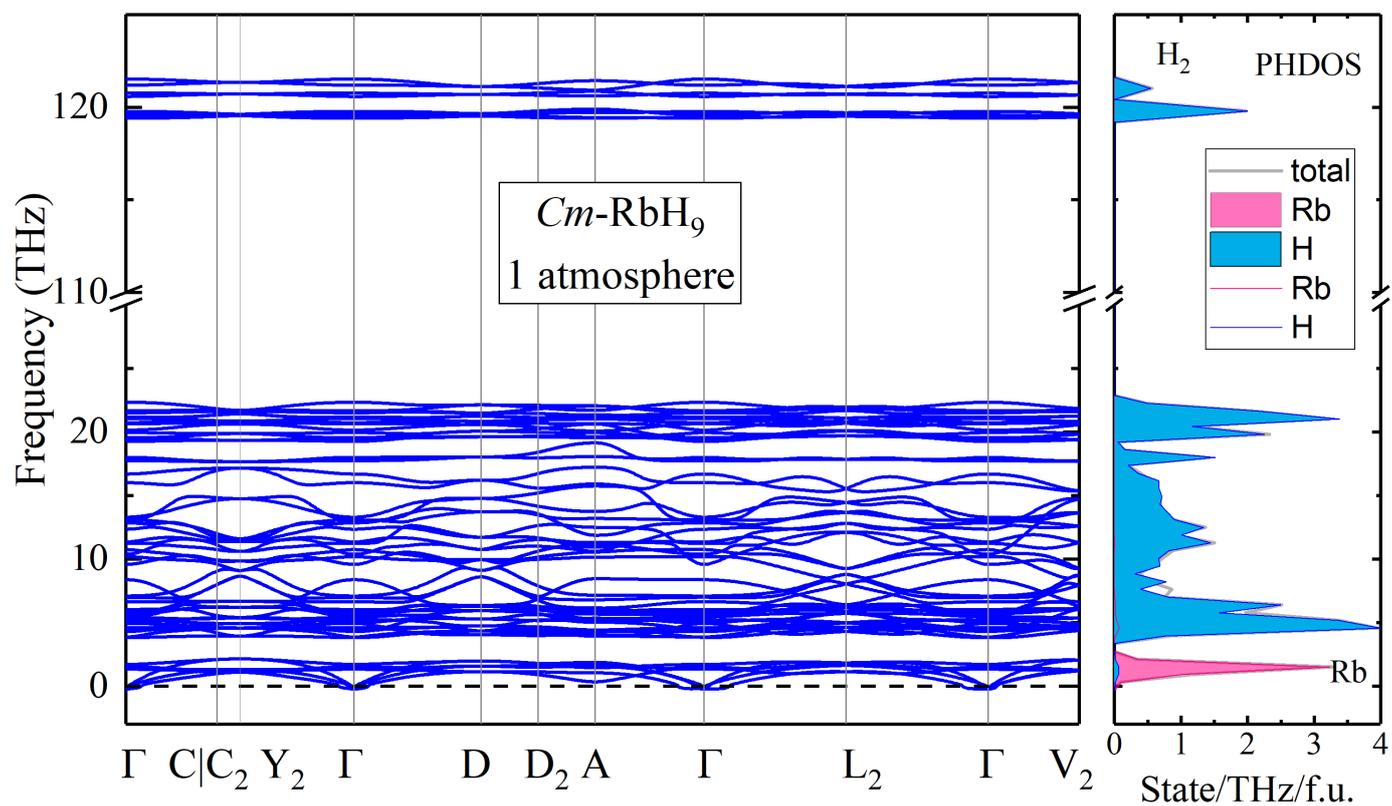

**Figure S51.** Phonon band structure and density of states in pseudo tetragonal RbH$_9$ at ambient pressure (0 K) calculated in the harmonic approximation. The compound is dynamically stable.

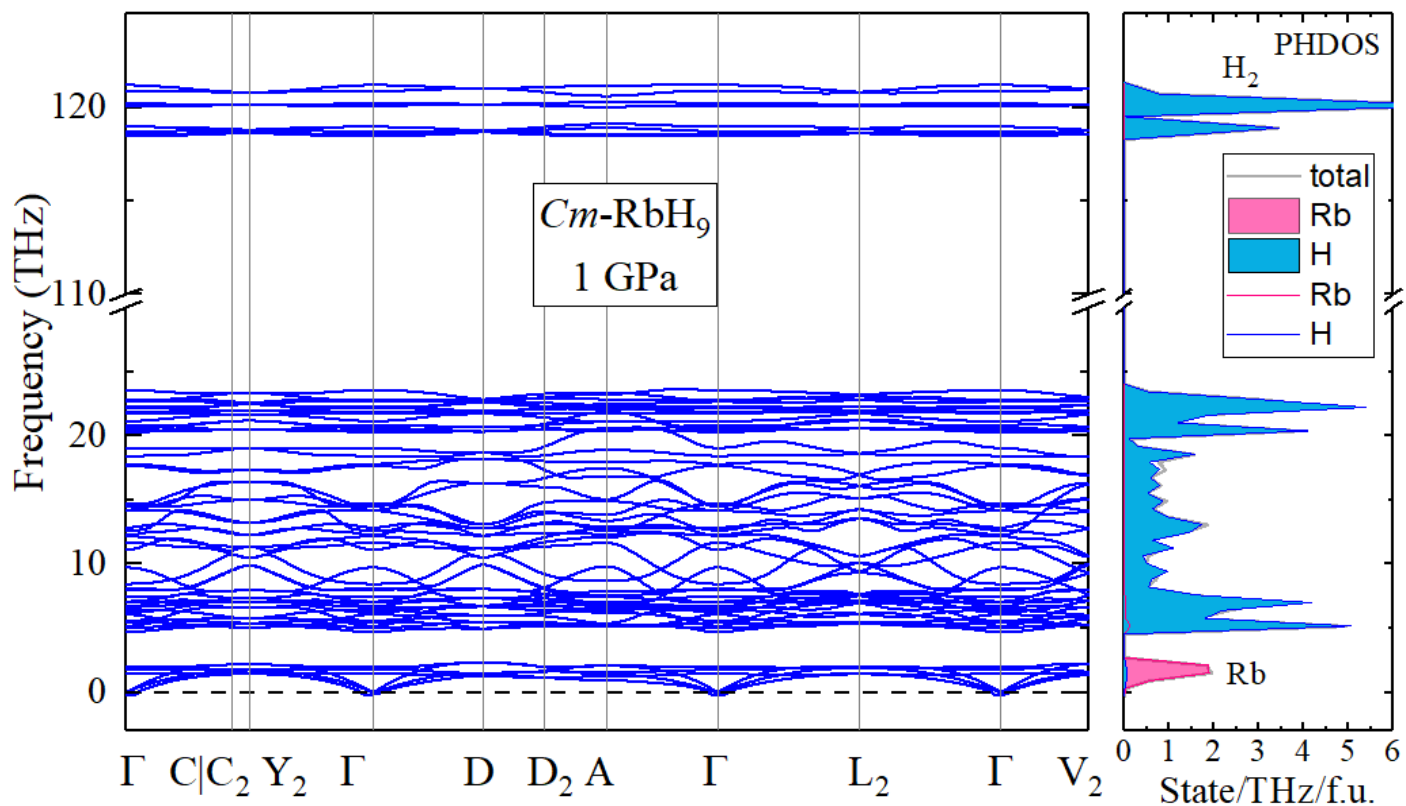

**Figure S52.** Phonon band structure and density of states in pseudo tetragonal RbH$_9$ at 1 GPa (0 K) calculated in the harmonic approximation. The compound is dynamically stable at 1 GPa.

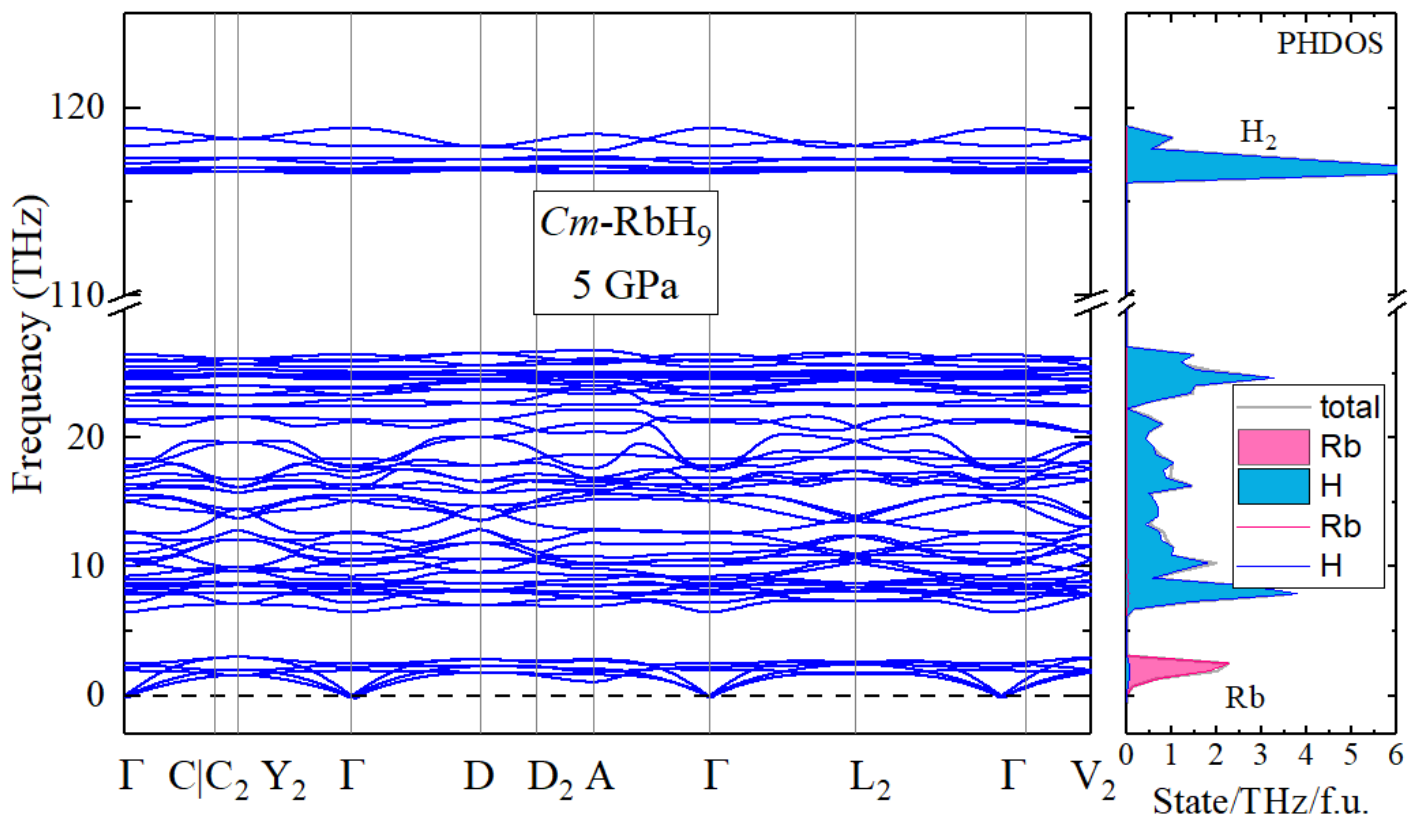

**Figure S53.** Phonon band structure and density of states in pseudo tetragonal RbH$_9$ at 5 GPa (0 K) calculated in the harmonic approximation. The compound is dynamically stable at 5 GPa.

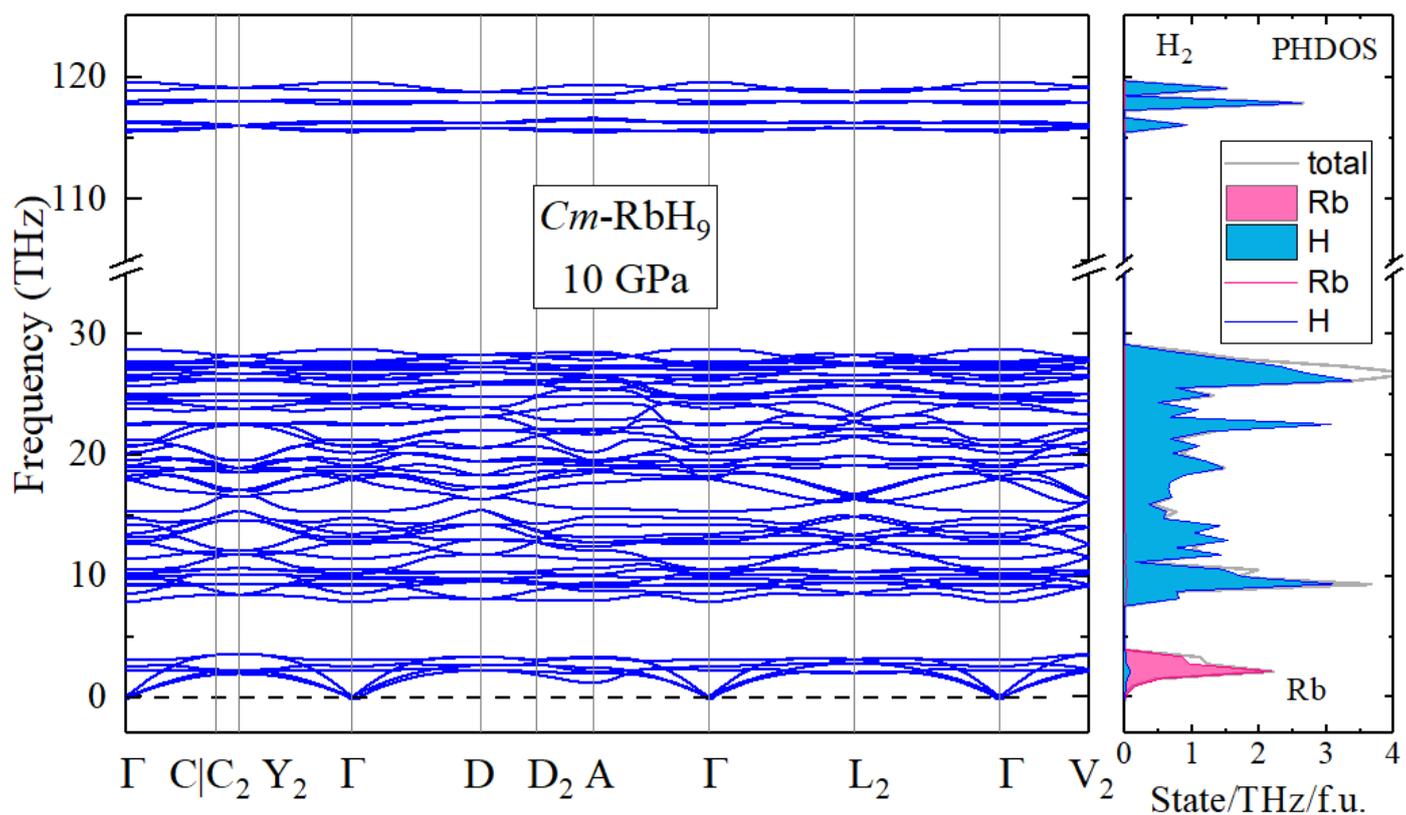

**Figure S54.** Phonon band structure and density of states in pseudo tetragonal RbH$_9$ at 5 GPa (0 K) calculated in the harmonic approximation. The compound is dynamically stable at 5 GPa.

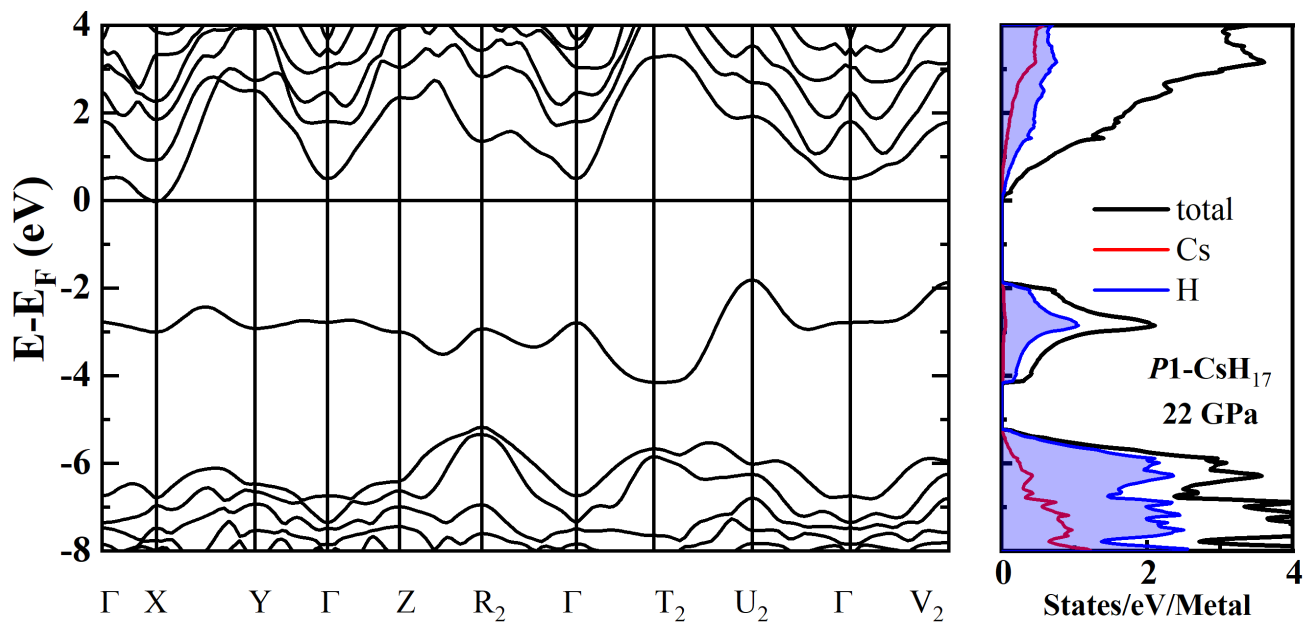

**Figure S55.** Electronic band structure and density of states of $P1$-CsH$_{17}$ at 22 GPa (0 K). This structure is a prototype of CsH$_{15+x}$ found in this work. The band gap is about 2.0 eV (620 nm). It is interesting to note that Cs and Rb hydrides have an electronic band structure containing an intermediate isolated band. Similar band structure is observed for many van der Waals layered metals (TaS$_2$, AlCl$_2$ etc.).

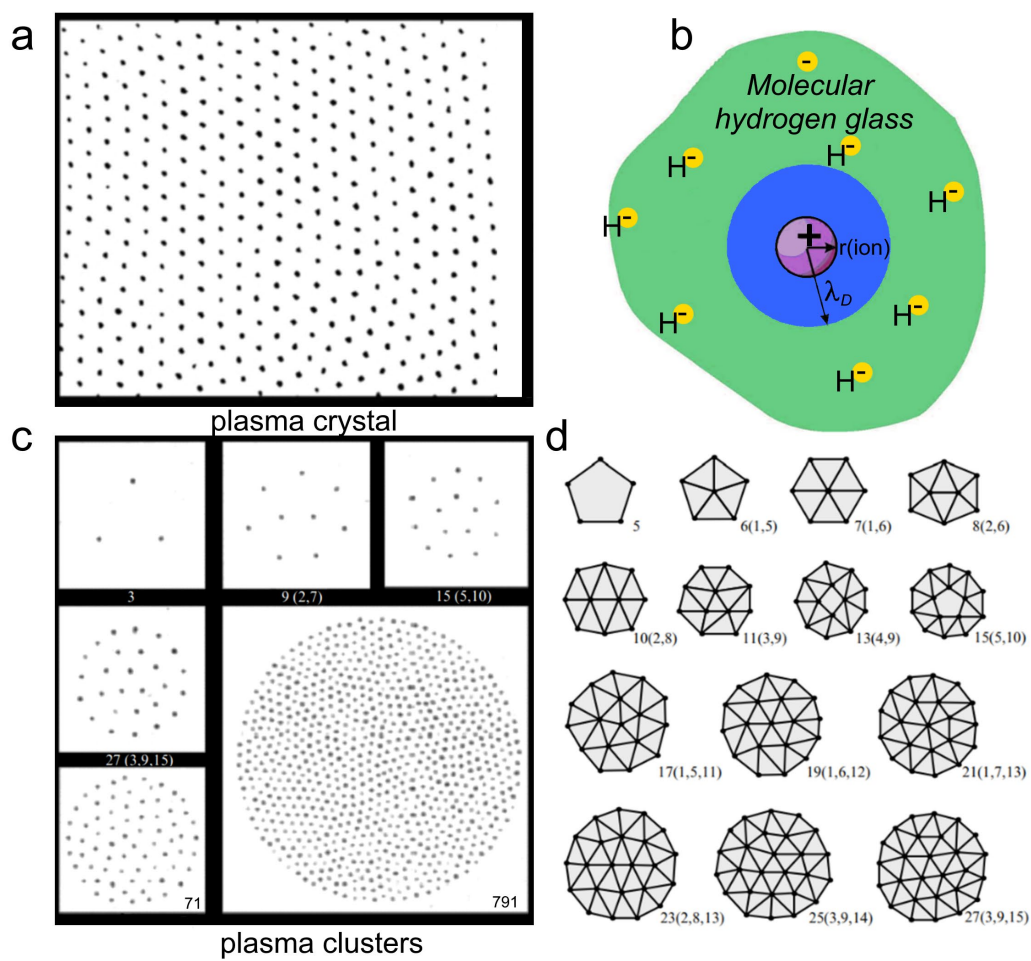

**Figure S56.** Plasma crystals and clusters obtained in microgravity conditions on the International Space Station [34]. This gives us a theoretical analogy for the behavior and structure of "hydrogen sponge" type compounds: CsH$_{15+x}$, SrH$_{22}$ etc. (a) Plasma crystal. (b) Debye sphere of hydrogen ions and polarized H$_2$ molecules surrounding large ions of alkali and alkaline earth metals (e.g., Cs$^+$) in molecular polyhydrides with glass-like H-sublattice. (c) Dust cluster structures consisting of different numbers of particles. Typical interparticle spacing is 300-700 μm. (d) Typical shell configurations of dust clusters composed of different numbers of particles.

There is an analogy between dusty plasma and the structure of compressed "hydrogen sponges" ($XH_n$, where n > 12):

(1) there is a large difference in the masses of the dust particle and ions that make up the plasma, just like the difference in mass between hydrogen and heavy alkali atoms that reaches 100 or more.

(2) Due to the large difference in electronegativity, charge transfer from heavy atoms (e.g., Cs, Rb, Sr…) to the hydrogen sublattice is usually observed. In a similar way, the charging of initially neutral dust particles in the plasma occurs.

(3) Plasma ions are in a state of active movement, just like hydrogen in higher polyhydrides, due to its high diffusion coefficient, hydrogen always moves from one unit cell to another.

(4) Finally, dust particles form a highly symmetrical "crystal" structure despite the chaotic movement of ions in the plasma, just as many polyhydrides have a highly symmetrical sublattice of heavy atoms, while the hydrogen sublattice is disordered [35].